\documentclass{llncs}
\pdfoutput=1
\usepackage[numbers]{natbib}
\usepackage{amsfonts, amsmath,amscd}
\usepackage{tikz-cd}
\usepackage{algorithm}
\usepackage{algorithmic}
\DeclareRobustCommand{\stirling}{\genfrac\{\}{0pt}{}}
\newcommand*\preffindex[1]{\overline{#1}}
\newtheorem{observation}{Observation}

\title{Differential information theory}
\author{Pieter Adriaans}

\institute{ILLC\\
University of Amsterdam\\
\email{pieter\ @\ pieter\ -\ adriaans\ .\ com}}

\date{}
\begin{document}

\maketitle

\begin{abstract}
This paper presents a new foundational approach to information theory based on the concept of the \emph{information efficiency} of a recursive function, which is defined as the difference between the information in the input and the output. The theory allows us to study \emph{planar representations} of various infinite domains. \emph{Dilation theory} studies the information effects of recursive operations in terms of \emph{topological deformations of the plane}. I show that the well-known \emph{class of finite sets of natural numbers} behaves erratically under such transformations. It is subject to phase transitions that in some cases have a fractal nature. The class is \emph{semi-countable}: there is no intrinsic information theory for this class and there are no efficient methods for systematic search.  

There is a relation between the \emph{information efficiency} of the function and the \emph{time} needed to compute it: a deterministic computational process can destroy information in linear time, but it can only generate information at logarithmic speed. Checking functions for problems in $NP$ are information discarding. Consequently, when we try to solve a decision problem based on an efficiently computable checking function, we need exponential time to reconstruct the information destroyed by such a function. At the end of the paper I sketch a systematic taxonomy for problems in $NP$. 
\end{abstract}
keywords: Information, computation, complexity theory, Information efficient, dilation theory, semi-countable sets, $NP$ completeness.

 \section{Introduction}
 Philosophy of Information is a relatively young discipline that attempts to rethink the foundations of science from the perspective of information theory  \cite{Adri20}.  Despite impressive breakthroughs in the twentieth century, information theory is a continent that is still largely unexplored. Some of the more intriguing white spots on the map are: 
 \begin{enumerate}
 \item The absence of a unified theory of quantitative information measurement. 
   \item The absence of a  theory that explains the interaction between information and computation. 
 \item The need for a set of adequate tools that allows us to give an exact information theoretical analysis of decision problems in the class $NP$. 
  \item Our lack of understanding of the information-theoretical qualities of multidimensional spaces.
  \item Our lack of understanding of the behaviour in the limit  of the mathematical functions we use to measure information. 
 \end{enumerate}
 
In this paper we develop information theory from first principles. First as a purely \emph{descriptive} mathematical theory and consequently as a \emph{generative} theory dealing with physical systems evolving in space and time. \emph{Differential Information Theory} (DIT) studies the increase or decrease of information during computational processes. DIT moves away from algorithmic computing as defined by Turing machines and studies the way information flows through functions on natural numbers.  It is based on two very general constraints: 
\begin{itemize}
\item it studies recursive functions defined on the set of natural numbers and
\item it measures the information content of a natural number in terms of its logarithm:
\begin{equation}\label{LOGINF}
I(x)=\log x
\end{equation}
\end{itemize}
A cornerstone of DIT is the strict separation between two ontological domains: 
\begin{itemize}
\item The mathematical world of \emph{numbers} and \emph{functions}.
\item The physical world of \emph{space} and \emph{time}.
\end{itemize}
DIT studies the interaction between these domains: the creation and destruction of information through computation in space and time. 

An advantage of differential information theory is the fact that recursive functions are defined axiomatically: we can really follow the creation and generation of information through the axioms from the very basic principles. Central is the concept of an \emph{information theory}: an efficiently computable bijection between a data domain and the set of natural numbers. In this way we can estimate the computational complexity of elementary recursive functions such as addition, multiplication and polynomial functions. 

As soon as we allow for efficiently computable bijections to more complex domains, like multi-dimensional datasets or finite subsets of infinite sets, the behaviour of these information theories is non-trivial. In the main body of the paper I investigate efficiently computable planar representations of infinite data sets. These representations are inherently inefficient. The Cantor packing functions, for example, \emph{map} the set of natural numbers to the two dimensional plane, but they do not \emph{characterise} it. Information about the topological structure is lost in the mapping and there exist many other efficiently computable  bijections. As a consequence there is no precise answer to the question: what is the exact amount of information of a point in a discrete multi-dimensional space? 
x
Since Cantor's work, planar representations of infinite data sets play an important role in proof theory, so it is worthwhile to attempt a deeper analysis of their structure. One set that is of particular interest is the set of finite sets of natural numbers, denoted as $\mathfrak{P}(\mathbb{N})$, in contrast to the classical power set of $\mathbb{N}$ which is denoted as $\mathcal{P}(\mathbb{N})$. It is generally assumed that we need the axiom of choice to prove this set to be countable. I show that this set is \emph{semi-countable}\footnote{The term was suggested to me by Peter Van Emde Boas.}: there are various ways to count this set but, dependent on the initial choices we make, the corresponding information measurement generated by the choices vary unboundedly. The proof technique I use is called \emph{dilation theory}\footnote{The term was suggested to me by David Oliver}: the systematic study of the information effects of recursive operations on sets of numbers in terms of topological deformations of the discrete plane. 

The recursive functions generate dilations of the discrete plane of decreasing density and, consequently, increasing information efficiency. For semi-countable sets there is an endless sequence of efficiently computable phase transitions from planar to linear representations: from density one, via intermediate densities, to sparse sets, to linear representations with density zero. Close to the linear representations the sets have a \emph{fractal} structure Since the linear dilations can be embedded in the plane the cycle phase transitions is unbounded. Consequently  for these sets there is no intrinsic measurement theory, no intrinsic definition of density, no intrinsic notions of typicality or neighbourhood. 

This special nature of semi-countable sets is important because several notoriously difficult open problems in mathematics are defined on these domains. One example is factorisation, the other is the $P$ vs. $NP$ problem. The subset sum problem is known to be $NP$-complete. It is defined on finite sets of natural numbers and via efficiently computable transformations the whole class of $NP$ complete problems shares the same semi-countable data structure. This implies that none of the well-known concepts like information, density, neighbourhood or typicality can be used unconditionally in attempts to solve these problems. A proof-theoretical analysis of this class is complex and there are many open questions. I show, however, that there are `easier' problems in $NP$, based on countable domains, that are not $NP$-complete and that are not in $P$. Based on this analysis I give  a taxonomy of problems in $NP$ based on three dimensions: expressiveness of the domain, density of the domain, information efficiency of the checking function (see table \ref{NPTAX} in section \ref{TAXDECPRO}). 

An overview of the structure of the paper: 
\begin{enumerate}
\item Introduction
\item Differential Information Theory
\item Space and Time
\item Dilation Theory
\item Semi-countable Sets
\item The class $NP$
\item A taxonomy for decision problems
\item Conclusion
 \end{enumerate}
Additional material can be found in the appendices: 
 \begin{itemize}
 \item Section \ref{INFSTRUDADO}: The information structure of data domains
 \item Section \ref{A1}:  Computing the bijection between $\mathfrak{P}(\mathbb{N})$ and $\mathbb{N}$ via $\mathbb{N}^2$ using the Cantor pairing function: an example
\item Section  \ref{SCALEFSSPROB}: Scale-free Subset Sum problems
\item Appendix: The entropy and information efficiency of logical operations on bits under maximal entropy of the input
 \end{itemize}
 
\section{Differential Information Theory}
$\mathbb{N}$ is the set of \emph{natural numbers} and $\mathbb{R}$ the set of \emph{real numbers}. Let $|s|$ be the cardinality of the set $s$. We have $|\mathbb{N}|= \aleph_0$ and $|\mathbb{R}|= \aleph_1$. We define: 
\begin{equation}
\mathcal{P}(\mathbb{N}) = \{x| x \subseteq \mathbb{N}\}
\end{equation}
\begin{equation}
\mathfrak{P}(\mathbb{N}) = \{x| x \subseteq \mathbb{N} \wedge |x| \in \mathbb{N}\}
\end{equation}
Here $\mathcal{P}(\mathbb{N})$ is the standard powerset of $\mathbb{N}$ with cardinality $|\mathcal{P}(\mathbb{N})|= \aleph_1$ and $\mathfrak{P}(\mathbb{N})$ is the countable set  of all finite subsets of $\mathbb{N}$ with cardinality $|\mathfrak{P}(\mathbb{N})|= \aleph_0$. 
The set of natural numbers $\mathbb{N} $ is the smallest set that satisfies Peano's axioms:
\begin{enumerate}
\item Zero is a number.
\item If a is a number, the successor of a is a number.
\item Zero is not the successor of a number.
\item Two numbers of which the successors are equal are themselves equal.
\item (induction axiom) If a set K of numbers contains zero and also the successor of every number in K, then every number is in K.
\end{enumerate}

If we use $S(x)$ as the successor of $x$ we can define addition recursively
\begin{align}
&a + 0 = a \\
&a + S(b) = S(a + b)
\end{align}
This allows us to define the increment operation: $x + 1= S(x)$. 

\subsection{Discrete Euclidean Spaces}\label{DISCEUCLSPAC}
\begin{definition}\label{TAXCABDEF}
The \emph{taxicab distance} between two points ${\overline{x}}= (x_1,\dots, x_k)$ and ${\overline{y}}= (y_1, \dots ,y_k)$ in a $k$-dimensional space $\mathbb{N}^k$  is: 
\begin{equation}\label{TAXICAB}
d({\overline{x}}, {\overline{y}}) = |x_1 - y_1 | + \dots + |x_k - y_k | 
\end{equation}
\end{definition}
\begin{definition}
The \emph{volume} of a set  $S \subset \mathbb{N}^k$ is: 
\begin{equation}
 V(S) = |S|
 \end{equation}
\end{definition}
The \emph{origin} $\overline{0}=(0_1, 0_2 , \dots , 0_k)$ of a $k$ dimensional space  $\mathbb{N}^k$ has volume $V(\overline{0})=1$. The concept of a \emph{successor} can be generalised to multidimensional spaces.
\begin{definition}\label{SUCCSPAC}
 The \emph{successor space} of the origin $\overline{0}$ in a $k$-dimensional space is:
\begin{equation}
S_{1}^k= S^k(\overline{0}) = \{\overline{x}| d(\overline{x},\overline{0}) \leq 1\}
\end{equation}
Note that $V(S_1^k) = V(S(\overline{0}))= k+1$. In general: 
\begin{equation}
S_n^k=  \{\overline{x}| d(\overline{0},\overline{x}) \leq n\}
\end{equation}
 The \emph{boundary or increment space} is:
\begin{equation}\label{BOUNDARY}
B(S_n^k)= S_n^k - S_{n-1}^k
\end{equation}
By the definition of the binomium we have: 
\begin{align}\label{VOLBIN}
& V(S_n^k)= {n+k \choose k} = {n+k-1 \choose k} + {n+k-1 \choose k-1} \\
&= V(S_{n-1}^k) + V(B(S_n)) \nonumber \\
& = V(S_{n-1}^k) + V(S_{n-1}^{k-1}) \nonumber 
\end{align}
\end{definition}

For each point $(x_1,\dots,x_k)$ in a $k$-dimensional space there are $k$ different increment operations: \[(x_1 + 1,x_2\dots,x_k), (x_1,x_2 + 1, \dots,x_k),\dots, (x_1,x_2\dots,x_k + 1)\]
\begin{definition}
The set of \emph{direct successors} of point $\overline{x} =(x_1,x_2,\dots,x_k)$ is: 
\[\{(x_1 + 1,x_2\dots,x_k), (x_1,x_2 + 1, \dots,x_k),\dots, (x_1,x_2\dots,x_k + 1))\}\]
We call this set the \emph{successor clique} of the \emph{root point} $\overline{x}$.
\end{definition}
Note that all successor points are located in the boundary space of dimension $k-1$. None of these points are direct successors of each other, and the taxicab distance between any two points in the successor clique is always $2$ via the root point, which is a motivation to study them as a clique. 

Using equation \ref{VOLBIN} a discrete euclidean space of dimension $k$ can be mapped to the natural numbers by a set $k!$ of of unique and isomorphic polynomial functions $\pi^k: \mathbb{N}^k \rightarrow \mathbb{N}$ of the following form: 
\begin{equation}\label{CANTPOLMULTII}
\pi^1(x) =  x
\end{equation}
\begin{equation}\label{CANTPOLMULTIII}
\pi^k(x_1,x_2,\dots,x_k) = V(S^k_{(x_1 + x_2 +\dots + x_k - 1)})) + \pi^{k-1}(x_2,\dots,x_k)
\end{equation}
The two $\pi^2$ equations are known as the Cantor pairing functions:
\begin{equation}\label{CANTPAIRFUNC}
\pi^2(x,y)=  V(S^2_{(x_1 + y - 1)})) + \pi^1(y)= \frac{1}{2}(x+y+1)(x+y)+y
\end{equation}
 
\subsection{The logarithmic function}

The choice for the logarithmic function is motivated by the fact that it characterizes the additive nature of information \emph{exactly}. There is a specific interaction between multiplication and addition, that satisfies our intuition about the behaviour of information. First we want information to be \emph{additive}: if we get two messages that have no mutual information we want the total information of the set of messages to be the sum of the information in the separate messages: 

\begin{definition} [Additivity Constraint:]
\[I(m × n) = I(m) + I(n)\]
\end{definition}

We want higher numbers to contain more information: 

\begin{definition} [Monotonicity Constraint:]
\[I(m) \leq I(m + 1)\]
\end{definition}

We want to have a unit of measurement: 

\begin{definition} [Normalization Constraint:]
\[I(a) = 1\]
\end{definition}

The following theorem is due to R\'{e}nyi \cite{Ren61}.

\begin{theorem}\label{RENYI}
The logarithm is the only mathematical operation that satisfies additivity, monotonicity and normalisation.
\end{theorem} 

 The log operation associated with information works as a \emph{type conversion}. In physics we cannot add seconds to meters, but if we are allowed to \emph{multiply} seconds with meters then we have license to add the \emph{information in our time measurements} to the \emph{information in our distance measurements}. If the number $x$ represents a measurement in unit $u$ in numbers, then the logarithm measures the \emph{information} in that measurement in \emph{bits}.  Exponentiation reverses this conversion: if $f(x)= x \ s$ in \emph{seconds} then $I(f(x))= \log_2 f(x) \ b$ in \emph{bits}, and $f(x)= 2^{I(f(x))} \ s$ in \emph{seconds} again.  Just like in  physics we need to maintain administration of the type conversions in derivations using information theory.

\subsection{Differential information of functions}\label{INFRECFUN}
DIT studies recursive functions defined on natural numbers $\mathbb{N}^+$. We define:
\begin{definition} [Information in Natural numbers]
For all $n \in \mathbb{N}^+$: 
\begin{equation}\label{INFODEF}
I(n) = \log_2 n
\end{equation}
\end{definition}
The information for zero is not defined since we use the logarithm to measure the information.   The information in the successor of zero is $\log_2 1 = 0$. The \emph{Differential Information} of a function is the difference between the amount of information in the input of a function and the amount of information in the output.  We use the shorthand $f(\overline{x})$ for  $f(x_1,x_2,\dots,x_k)$: 

\begin{definition}[Differential Information of a Function]\label{EFFFUNCTION}
Let $f: \mathbb{N}^k \rightarrow \mathbb{N}$  be a function of $k$ variables.  We have:
\begin{itemize} 
\item the \emph{input information} $I(\overline{x})$ and 
\item the \emph{output information}   $I(f(\overline{x}))$. 
\item The differential information of the expression $ f(\overline{x})$ is  
\begin{equation}\Delta(f(\overline{x}))= I(f(\overline{x})) - I(\overline{x})) \ bits \end{equation}
\item A  function $f$ is \emph{information conserving} if $\Delta(f(\overline{x}))=0$ i.e. it contains exactly the amount of information in its input parameters, 
\item it is \emph{information discarding} if  $\Delta(f(\overline{x}))<0$ and 
\item it has \emph{constant information } if  $\Delta(f(\overline{x})) = 0$. 
\item it is \emph{information expanding} if  $\Delta(f(\overline{x}))>0$. 
\end{itemize}
\end{definition}

\subsection{Elementary Arithmetical Operations}
We can easily compute the differential information of elementary arithmetical operations: 
\begin{itemize}  \label{BASICRULES}

\item \emph{Addition of different variables is information discarding}.  In the case of addition we know the total number of times the successor operation has been applied to both elements of the domain: for the number $c$ our input is restricted to the tuples of numbers that satisfy the equation $a + b =c $ $(a,b,c \in \mathbb{N})$. Addition is information discarding for numbers $>2$. This can be computed using equations \ref{LOGINF}:
 \begin{equation}\label{EA1}
 \Delta(x + y)= I(x+y) -I(\{x,y\})= \log (x + y) -(\log x + \log y) < 0 
 \end{equation}
 \item 
The incremental growth of the information in numbers is described by the Taylor series for $\log (x+1)$:   
\begin{equation}\label{MEASINFGROW}
I(x+1) = \log (x + 1) =  \sum_{n=1}^{\infty}(-1)^{n-1}\frac{x^n}{n} 
\end{equation}
This gives for the information efficiency of the increment operation: 
 \begin{align}\label{EAINC}
& \Delta(x + 1)= I(x+1) -I(\{x,1\})=\\
&\log (x + 1) -(\log x + \log 1) =  \log (x + 1) - \log x > 0 \nonumber
 \end{align}
 We have $\lim_{x \rightarrow \infty}\Delta(x+1)=0$. The amount of incremental information generated by a counter slowly goes to zero in the limit. 
 
\item \emph{Addition of the same variable has constant information}. It measures the reduction of information in the input of the function as a constant term:
\begin{equation}\label{EA2}
\Delta(x + x)=  \log 2x -\log x  = \log 2 = 1
\end{equation}

  \item  \emph{Multiplication of different variables is information conserving.} We call this operation \emph{extensive}: it conserves the amount of information but not the structure of the generating sets of numbers. In the case of multiplication  $a \times b =c $ $(a,b,c \in \mathbb{N})$ the set of tuples that satisfy this equation is much smaller than for addition and thus we can say that multiplication carries more information. If $a=b$ is prime (excluding the number $1$) then the equation even identifies the tuple.  Multiplication is information conserving for numbers $>2$: 
\begin{equation}\label{EA3}
 \Delta(x \times y) = I(x \times y) - I(\{x,y\})=\log (x \times y) -(\log x - \log y) = 0 \end{equation} 
 \item \emph{Multiplication by the same variable is information expanding}. It measures the reduction of information in the input of the function as a logarithmic term. For $x>1$ we have:
\begin{equation}\label{EA4}
\Delta(x \times x) = \log (x \times x) -\log x  = \log x > 0
\end{equation}
\end{itemize}

\subsubsection{Information in sets of numbers}
 Equation \ref{EA3} is the basis for our theory of information measurement. The function $I_k: \mathbb{N}^k \rightarrow \mathbb{R}$ defines the information in a finite set of numbers $s \subset (\mathbb{N}^+)^k$ as: 

\begin{equation}\label{INFSET}
I_k(s)=\sum_{i \in s} \log_k i \ bits
\end{equation}
This implies that the logarithm is a \emph{unification operation}. It defines a notion of extensiveness that is independent of the dimension of the space it is applied to, which opens the door for the development of a general theory of information and computation. We can use this to define the standard measurement of the amount of information in a set of numbers in bits with the function $I: \mathfrak{P}(\mathbb{N}^+) \rightarrow \mathbb{R}$ as: 

\begin{equation}\label{INFSETGEN}
I(s)=\sum_{i \in s} \log_b i \ bits
\end{equation}

\subsubsection{Differential information in polynomial functions}
Equations \ref{EA2} and \ref{EA4} give the foundation of a differential information theory  for polynomial functions. With these functions the information in a variable can be inflated to any size at little cost:

\begin{equation}\label{DEFPOLI}
\Delta(cx^k)= \log_2 cx^k - \log_2 x =   \log_2 c + (k-1)\log_2 x   \ bits
\end{equation}
For a polynome $\pi^k$ on $\mathbb{N}^k$ we have: 
\begin{align}\label{DELPOL2}
& \Delta(\pi^k(x_1,x_2,\dots,x_k)) = \\
&\log_2 \pi^k(x_1,x_2,\dots,x_k) - (\log_2 x_1 + \log_2 x_2 + \dots + \log_2 x_k) \nonumber \\
&\log_2 \frac{ \pi^k(x_1,x_2,\dots,x_k) } {x_1  x_2 \dots x_k}  \ bits \nonumber 
\end{align}
This is the information in the ratio of the value of the function $\pi^k$ at point $(x_1,x_2,\dots,x_k)$ and the volume of the $k$-dimensional space up to and including the point itself. Polynomial functions are the first class of an infinite set of recursive functions that produce highly compressible natural numbers on the basis of small amounts of information. These classes are studied in \emph{Descriptive Information Theories} such as Kolmogorov complexity \cite{LiVi19}.

\subsection{Differential Information of  Recursive Functions} 
An advantage for differential information theory is the fact that recursive functions are defined axiomatically: we can really follow the creation and generation of information through the axioms from the very basic principles.  Given definition \ref{EFFFUNCTION} we can construct a theory about the flow of information in computation. 

\subsubsection{Primitive recursive functions}\label{PRIMRECFUNC}
For  primitive recursive functions we follow \cite{Odi16}.  

\begin{itemize}
\item \emph{Composition of functions} is information neutral: 
\begin{align} \label{INFEFFPRIMREC1}
&\Delta(f(g(x))) + \Delta(g(x)) \nonumber \\ 
&= \log f(g(x)) - \log g(x) + \log g(x) - \log x \nonumber \\
&=\log f(g(x)) - \log x 
\end{align}

\item  The \emph{constant function} $z(n)=0$ carries no information $z^{-1}z(n)= \mathbb{N}$.  
\item The differential information of $S(0)$ is not defined. 
\item The \emph{successor function} expands information for values $>1$:
 \begin{equation} \label{INFEFFPRIMREC2}
\Delta(s(x))= I(s(x)) - \log x - \log 1=  \log (x + 1) -\log x = \epsilon > 0
\end{equation}
\item The \emph{projection function} $P^{i}_{n}((x_1,x_2,\dots,x_n) = x_i$, which returns the i-th argument $x_i$,  is information discarding. Note that the combination of the index $i$ and the ordered set $(x_1,x_2,.., x_n)$ already specifies $x_i$ so: 
 \item The \emph{successor function} expands information for values $>1$:
 \begin{equation} \label{INFEFFPRIMREC3} 
 \Delta(P^{i}_{n}(x_1,x_2,.., x_n)=  I(x_i) - \log i - I(x_1,x_2,\dots, x_n) <  0\end{equation}
\item \emph{Substitution.} If $g$ is a function of $m$ arguments, and each of $h_1,\dots,h_m$  is a function of $n$ arguments, then the function $f$:
 \begin{equation} \label{INFEFFPRIMREC4} 
f(x_1,\dots,x_n)=g(h_1(x_1,\dots,x_n),\dots,h_m(x_1,\dots,x_n))
 \end{equation}
is definable by composition from $g$ and $h_1,\dots,h_m$. We write $f=[g \circ h_1,\dots,h_m]$, and in the simple case where $m=1$  and $h_1$  is designated $h$, we write $f(x)=[g \circ h](x)$. Substitution is information neutral: 
 \begin{align} \label{INFEFFPRIMREC5} 
&\Delta( f(x_1,\dots,x_n)) \nonumber \\ 
& =\Delta(g(h_1(x_1,\dots,x_n),\dots,h_m(x_1,\dots,x_n))) =  \nonumber \\
&I(f(x_1,\dots,x_n)) - I(x_1,\dots,x_n)
 \end{align}

Where $I(f(x_1,\dots,x_n))$ is dependent on $\Delta(g)$ and $\Delta(h)$. 
\item \emph{Primitive Recursion.} A function $f$ is definable by primitive recursion from $g$ and $h$ if $f(x,0)= g(x)$ and $f(x,s(y))=h(x,y,f(x,y))$.  Primitive recursion is information neutral: 
 \begin{equation} \label{INFEFFPRIMREC6} 
\Delta(f(x,0))= \Delta(g(x)) = I(g(x) - I(x)
 \end{equation}
which is dependent on $ I(g(x)$ and 
 \begin{equation} \label{INFEFFPRIMREC7} 
\Delta(f(x,s(y)))=\Delta(h(x,y,f(x,y))) = I(h(x,y,f(x,y))) - I(x) - I(y)
 \end{equation}
which is dependent on $I(h(x,y,f(x,y)))$.
 \end{itemize}  
  
 Summarizing: the primitive recursive functions have one information expanding operation, \emph{increment}, one information discarding operation, \emph{choosing}, all the others are information neutral. With the primitive recursive functions we can define everyday mathematical functions like addition, subtraction, multiplication, division, exponentiation etc. 
  
 \subsubsection{General recursive functions}
In order to get full Turing equivalence one must add the $\mu$-operator.  It is defined as follows in \cite{Odi16}:

\begin{definition}\label{DEFMU} For every 2-place function $f(x,y)$ one can define a new function, $g(x) = \mu y[f(x,y)=0 ] $, where $g(x)$ returns the smallest number y such that $f(x,y) = 0.$  
\end{definition}
$\mu$ is a partial function. One way to think about $\mu$ is in terms of an operator that tries to compute in succession all the values $f(x,0)$, $f(x,1)$, $f(x,2)$, ... until for some $m$ $f(x,m)$ returns $0$, in which case such an $m$ is returned. In this interpretation, if $m$ is the first value for which $f(x,m) = 0 $ and thus $g(x) = m$, the expression  $\mu y[f(x,y)=0 ]$ is associated with a routine that performs exactly  $m$ successive test computations of the form $f(x,y)$ before finding $m$. The following theorem holds: 

\begin{theorem}\label{INFEFFRECFUNC}
1) The information efficiency of primitive recursive functions is defined for $\mathbb{N}^+$. 2) The information efficiency of general recursive functions is not defined. 
\end{theorem}
Proof: 1) See results in paragraph \ref{PRIMRECFUNC}: equations \ref{INFEFFPRIMREC1} to  \ref{INFEFFPRIMREC7}.  2) Since the $\mu$-operator is unbounded $m$ can have any value. 
\begin{equation}
\Delta(g(x))= \log m - \log x \ bits
\end{equation}
There is no direct relation between $x$ and $m$. Consequently the descriptive complexity of the operator in the context of primitive recursive functions is unbounded. $\Box$

\subsection{Descriptive Complexity} 
\begin{definition}[Descriptive complexity]\label{INFTHEOR}
Let $A$ be an infinite set of objects. An \emph{information theory} for $A$ is an efficiently computable bijection $f: \mathbb{N} \rightarrow A$. For any $x \in A$ the \emph{descriptive complexity} of $x$ given $f$ is: 
\[I(x|f)= \log_2(f^{-1}(x))\]
\end{definition}

\begin{example}\label{EX1}
The set $\mathbb{Z}$ can easily be mapped onto $\mathbb{N}$ by mapping the even numbers to positive numbers and the uneven to negative numbers:

\[
g(x)=
\begin{cases}
0 &\text{if } x = 0 \\
2x  &\text{if } x > 0 \\
-2x -1  &\text{if } x < 0
\end{cases}
\]
Given the fact the $g$ is an efficiently computable bijection we now have an efficient information theory for $\mathbb{Z}$:

\begin{equation}I(x|g) = \log(g^{-1}(x))\end{equation}

The information efficiency of this function is for positive $x \in \mathbb{Z}$: 
  \[ \Delta(f(x)) = \log_2 g^{-1}(x) - \log_2 x = 1 \] 
  This reflects the fact that by taking the absolute value of  $x \in \mathbb{Z}$ we lose exactly $1$ bit of information about $x$. This bit is coded by the function  $g$ as it multiplies the absolute value of a positive element of $\mathbb{Z}$ by a factor two  and uses the interleaving spaces to code the negative values. There are many other possible informarion theories for $\mathbb{Z}$ and, as we shall see, designing the right information theory for a set is in mnay cases non-trivial.
 \end{example}

 We get a theory about compressible numbers when we study information theories for subsets of $\mathbb{N}$. If $f$ is a surjection then $x$ is an index of the value $f(x)$.  Let A be a subset of the set of natural numbers $\mathbb{N}$. For any $n \in \mathbb{N}$ put $A(n)=\{1,2,\ldots,n\} \cap A$. The \emph{index function} of $A$ is $i_A(j)=n$, where $n=a_j$ the $j$-th element of $A$.  The \emph{compression function} of $A$ is $c_{A}(n)=|A(n)|$. Note that the compression function  is the inverse of the index function: together they define a bijection between $A$ and $\mathbb{N}^+$. We can use these functions to \emph{count} the elements of $A$. The density of a set is defined if in the limit the distance between the index function and the compression function does not fluctuate. 

\begin{definition}
Let A be a subset of the set of natural numbers $\mathbb{N}$ with  $c_{A}(n)$ as compression function. The lower asymptotic density $\underline{d}(A)$ of $A(n)$ in $n$ is defined as: 

\begin{equation}
\underline{d}(A) = \liminf_{n \rightarrow \infty}  \frac{c_{A}(n)}{n} 
\end{equation}

We call a set \emph{dense} if  $\underline{d}(A)  > 0$. The upper asymptotic density  $\overline{d}(A)$ of $A(n)$ in $n$ is defined as: 

\begin{equation}
\overline{d}(A) = \limsup_{n \rightarrow \infty}  \frac{c_{A}(n)}{n} 
\end{equation}

The natural density  $d(A)$ of $A(n)$ in $n$ is defined when both the upper and the lower density exist as: 

\begin{equation}\label{DENSITY}
d(A) = \lim_{n \rightarrow \infty} \frac{c_{A}(n)}{n} 
\end{equation}
\end{definition}

With these definitions we can, for any subset $A$ of any countably infinite set  $\mathcal{A}$, estimate the density based on the density of the index set of $A$. The density of the even numbers is $\frac{1}{2}$. The density of the primes is $0$. Note that $d$ is not a measure. There exist sets $A$ and $B$ such that $d(A)$ and $d(B)$ are defined, but $d(A \cup B)$ and $d(A \cap B)$ are not defined. 
The sequence $ A=\{a_1<a_2<\ldots<a_n<\ldots; n\in\mathbb{N}\}$  has a natural density $\alpha \in [0,1]$ iff  $d(A) = \lim_{n \rightarrow \infty} \frac{n}{c_{A}(n)}$ exists \cite{NIVEN51}. There is a close relation between the density of a computable set and compressibility:

\begin{definition}[Asymptotic decay]\label{ASYMPTOTICDECAY}
When the density of a recursive set $A(n)$ is defined then the asymtotic decay function $f$ is defined by: 
\[\lim_{n\to\infty}\frac{c_{A}(n) f(n)}{n}=1\]
In the limit every $n \in A$ can be compressed by $\log f(n)$ bits: $\log a(n) + \log f(n) \approx \log n$.   
\end{definition}

The decay of the set of primes is $\ln n$ which in the limit gives a compression of $\log \ln n $ bits for each prime.  

\begin{lemma}[Incompressibility of the set $\mathbb{N}$] \label{INCOMPRESSIBILITY}
The density of the set of numbers compressible by more than a constant factor is zero in the limit.
\end{lemma} 
Proof: Suppose $A(n)$ is the set of compressible numbers. If the density is defined, the asymptotic decay function is defined: \[\lim_{n\to\infty}\frac{c_{A}(n) f(n)}{n}=1\] which gives $\log c_{A}(n) + \log f(n) \sim \log n$. Members of $A(n)$ can be compressed by a factor $\log f(n)$. If $\log f(n)$ is not bound by a constant then $f(n)$ is growing strictly monotone which gives:
 \[\lim_{n\to\infty}\frac{c_{A}(n)}{n}=\frac{1}{f(n)}=0\] $\Box$

The set of decimal numbers that start with a $1$ has no natural density in the limit. It fluctuates between $1/9$ and $5/9$ indefinitely, but this set can be easily described by a total function with finite information. The countable subset of the Cantor set also has no defined density. There are also sets that have density $1$ with infinite information: take the set of natural numbers in binary representation and scramble for each $k$ all the numbers with a representation of length $k$. In such a set all the numbers keep the same information content up to one bit, but the set has no finite description.

\section{Space and Time}
In the platonic world of mathematics time does not exist. There is an abstract notion of space and dimensionality represented by sets like $\mathbb{N}^k$  and $\mathbb{R}^k$ but these structures are atemporal. Our experience of time is by nature \emph{local}: the past is forever beyond our reach and the future is presented to us in a sequential mode. As soon as we introduce the concept of time our concept of space also changes in the sense that the notion of locality associated with time is essentially spacial. The phrase ``travelling through" time already suggests this. 

Time in our approach is linear, directed and discrete. It is modeled using the \emph{Successor Function} and its domain is the set of natural numbers $\mathbb{N}$. The successor function sends a natural number to the next one: $S(x) = x+1$. This gives the following elementary units of measurement: 
\begin{itemize}
\item Time - time steps (s)
\item Information - bit (b)
\end{itemize}

We can now develop differential information theory along the lines of classical mechanics. For the moment we will suppose time and space to be discrete. Throughout the rest of this paper we will use the logarithm base $2$ . Our unit of information will be the \emph{bit} with the symbol $b$. In some cases it makes sense to use the natural logarithm $\ln x$ with the exponential function $f(x) = e^x$ as our reference. In this case the unit of measurement is the \emph{gnat}.

Movement through time is deterministic. The average information generated by our measurement of time $I(t)$ is its \emph{information velocity}:
\begin{equation}\label{INFVELAV}
v= \frac{I(t)}{t} \ b/s
\end{equation}  
It is measured in bits per time step, where a natural number codes the time step. The concept of a time step is an abstract notion. In the context of physical systems in the real world we might read it as ``second" in line with the SI-system. The differential information of  $I(t)$ is  (by definition \ref{EFFFUNCTION}): 
\begin{equation}\label{INFGEN}
\Delta(I(t)) = \log I(t) - \log(t) = \log \frac{I(t)}{t}= \log v \ b^2/s
\end{equation}
The \emph{information speed} is $|v|$, the average \emph{information acceleration} is: 

\begin{equation}
a= \frac{v}{t} \ b/s^2
\end{equation}
The \emph{information velocity} of the time is given by the \emph{differential information in bits per time step}:   
  \begin{equation}\label{INFVEL}
 \Delta(I(S(x))))= \log (x + 1) - \log x = \log \frac{x+1}{x}= \log v \  b^2/ s
 \end{equation} 
 
 The \emph{information acceleration} is: 
 
  \begin{equation}\label{INFACC}
 \Delta(I(v(S(x))))=\log v(x+1) - \log v(x)  = \log \frac{t(t+2)}{(t+1)^2}= \log a \ b^2/ s^2
 \end{equation} 
 Important for an understanding of the nature of time in the context of information theory is that the production of information by linear movement in time of a deterministic system \emph{decays logarithmically}.  The decay can be approximated using equation \ref{MEASINFGROW}. 

The introduction of time in our model of the world implies a shift from pure mathematics to \emph{physics} and has some repercussions. We leave the realm of pure recursive functions and enter a universe of \emph{computational systems}: agents that generate information by moving through space. Even in a one-dimensional space such an agent can move in two directions with gives rise to the concept of \emph{uncertainty}: we might not know where the agent will be in some future. With the notion of uncertainty  the concepts of \emph{determinism} and \emph{indeterminism} arise as well as the possibility of studying \emph{probability distributions} over future events. The introduction of time marks a shift from pure \emph{descriptive} theories of information (like Kolmogorov theory where time is irrelevant), to \emph{generative} theories (like Shannon information theory that studies the generation of information by non-deterministic systems). In this context equation \ref{LOGINF} gets a more specific information theoretical interpretation known as the Hartley function\footnote{The historical background of this definition is discussed extensively in \cite{Adri20}. Formula \ref{HARTLEY} is related to Boltzmann's entropy equation, $S=k \log W$, for an ideal gas, that gives, in Planck's words: the logarithmic connection between entropy and probability. Here $S$ is the entropy, $k$ is the Boltzmann constant and $W$ is the number of microstates of the gas that correspond to the observed macrostate of the gas. The formula statistically links the information we have about the observed state, to the information we lack about the specific microstate of the ideal gas. The logarithmic function is necessary to capture the specific \emph{extensive} nature of entropy as a measure which is specified by theorem \ref{RENYI}: if we join two ideal gases with entropy $S$ we get $k \log (W \times W) = 2S$, twice the amount of entropy from the product of the sets of possible microstates $W \times W$. }: 
\begin{equation}\label{HARTLEY}
H_0(A)= \log_2 |A| \ b
\end{equation}
Originally the function is defined as the information that is revealed when one picks a random element from $A$ under uniform distribution \cite{Adri20}. Since DIT is a non-stochastic theory we will assume a more simple modal interpretation: it gives the information generated when we make a choice from a set of $n$ possibilities. 

\subsection{Shannon Information: Monotone Non-deterministic Walks in Discrete Euclidean Space}\label{EUCLIDSPAC}
A walk in a $k$ dimensional space is  a sequence of neighbouring points: 
 \[w=(\overline{s_0},\overline{s_1}, \dots, \overline{ s_{n}})\]
\begin{definition}
A walk is \emph{monotone increasing} if the distance from the origin $\overline{0}$ is increasing with each step.
\end{definition}
Suppose an agent makes a monotone increasing non-deterministic walk $w$  of length $n$ in a discrete space $\mathbb{N}^k$  starting at the origin $\overline{0}$: we have: $d({\overline{0}}, {\overline{s_n}})=n$. Non-deterministic walks in space have their own information dynamics, since they involve agents making \emph{choices}. The amount of information generated per choice is given by the Hartley function in equation \ref{HARTLEY}
The total choice information generated by the walk is: 
\begin{equation}\label{INFWALK}
I_c(w)= n \log k \ b
\end{equation} 

A point $\overline{s}= (s_1, \dots, s_k)$ in a $k$-dimensional discrete space $(\mathbb{N}^{+})^{k}$ defines a $k$ dimensional sub space with \emph{volume}: $V(\overline{s})= s_1 \times \dots \times s_k$. The \emph{information} in the point is computed using equation \ref{INFSET} and is equal to the information in the volume:
\begin{equation}\label{INFVOL}
I(\overline{s})=\sum_{s_i \in \overline{s}} \log s_i = \log \prod_{s_i \in \overline{s}} s_i = \log V(\overline{s}) \ b
\end{equation}
The \emph{distance} of the point from the origin is measured in terms of  the taxicab distance. We'll call this \emph{distance} $d$ measured in \emph{steps}:
\begin{equation}\label{DISTPOINT}
 d(\overline{0},\overline{s}) = d \ s 
 \end{equation} 
  The \emph{Shannon entropy} of a walk from $\overline{0}$ to $\overline{s}$ is: 
\begin{equation}\label{SHANENT}
H(\overline{s})=- \sum_{s_i \in \overline{s}} \frac{s_i}{d}\log \frac{s_i}{d}  \ b/s
\end{equation}
This is essentially an x of equation \ref{INFSET} normalised by the length of a monotone walk from the origin. Note that many walks end in the same location and, consequently, have the same entropy.
 
If we restrict ourselves to \emph{monotone increasing} walks in $\mathbb{N}^2$, then the walks code \emph{bitstrings}. At every point $(x,y)$ in the space the agent has \emph{two} choices. The information generated by the walk is $d\log 2 \ b $. The space of all possible walks of length $d$ is characterised by equation \ref{VOLBIN}: 

\begin{align}\label{VOLBINSHAN}
& V(S_d^2)= {d+2 \choose 2} = {d+2-1 \choose 2} + {d+2-1 \choose 2-1} \\
&= V(S_{d-1}^2) + V(B(S^1_d)) \nonumber \\
& = V(S_{d-1}^2) + V(S_{d-1}^{1}) \nonumber 
\end{align}

The uncertainty area of our agent walking for distance $d$ is the  \emph{boundary} space given by equation \ref{BOUNDARY}: 

\begin{equation}\label{BOUNDARYSHAN}
B(S_d^2)= S_d^2 - S_{d-1}^2 \end{equation}
This gives: 
\begin{equation}
V(B(S_d^2))= V(S_d^2) - V(S_{d-1}^2)= \frac{1}{2}d(d+1) - \frac{1}{2}d(d-1)  = d  
\end{equation} 

This is the size of the set of points that lie on the \emph{counter diagonal}: $\{(x,y)|x + y = d \}$. The entropy of the point $(x,y)$ at the counter-diagonal can be computed using equation \ref{INFVEC}: 
\begin{equation}\label{INFVEC}
H((x,y))= -  \frac{x}{d}\log \frac{x}{d} -   \frac{y}{d}\log \frac{y}{d} \ b/s
\end{equation}
The number of monotone walks ending in $(x,y)$ is: ${d \choose x }$, which gives a density distribution of the counter diagonal $x+y=d$. The corresponding \emph{probability mass function} is: 
\begin{equation}\label{PROBMASS}
f(k,n,p)= Pr(k;n,p) = Pr(X=k) = {n \choose k} p^k(1-p)^k
\end{equation}
In this interpretation the walk is a Bernoulli process where $p$ is the probability that the agent moves in the direction of the $x$-axis, and $1-p$ the probability that it moves in the $y$-axis. This allows us to give a \emph{stochastic} interpretation of equation \ref{INFVEC}. The expected position of the agent after a walk of length $n$ is $(px, (p-1)y)$, where $n=x+y$. 

\subsubsection{Random Walks in Graphs}
Graphs are dimensionless spaces. Let $G= (V,E)$ be a graph, where $V$ is the set of \emph{vertices} and $E \subseteq V^2$ is the set of \emph{edges}.    In a \emph{directed graph} or \emph{digraph} the edges have orientations and are coded by ordered pairs $E \subseteq \{(x,y) | (x,y) \in V^2 \wedge x \neq y\}$. A finite \emph{walk} is: 
\[w=(e_1,e_2, \dots, e_n)\] 
The associated sequence of vertices $(v_0,v_1,\dots,v_{n-1}, v_n)$ has for each pair of vertices $(v_i,v_{i+1})$  an edge $v_i=\{e_i,e_{i+1}\}$ that connects them.  

The \emph{length} of this walk is $\sl{l}(w)=n$.The \emph{degree} of a vertex $v$, noted as $deg(v)$, is the number of vertices that are incident to it.  If each vertex has the same degree  then the  graph is \emph{regular}. 

\begin{lemma}\label{STOCHINF}
The  \emph{choice} information generated by a walk $w$ in a graph, measured in bits, is: 
\begin{equation}\label{INFWALKSTOCH}
I_c(w)= \sum_{i=1 }^{n} \log_k deg(v_i) \ b
\end{equation}
\end{lemma}
Proof: The information generated by a single step in the process starting from vertex $v$ is $\log deg(v)$ by formula \ref{HARTLEY}. We choose a specific edge from a set of $deg(v)$ possibilities.  The total information is given by the  the sum of the information in the individual of choices. $\Box$

Note that this is a generalisation of equation \ref{INFWALK} to spaces with \emph{irregular} dimensions. We can develop an information theory for walks in graphs along the same lines as the one for walks in euclidean spaces in paragraph \ref{EUCLIDSPAC}. 

\subsection{Processes and information}
 
\begin{figure}[ht!]
\centering
\includegraphics[width=120mm]{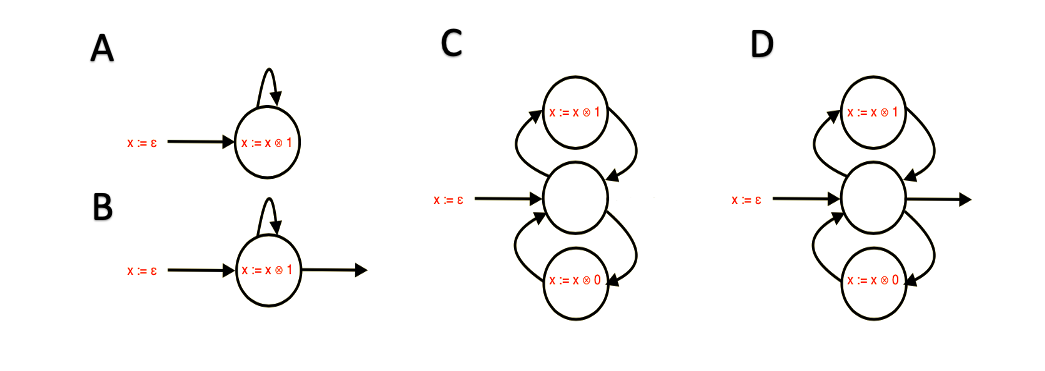}
\caption{Processes that generate binary strings.\label{ELEMCOUNTPROC}}
\end{figure}

In figure \ref{ELEMCOUNTPROC} four basic processes that produce strings are illustrated in terms of their generating finite automata. Time for these automata is discrete and measured as the number of transitions.   

\begin{itemize}
\item Automaton $A$ is \emph{deterministic}. It has no stop criterion and only describes one object: an infinite sequence of ones $1^{\infty}=111111\dots$. This object is \emph{transcendent}. It is not accessible in our world. and its information content cannot be measured.  Differential information theory however can measure the growth of information during its execution as a function of the time: 
\begin{equation}\label{INEFFTIME}
\Delta(A_{t_i,t_j}) = I(output_{t_i})- I(input_{t_j}) = \log_2 {i} -\log{j} 
\end{equation}
At $t_0$ we have the empty word $\varepsilon$ for which the information is undefined. so equation \ref{INEFFTIME} does not apply.  At $t_{1}$ we have the unit string $''1''$ with $\log_2 1= 0$ bits of information.  The information velocity of automaton $A$ is equal to the runtime and is given by equation \ref{INFVELAV}. The production of information by automaton $A$ \emph{decays logarithmically} can be approximated using equation \ref{MEASINFGROW}. 
\item Automaton $B$ is \emph{non-deterministic}. At any stage of its execution it has two choices: 1) write another one or 2) quit. Automaton $B$ describes the set $\{1\}^{\ast}$ that is closed under concatenation. This set is \emph{countable} in a natural manner according to the correspondence:
\[(0, \varepsilon ), (1,1), (2,11), (3,111), (4,1111), \ldots \]
We can estimate the amount of information in the elements of $\{1\}^{\ast}$. A finite sequence of ones of length $k$ is the result of $k$ binary decisions.  The last decision, by necessity, is taken only once. It is a meta decision that does not  generate any information: it is merely the decision to bring the string into existence as an object. The amount of information in the string is given by equation \ref{INEFFTIME} as $\log_2 k$, witnessing the fact that the decision not to end the process was taken $k$ times. 
\item Automaton $C$ is also \emph{non-deterministic} but in a conceptually different way since it has no stop criterion. Instead it generates an infinity of infinite binary strings. We could call the class of objects $\{0,1\}^{\infty}$. All its elements are \emph{transcendent}.  The amount of choice information generated at a certain point of time by $C$ can be computed using equation \ref{INFWALKSTOCH}.This allows us to to apply DIT to compute the growth of information at a certain time interval: 
\begin{equation}\label{INEFFTIMEHART}
\Delta(C_{t_i,t_j}) = I(output_{t_i})- I(input_{t_j}) = i\log_2 {2} -j\log{2}= i - j 
\end{equation}
\item Automaton $D$ again is \emph{non-deterministic} but  it has a stop criterion. The set of strings that it describes is $\{0,1\}^{\ast}$, which is closed under concatenation. The set  $\{0,1\}^{\ast}$ is \emph{countable}  We can define a bijection between  ${\mathbb N}$ and $\{0,1\}^{\ast}$ according to the correspondence:
\[(0, \varepsilon ), (1,0), (2,1), (3,00), (4,01), \ldots \]
We can measure the amount of information in elements of  $\{0,1\}^{\ast}$.  At any stage, except the last step, the automaton makes a choice between two options (write a zero or write a one). When we apply  equation \ref{INFWALKSTOCH} it adds one bit of information at each generative step. The maximum amount of information in a binary string of length $k$ is given by equation \ref{INEFFTIMEHART} as $k$ bits. 
\end{itemize}

 \subsection{Turing Machines}

Algorithmic computation is the sequential manipulation of a finite set of discrete symbols, local in space and time, according to a finite set of fixed rules. There are many abstract systems that incorporate these ideas, but the computational power of all these systems is equivalent. We can restrict ourselves to the study of \emph{Turing machines} that work in discrete one-dimensional space (a tape with cells) and time, using only three symbols: 0, 1 and b (for blank). A Turing machine has a read-write head and a finite set of internal states. The computational behaviour is given by the state transition table. A move of a Turing machine consists of reading a symbol in a cell, overwriting it with a symbol and moving the tape one cell in one or the other direction. 

A \emph{Turing machine} (TM) is described by a 7-tuple \[M=(Q,\Sigma,\Gamma,\mathfrak{D},q_0, B,F)\]
\begin{itemize}
\item Here $Q$ is the finite set of states, 
\item $\Sigma$ is the finite set of input symbols with $\Sigma \subset \Gamma$, 
where $\Gamma$ is the complete set of tape symbols, 
\item $\mathfrak{D}$ is a transition function such that $\mathfrak{D} (q,X)= (p,Y,D)$, if it is defined, where: 
\begin{enumerate} 
\item $q \in Q$ is the current state, 
\item $X \in \Gamma$ is the symbol read in the cell being scanned, 
\item $p \in Q$ is the next state, 
\item $Y \in \Gamma$ is the symbol written in the cell being scanned,  
\item $D \in \{L,R\}$ is the direction of the move, either left or right,
\end{enumerate} 
\item $q_0 \in Q$ is the start state,
\item $B \in \Gamma - \Sigma$ is the blank default symbol on the tape and 
\item $F \subset Q$ is the set of accepting states. 
\end{itemize}
A \emph{move} of a TM is determined by the current content of the cell that is scanned and the state the machine is in. It consists of three parts:
\begin{enumerate}
  \item Change state
  \item Write a tape symbol in the current cell
  \item Move the read-write head to the tape cell on the left or right
\end{enumerate}

A \emph{nondeterministic Turing machine} (NTM) is equal to a deterministic TM with the exception that the range of the transition function consists of sets of triples: \[\mathfrak{D} (q,X)= \{(p_1,Y_1,D_1), (p_2,Y_2,D_2), ... ,(p_k,Y_k,D_k)\}\]

If Turing machine $T_i$ produces output $y$ on input $x$ we write $T_i(x)=y$. Naturally we have: 
\begin{equation}\label{DELTUR}
\Delta(T_i(x))= I(y) - I(x)
\end{equation}
The exact value of equation \ref{DELTUR} depends on our specific measurement theory for binary strings. It is possible to define universal Turing machines that can emulate any other Turing machine: 

\begin{equation}\label{UNIVTUR}
U(\preffindex{i}x)= T_i(x)=y
\end{equation}
Here $\preffindex{i}$ is the self delimiting code for this string $i$  in the format of the universal Turing machine $U$.  A general analysis of the information production by Turing machines is given in \cite{AB2011}.  

\subsection{Turing machines and Differential Information}
Suppose we have a way to measure the descriptive complexity of a working Turing machine during its execution. One could think of the minimal number of bits needed to be transported over a telephone line in order to decide whether the configuration of Turing machine $A$ is equal to that of $B$ at time $t_i$ of their execution. Using this theory we could study the creation and destruction of information during the execution of a program.The basic equation of the theory of computation is: 

\begin{equation}\label{BASEQ}
program(input)=output
\end{equation}

We specify a $program$, that operates on a certain $input$ and after some finite time this combination produces an $output$. The \emph{Information Efficiency} of a program is measured as: 

\begin{equation}\label{INEFF}
\Delta(program(input)) = I(output) - I(input)
\end{equation}

Here $I(x)$ is the information in the object $x$. Note that the complexity of the program is irrelevant for measuring the information efficiency. The following thought experiment might help. 

\begin{example}
Imagine a closed room containing, amongst other objects (a sofa, some books, some random strings), a Turing machine, with a binary string on the input tape at some time t of a computational process. We ask ourselves: what is the difference in descriptive complexity of the total room at time t and at time t+c given the information that the only change in the room is the fact that c computational steps have taken place? We will abstract from issues like the energy that our machine is using during its computations. In this case the descriptive complexity of the invariant part of the room itself, including the specific complexity of the selected Turing machine, is irrelevant. The only things we need to know to estimate the in- or decrease of information in the room are: 
\begin{itemize}
\item The change in the content of the tape
\item The change in the place of the read-write head 
\item The change in the internal state of the machine
\end{itemize}

By taking two snapshots of the room at different times we can eliminate some of the problems of descriptive complexity theory.   We do not need to compute the full descriptive complexity of the room at time $t$, since we subtract this descriptive complexity of the contents that have not changed again at time $t+c$. Without loss of generality we get a pure measurement based on the contents of the tape only if we assume that the read-write head is at the end of the string created so far and the internal state of the machine is the same at the moments we take the measurements.  
\end{example}

In this way our differential measurement is insensitive to a specific Turing machine we use and computable in sufficiently rich and relevant paradigmatic cases.  

\begin{example} [Information generation by means of unary counting]
A unary number is one that exists only of ones or zeros. All a computer needs to know to create such a number is a program that writes a sequence of symbols and a counter that specifies the length of the sequence. The length of a unary number of length $k$ be expressed in $\log_2 k$ bits. Suppose the Turing machine simply writes an infinite string of ones. At $t_0$ the tape is empty. At $t_i$ the tape contains $i$ ones. The naive descriptive complexity at $t_{(0+i)}  = log_2 i$, In general, for this program, the  complexity at time $t_{(i + j)}$ compared to the complexity at $t_i$ is $\log_2 j  - log_2 i$. Note  that the read-write head is always at the end of the generated string, so this is an invariant aspect of the process. Unary counting processes generate information at logarithmic speed. 
\end{example}

\begin{example}(Information destruction by means of erasing strings)
Suppose the tape of the Turing machine at $t_0$ contains a finite  string of length $k$ with maximal information and the program simply overwrites the cells by blanks.  When the program is ready erasing the string there are $k$ bits less information in the total system. 
\end{example}

Important is the insight that deterministic processes can destroy information at relatively high speed and only generate information at exponentially slow speed. The production of information by Turing machines (deterministic or non-deterministic is constrained by the \emph{asymmetry theorem} \ref{ASSYM}. A deterministic machine can only generate information at logarithmically decaying speed, but it can destroy information at linear time. 

\begin{theorem} 
\begin{itemize}\label{ASSYM}
\item The \emph{Maximum Differential Information} generated by  a \emph{deterministic} Turing machine $T(t)$ is $\log_2 t$, where $t$ is the computation time. It is reached when $T$ is the incremental function (i.e. a simple unary counter):
\begin{equation}\label{EXPTIME}
\Delta(T(t))= I(T(t)) - I(T(1)) = \log_2 t
\end{equation}
\item The \emph{Maximum Information} generated by of a \emph{non-deterministic} Turing machine is $t$.
\begin{equation}\label{LINTIME}
\Delta(T(t))= I(T(t)) - I(T(1)) =  t
\end{equation} 
\item The \emph{Maximum Negative Differential Information} discarded by  a \emph{deterministic} Turing machine $-t$: 
\begin{equation}\label{LINNEGTIME}
\Delta(T(t))= I(f(t)) - I(T(1)) =  n - t - n = -t
\end{equation} 
\end{itemize}
\end{theorem}
Proof: 
\begin{itemize}
\item A unary counter is most efficient since it expands the information with each time step. The value at time $t$ is given by equation \ref{INEFFTIME}. 
\item This is achieved if the function is a random bit generator.  The value at time $t$ is given by equation \ref{INEFFTIMEHART}. 
\item The maximum amount of information destroyed is reached when we erase $t$ bits from a maximum entropy string of length $n$ we get a new string of length $n-t$. 
\end{itemize}
$\Box$

An important other result in this context is what one could call the: 

\begin{lemma}[Collapse lemma] \label{COLL}
The descriptive complexity of the output of a deterministic computation after halting is at most one bit higher than the complexity at the start of the process. \end{lemma}
Proof: if we have the description of the system at the start, we only must code one bit of extra information telling us the process has finished. $\Box$

\begin{lemma}
Collorary: The descriptive complexity of deterministic computational processes during their execution is in general higher than after halting.
\end{lemma} 
Proof: immediate consequence of the asymmetry theorem \ref{ASSYM} and the collapse lemma \ref{COLL}. $\Box$

\begin{example}Take a program that writes a unary number (sequence of ones) of length $2^k$. The stop criterium for the program is easily specified in terms of $O(\log_2 k)$ bits, but during execution the program will sequentially generate all incompressible numbers of complexity $k-1$ bits. 
\end{example}

\begin{lemma}\label{MAXINFO}
The maximum amount of information generated by a polynomial time deterministic program during its execution working on input of length $x$ is $c log_2 x$ bits. 
\end{lemma}
Suppose the input of a program $p$ has length $x$ and the computation finishes within $x^c$ steps. Apply theorem \ref{ASSYM}. $\Box$

\subsection{Types of Computational Processes}
 
There are at least three fundamentally different types of computing (See Figure \ref{Computing-Classes}) :

\begin{figure}[ht!]
\centering
\includegraphics[width=70mm]{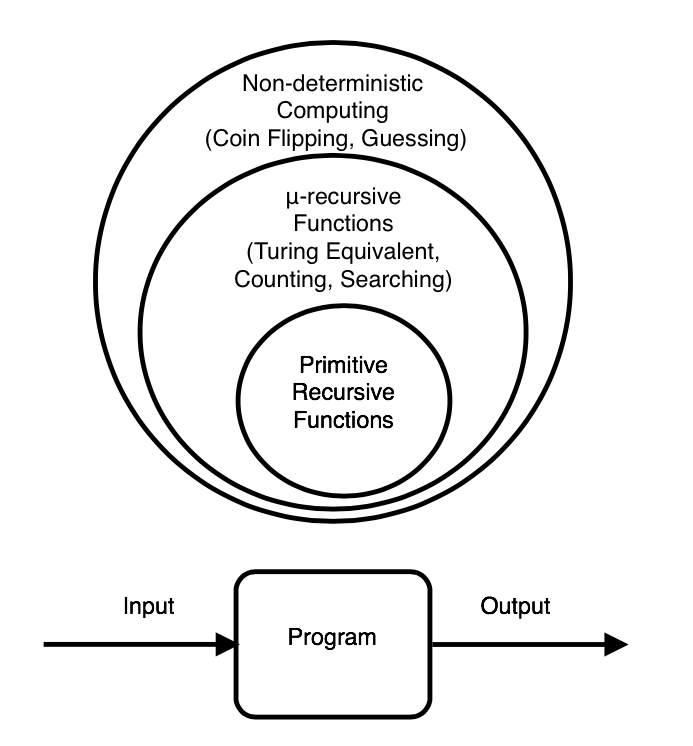}
\caption{Classes of Computing Systems \label{Computing-Classes}}
\end{figure}

\begin{itemize}
\item \emph{Elementary deterministic computing} as embodied in the primitive recursive functions. This kind of computing  does not generate information: the amount of information in the Output is limited by the sum of the descriptive complexity of the Input and the Program.  
\item \emph{Deterministic computing enriched with search} (bounded or unbounded) as embodied by the class of Turing equivalent systems, specifically the $\mu$-recursive functions. This type of computing  generates information at logarithmic speed: the amount of information in the Output is not limited by the sum of the descriptive complexities of the Input and the Program. 
\item \emph{Non-deterministic computing generates} information at linear speed. 
\end{itemize}

There is a subtle difference between systematic search and deterministic construction that is blurred in our current definitions of what computing is. If one considers the three fundamental equivalent theories of computation, Turing machines, $\lambda$-calculus and recursion theory, only the latter defines a clear distinction between construction and search, in terms of the difference between primitive recursive functions and $\mu$-recursive functions. The  set of primitive recursive functions consists of: the  constant function, the successor function, the projection function, composition and primitive recursion. With these we can define everyday mathematical functions like addition, subtraction, multiplication, division, exponentiation etc. In order to get full Turing equivalence one must add the $\mu$-operator. In the world of Turing machines this device coincides with infinite loops associated with undefined variables. The difference between primitive recursion and $\mu$-recursion formally defines the difference between \emph{construction} and \emph{search}. Systematic search involves an \emph{enumeration} of all the elements in the search space together with \emph{checking function} that helps us to decide that we have found what we are looking for (See appendix in paragraph \ref{MUOPRECURSIVE} for a discussion).

\subsection{Kolmogorov Complexity}
When we select a reference prefix-free universal Turing machine $U$ we can define the prefix-free Kolmogorov complexity $K(x)$  of an element  $ x \in \{0,1\}^{\ast}$ as the length $\mathit{l}(p)$ of the smallest prefix-free program $p$ that produces $x$ on $U$:

\begin{definition}\label{KOL}
$K_U(x|y)= \min_{i}\{\mathit{l}(\preffindex{i}):U(\preffindex{i}y)=x\}$ The actual Kolmogorov complexity of a string is defined as the one-part code:
$K(x)= K(x|\varepsilon)$
\end{definition}

 The basic reference for Kolmogorov complexity is \cite{LiVi19}.  It is possible to define a version of Kolmogorov complexity  in the context of differential information theory. 
 Suppose $A$ is the set of recursive functions and we have a computable function $f: \mathbb{N} \rightarrow A$  such that $f(i) = r_i$, where $i$ is the index of the recursive function $r_i$. The amount of descriptive information in a number $x \in \mathbb{N}$ conditional to a number $y$ is:   
\begin{equation}\label{KOLDIFF}
I_f(x|y)= min_i \{ \log_2 i | f(i)=r_i \wedge r_i(y)=x  \}
\end{equation}

The unconditional version  is $I_f(x)= I_f(x|0)$.  This definition has the same problems as standard Kolmogorov complexity: it is uncomputable, but can be approximated. It is also asymptotic, i.e. relative to the enumeration of recursive functions we choose. For discussion see \cite{Adri20}.

\subsection{Typical (random, incompressible) strings and numbers}\label{TYPSET}
 
As a direct consequence of lemma \ref{INCOMPRESSIBILITY} there must be an abundance of numbers that cannot be redefined in terms of recursive functions with more efficient representation. We will call these nubers, \emph{random} or \emph{typical}. When we think of string representations of these numbers their defining characteristic is that they are \emph{incompressible}. A number $x$ is compressible in some context if there exists a representation shorter than $\mathit{l}(x) =  \lceil \log_2 x+1 \rceil$. By the equality $\Sigma_{i=0}^{k} 2^i =2^k - 1$ there is at least one incompressible number in the set of numbers with length $k$, and at most $2^{k-c}$ numbers can be compressed to numbers of length $k-c$. The concepts of compressibility and density of numbers are related. Random strings are an important proof tool since they contain the maximum amount of information, exactly $k$ bits for a string of length $k$. Although we cannot prove that a binary string is random, we can specify some general characteristics: 
\begin{itemize}
\item Kolmogorov complexity tells us that, when we see some overt repeating pattern in the string it cannot be random, since the pattern can be used in a program generating the string. 

\item Shannon information tells us that, if there is an unbalance in the number of zeros and ones in a string, it cannot be random. Apparently, such a string was generated by a system of messages with less than maximal entropy. 
\end{itemize}

A useful concept is the notion of \emph{randomness deficiency}:
 
\begin{definition}
The \emph{randomness deficiency} of a string $x$ is $\delta(x) = \mathit{l}(x) - K(x)$. A string $s$ is \emph{typical} or \emph{random} if $\delta(x) \leq \log \mathit{l}(x)$. A string is \emph{compressible} if it is not typical.  
\end{definition}

\begin{lemma}\label{DENSITYI}
Almost all strings are typical: the density  of the set of compressible strings in the limit is $0$.
\end{lemma}
Proof: The set of finite binary strings is countable. The number of binary strings of length $k$ or less is $\Sigma_{i=0}^k 2^i = 2^{k+1}-1$ so the number of strings of length $k-d$, where $d$ is a constant is at most $2^{k-d+1}-1$. A string $s$ is compressible if $\delta(s) \leq c \log \mathit{l}(s)$, i.e. $K(s) <  \mathit{l}(s) - c \log \mathit{l}(s)$. The density of the number of strings that could function as  a program to compress a string $s$ in the limit is $\lim_{n \rightarrow \infty} (2^{k-c(\log k) +1}-1)/2^k = 0$. Since the upper density is zero, the lower- and natural density are defined and both zero. $\Box$

Lemma \ref{DENSITYI} in the world of strings is the counterpart of lemma \ref{INCOMPRESSIBILITY} in the world of numbers.  By the correspondence between binary strings and numbers these results also hold for natural numbers. The randomness deficiency of a number is $\delta(x) = \log_2 x - K(x)$.  Most numbers are typical, the density of the set of compressible numbers is $0$ in the limit. Note that the existence of an abundance of typical numbers is  established by a simple counting argument. If we measure the amount of information by means of the logarithm then most numbers will be incompressible simply because the number of smaller numbers that can be used to index compressible numbers decays exponentially with their information content. 

 A set is typical if it has no special qualities that distinguish it from most other sets. To pin the concept down, we ask the following question: Given a finite set $s$ of cardinality $|s|=n$, which subsets of $n$ are hard to find? 

 \begin{lemma} \label{TYPICAL}
 The typical  (most information rich) subsets of a set $s$ with $n$ elements are those that consist of a random selection of $\frac{n}{2}$ elements. The amount of information in such a subset approaches $n$ bits. 
\end{lemma} 
Proof: The binomial distribution is symmetrical so we expect that the class of the hardest subsets to be those with $\frac{n}{2}$ elements. We have the following limit: 
\begin{equation}\label{LEXPOW}
\lim_{n \rightarrow \infty}  \frac{\log_2
{
n \choose \frac{n}{2}
}
}{n} = 1
\end{equation}
This is also a hard upper bound since the conditional information in a subset of a set with $n$ elements can be characterised by a binary string of length $n$. $\Box$

\begin{figure}[ht!]
\centering
\includegraphics[width=80mm]{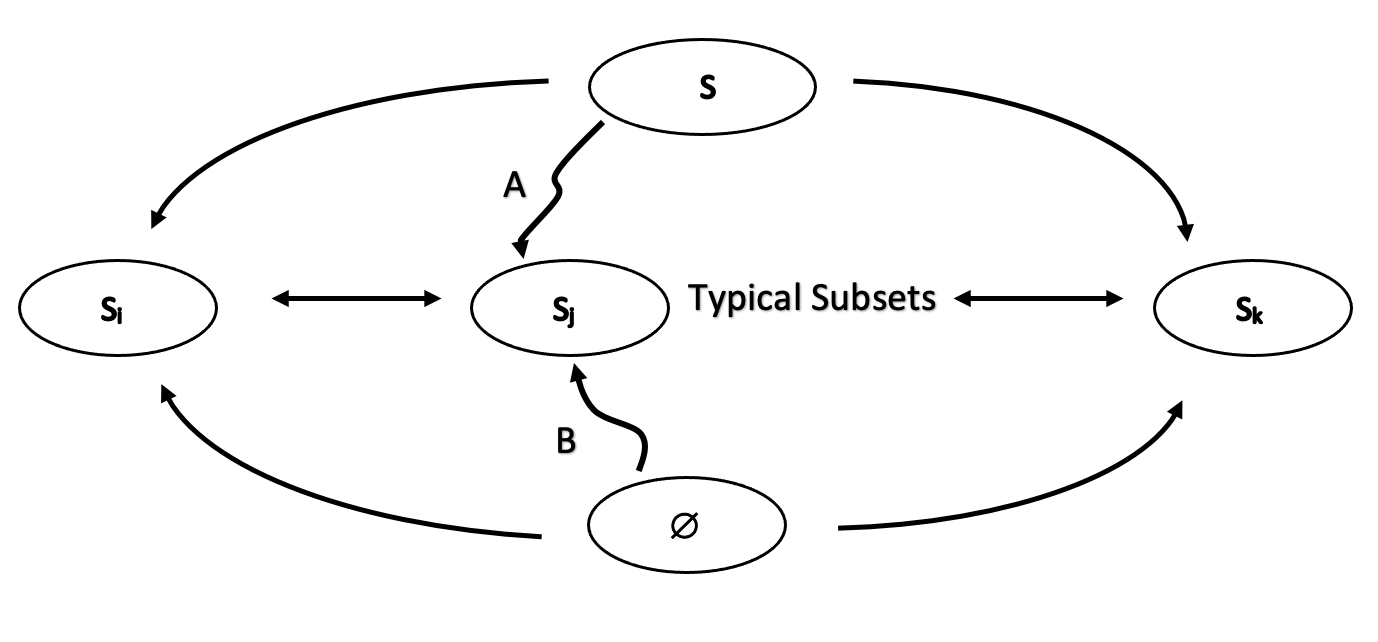}
\caption{Schematic overview of the location of \emph{typical sets} in terms of a cardinalility lattice ordered on the subset relation from top to bottom. Typical sets are placed in the middle of the lattice. In order to move over the lattice from sets with low conditional complexity sets like $S$ (via arrow $A$) and $\emptyset$ (via arrow $B$) to any typical set $S_i$, we have to make at least $n$ binary choices. An efficient deterministic algorithm cannot do this but a non-deterministic algorithm can make the choices in linear time.  \label{LOCTYPSET}}
\end{figure}

The location of typical sets in a lattice of subsets of a finite set $S$ is given in figure \ref{LOCTYPSET}.  On top we find the set $S$ with cardinaltity $n$, at the bottom the empty set $\emptyset$. The conditional information in these sets is constant for all finite sets: $I(S|S) = O(1)$, $I(\emptyset) = 0$. Exactly in the middle we find a dense layer of mostly typical sets with cardinality  $\frac{n}{2}$. The conditional information of a typical set $S' \subset S$ is $I(S'|S) \approx n$ bits. This makes typical sets \emph{inaccessible} for efficient deterministic algorithms:  

\begin{lemma}
There is no deterministic algorithm that generates a typical subset of a set $S$ in time polynomial to the cardinality of $S$. 
\end{lemma}
Proof: Immediate consequence of lemma \ref{TYPICAL} and lemma \ref{MAXINFO}. If the cardinality of $S$ is $n$ and $S'$ is a typical subset of $S$ then $I(S'|S) \approx n$. By lemma \ref{MAXINFO} the maximum amount of information produced by a deterministic algorithm running in time $n^c$ is $c \log n \ll n$. $\Box$ 

Another consequence of this analysis is: 

\begin{theorem}\label{INFNONMONOTONE}
Information is not monotone over set theoretical operations.
\end{theorem}
Proof: Subsets can contain more information than their supersets. Consider the set $S_{<n}$ of all natural numbers smaller than $n$. The descriptive complexity of this set is $O(\log n)$. Create a set $A$ by selecting $\frac{1}{2}n$ elements from $n$ under uniform distribution. This set is a typical subset with $I(A|N)\approx n$, according to definition \ref{CONDINFSUBS} and lemma \ref{TYPICAL}. The same holds for $B = S_{<n} -A$. Yet we have $S_{<n} = A \cup B$, so $I(A \cup B) = O(\log n) \ll I(A) \approx I(B) \approx n$. $\Box$

I make some observations: 

\begin{observation}
Typicality is a structural concept that is independent from the information in the elements of the generating set.
\end{observation}
\begin{observation}
Typical sets can be generated by non-deterministic algorithms in linear time. See figure \ref{LOCTYPSET}. 
\end{observation}

\begin{observation}
The notion of typicality is not defined for finite sets of natural numbers. This is an immediate consequence of the fact that $\mathfrak{P}(\mathbb{N})$ is semi-countable. 
\end{observation}

\section{Dilation Theory }\label{PLANINF}

\begin{figure}[ht!]
\centering
\includegraphics[width=40mm]{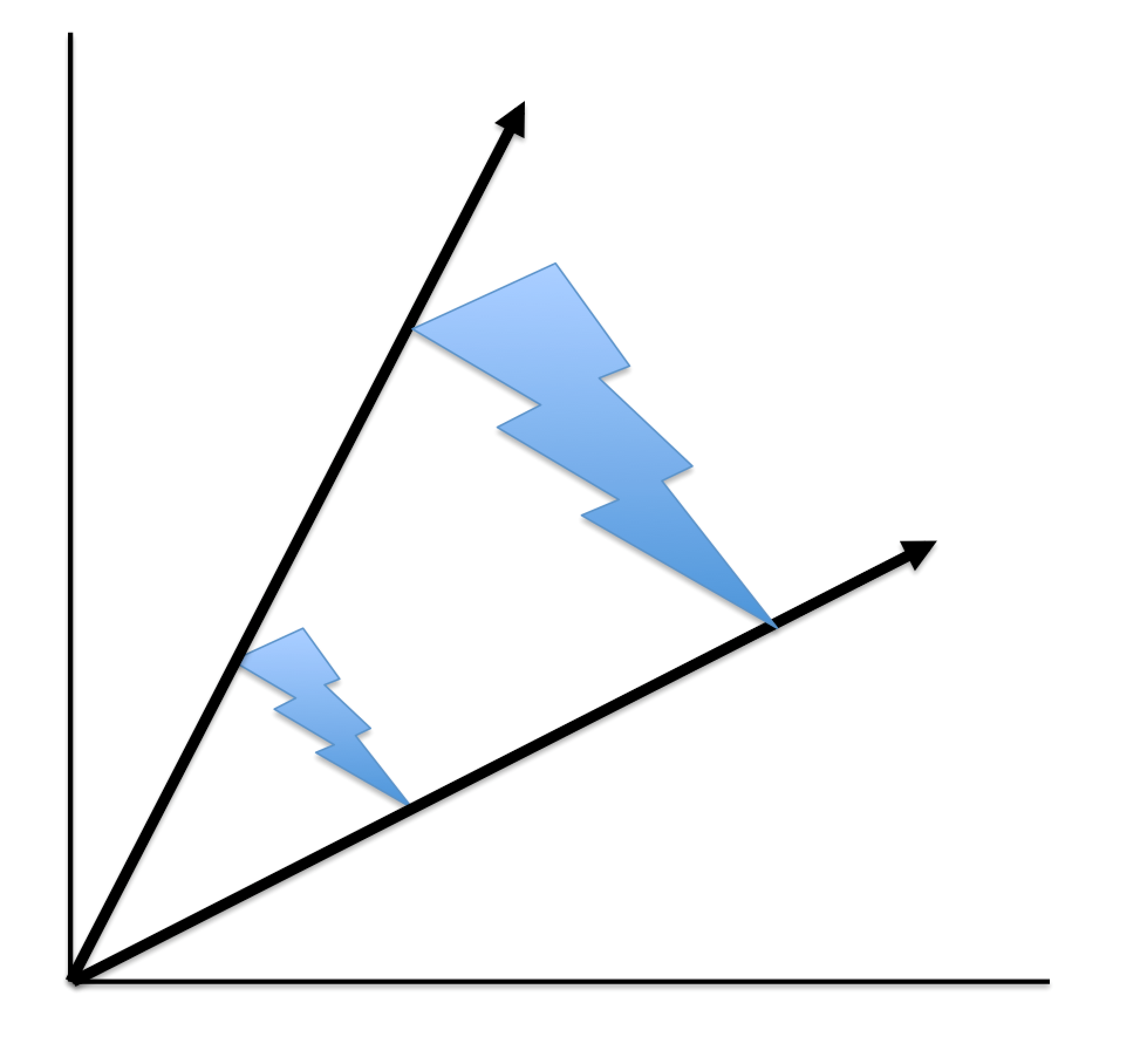}
\caption{Dilation of a complex geometric figure in two-dimensional Euclidean space. The shape and orientation of the image is preserved. The size varies.  \label{Dilation_in_Metric_space}}
\end{figure}

A dilation is a function $f$ from a metric space into itself that satisfies the identity $d(f(x),f(y))=rd(x,y)$ for all points $(x,y)$, where $d(x,y)$ is the distance from $x$ to $y$ and $r$ is some positive real number. Dilations preserve the shape of an object (See figure \ref{Dilation_in_Metric_space}). 

In a discrete space like $\mathbb{N}^2$ a dilation takes a different form. We need an alternative notion of distance: 
 
\begin{definition}
The \emph{taxicab distance} between two vectors ${\overline{a}}= (x_1,y_1)$ and ${\overline{b}}= (x_2,y_2)$ is: 
\[d({\overline{a}}, {\overline{b}}) = |x_1 - x_2| + |y_1 - y_2 |\]
\end{definition}

\begin{figure}[ht!]
\centering
\includegraphics[width=90mm]{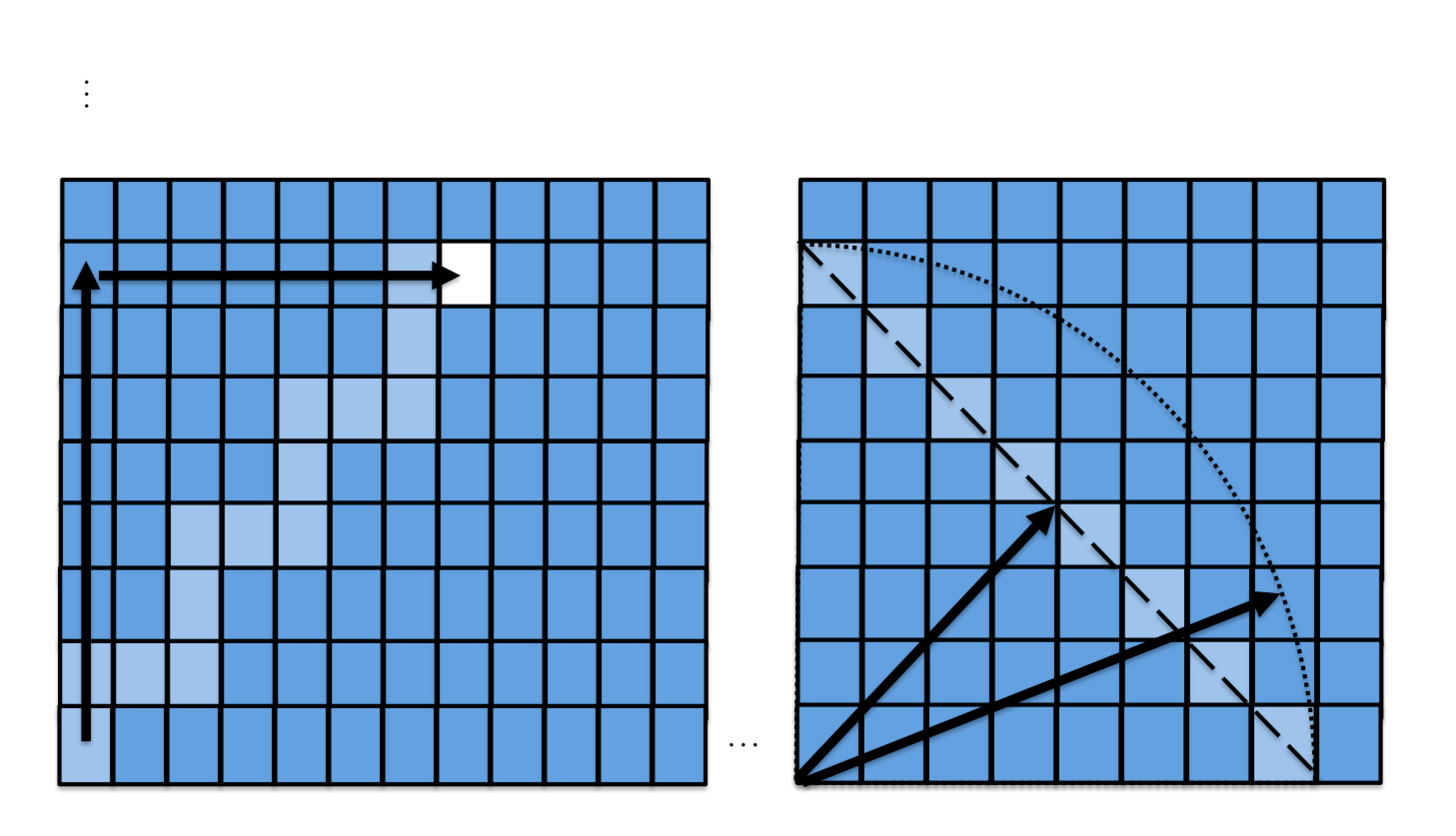}
\caption{An illustration of the taxicab distance. Unlike the situation in Euclidean space there is no shortcut to point  $(m,n)$.  \label{Euclidean_vs_Taxicab}}
\end{figure}

\begin{figure}[ht!]
\centering
\includegraphics[width=90mm]{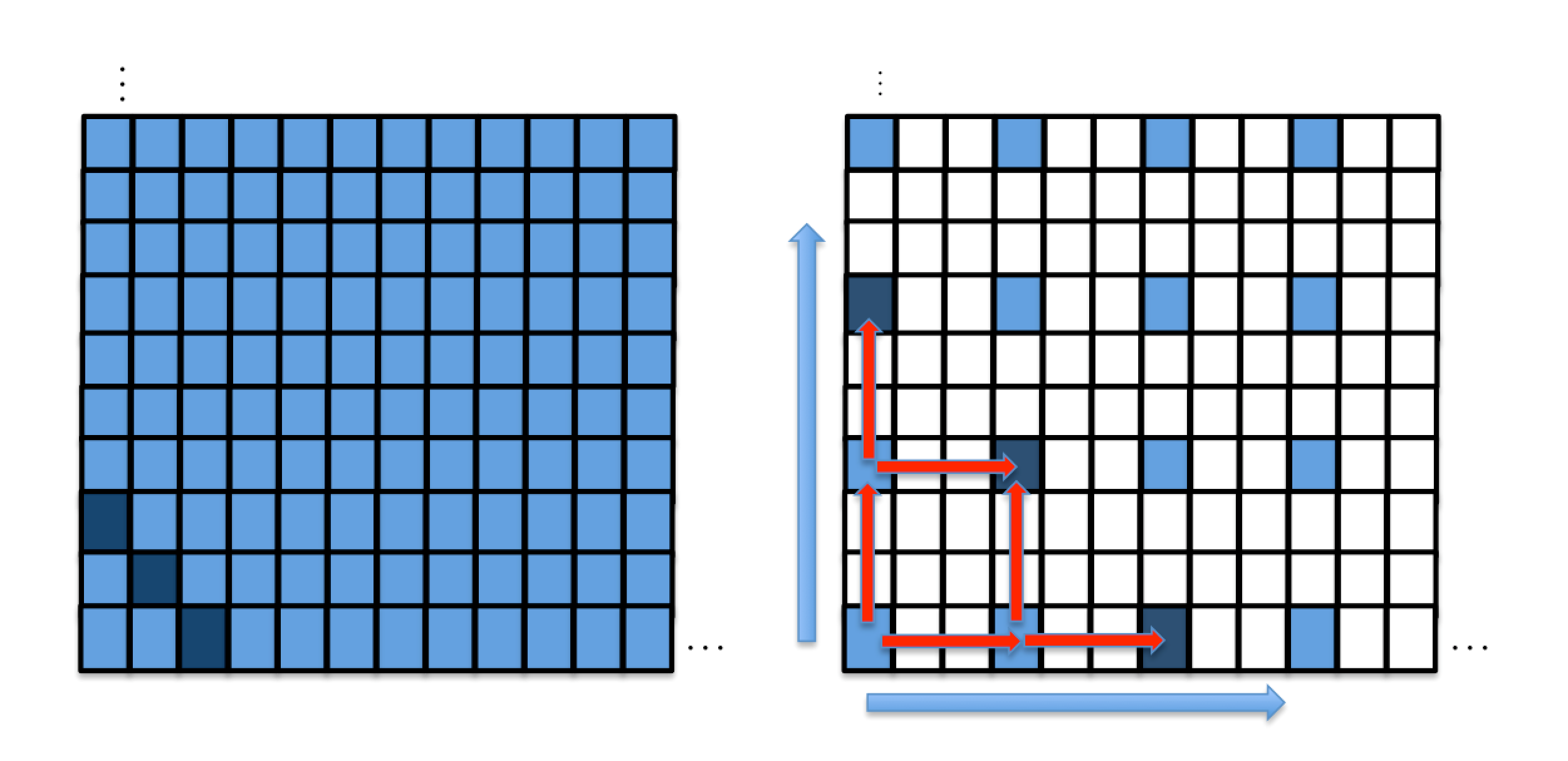}
\caption{Dilation with a factor $3$ over the discrete plane $\mathbb{N}^2$. The taxicab distance of points on the counter diagonals is preserved. The density of the set after a dilation by a factor $n$ is $\frac{1}{n^2}$, in this case $\frac{1}{3^2} = \frac{1}{9}$.   \label{Discrete_Dilations}}
\end{figure}

We define  ${\bf{0}}= (0,0)$. There are exactly $k+1$ points  ${\bf{a}}= (x,y)$,  with $d({\bf{a}}, {\bf{0})} =k$.  All points with the same distance to the origin are situated on the counter diagonal.The distance function in Euclidean space is $x^2 + y^2 = r^2$, a second degree equation that defines a semi-circle, while in discrete space we have $x + y = r$, a first degree equation that defines the counter diagonal (See figure \ref{Euclidean_vs_Taxicab}). 

\begin{lemma}\label{INFTHEORTAXIC}
Point  ${\bf{a}}= (x,y)$ with taxicab distance $k = x + y$ to the origin has entropy: \[H({\bf{a}}) = - \frac{x}{k} \log \frac{x}{k}  - \frac{y}{k} \log \frac{y}{k}\] 
This is also the entropy rate of stochastic walks over the plane in the directions $\frac{x}{y}$ or $\frac{y}{x}$.
\end{lemma}
Proof: There are exactly $n+m \choose m$ different ascending paths from the origin ${\bf{0}}= (0,0)$ to the point with coordinates $(x,y)$. These paths can be expressed as binary strings of length $x+y$, with $1$ for a step in the $x$ direction and $0$ for a step in the $y$ direction.  All strings associated with paths on the line $x=\frac{x}{y}$ have the same Shannon entropy:  $- \frac{x}{x+y} \log \frac{x}{x+y}  - \frac{y}{x+y} \log \frac{y}{x+y}$. $\Box$

I make some observations:  
\begin{enumerate}
\item Random strings with max entropy are located on the line $x=y$. 
\item Walks in the directions $x=c$ or $y = c$ have zero entropy in the limit.   
\item The set $\{0,1\}^k$ charaterizes all paths to points with taxicab distance $k$.
\item  This gives an embedding of the uncountable object $\{0,1\}^{\infty}$ in the discrete plane. The plane itself is countable but the number of ascending walks we can make on this plane is transfinite. 
\end{enumerate}

Using the taxicab distance we can define the notion of a discrete dilation (See figure \ref{Discrete_Dilations}):

\begin{definition}
A \emph{discrete dilation} is a function $f:\mathbb{N}^k \rightarrow \mathbb{N}^k$ from a discrete space into itself that satisfies the identity $d(f(x),f(y))=rd(x,y)$ for all points $(x,y)$, where $d(x,y)$ is the taxicab distance from $x$ to $y$ and $r$ is some natural number. 
\end{definition}

\begin{figure}[ht!]
\centering
\begin{tikzcd}
\mathbb{N}^2\arrow{r}{\pi^k} \arrow{d}{\rho_n}
        & \mathbb{N} \arrow{d}{\tau_n} \\
\mathbb{N}^2\arrow{r}{\pi^k}
&\mathbb{N} \end{tikzcd}
\caption{The effect of a dilation  $\rho_n: \mathbb{N}^2 \rightarrow \mathbb{N}^2$ given the fact that $\pi^k$ is a bijection. There is a correponding dilation $\tau_n: \mathbb{N} \rightarrow \mathbb{N}$.  \label{COMMUTE} }
\end{figure}

Note that a dilation of the discrete plane is also an injection. If the function $\pi^k: \mathbb{N}^k \rightarrow \mathbb{N}$ is a bijection and $\rho_n: \mathbb{N}^2 \rightarrow \mathbb{N}^2$ a dilation by a factor $n$ then  there will be a corresponding dilation $\tau_n: \mathbb{N} \rightarrow \mathbb{N}$. The relations are given in figure \ref{COMMUTE}. Both dilations will give a uniform density reduction to $\frac{1}{n^2}$.

\subsection{The Cantor pairing functions}
The set of natural numbers $\mathbb{N}$ can be mapped to its product set by the so-called Cantor pairing function $\pi: \mathbb{N} \times \mathbb{N} \rightarrow \mathbb{N}$ (and its symmetric counterpart) that defines a two-way polynomial time computable bijection:

\begin{equation}\label{CANTORPAIRING}
\pi(x,y) := \frac{1}{2}(x + y)(x + y + 1)+y
\end{equation}

\begin{figure}[ht!]
\centering
\includegraphics[width=90mm]{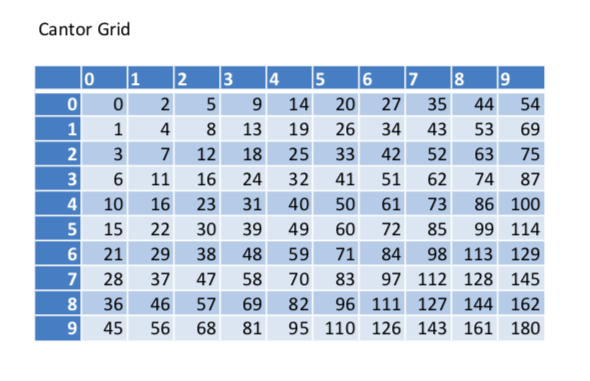}
\caption{An initial segment of the Cantor pairing function \label{Cantor_Function}}
\end{figure}

Note that the Cantor pairing function uses the taxicab distance to order points on the plane. If ${\bf{a}}= (x,y)$,  with $d({\bf{a}}, {\bf{0}}) =k$ then: 

\begin{equation}\label{CANTORPAIRINGTAXICAB}
\pi({\bf{a}})= \pi(x,y) =   \frac{1}{2}(k)(k + 1)+y = y + \sum_{1}^{k}i  
\end{equation}

Consquently lemma  \ref{INFTHEORTAXIC} is relevant for an understanding of the information theoretical behaviour of the function. The Fueter - P\'{o}lya theorem \cite{FP23} states that the Cantor pairing function and its symmetric counterpart $\pi'(x,y)=\pi(y,x)$ are the only possible quadratic pairing functions.  Note that there exist many other polynomial time computable bijective mappings between $\mathbb{N} \times \mathbb{N}$ and $\mathbb{N}$ (e.g. Szudzik pairing)\footnote{See http://szudzik.com/ElegantPairing.pdf, retrieved January 2016.} and $\mathbb{N}^k$ and $\mathbb{N}$.  A segment of this function is shown in figure \ref{Cantor_Function}.

\begin{figure}[ht!]\label{CANTORPLANEEFF}
\centering
\includegraphics[width=90mm]{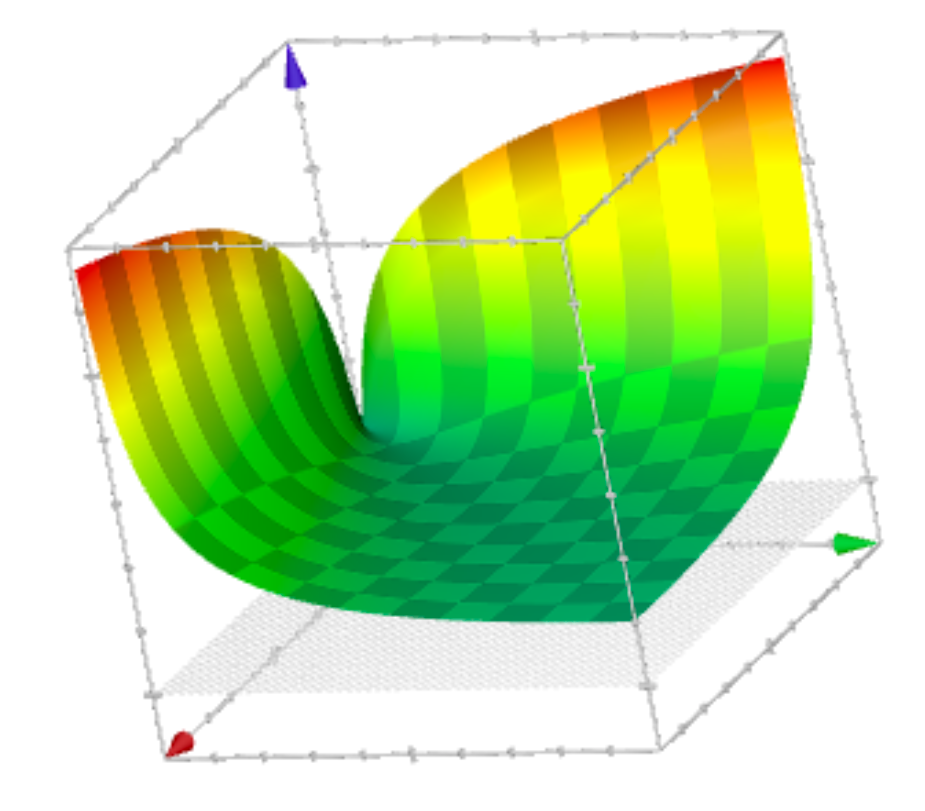}
\caption{The information efficiency of the Cantor pairing function, $0< x < 10^9$, $0< y < 10^9$, $-1< z < 7$. The shaded area is the $z=0$  surface. The Cantor index values stay at least one bit above this plane, because of symmetry. Close to the lines $x=0$ and $y=0$ the Cantor index contains orders of magnitude more information than the input, which indicates that the Cantor bijection is inefficient close to the edges of the space.  \label{Cantor_Efficiency}}
\end{figure}

The information efficiency of this function is: 

\begin{equation}\label{CantEff}
\Delta(\pi(x,y)) = \log (\frac{1}{2} (x+y+1) (x+y) + y) - \log x - \log y
\end{equation}

For the majority of the points in the space $\mathbb{N}^2$ the function $\pi$ has an information efficiency close to one bit. This total lift of the graph with one bit is easily explained by the symmetry of the plane over the diagonal $x=y$. There are two ordered pairs $(a,b)$ and $(b,a)$ but there is only one set $\{a,b\}$. In the formula $\Delta(\pi(x,y)) = \log (\frac{1}{2} (x+y+1) (x+y) + y) - \log x - \log y$ the input is an ordered pair $(x,y)$, but the order of the terms $- \log x - \log y$ is irrelevant. Consequently the input always contains at least one bit of information more than the balance factor $- \log x - \log y$. The pairing function defines what one could call: a  \emph{discontinuous folding operation over the counter diagonals}. On the line $y=0$ we find the images $1/2x(x-1) =\Sigma_{i}^{x} i $.  Equation \ref{CantEff} can be seen as the description of an information topology. The Cantor function runs over the counter diagonals and the image shows that the information efficiencies of points that are in the same neighborhood are also close. The coding given by the Cantor pairing function is maximally efficient near the diagonal $x=y$ and unboundedly inefficient near the edges $x=0$ and $y=0$. This illustrates the fact that $\mathbb{N}^2$ is an object that is fundamentally different from $\mathbb{N}$. Objects of a dimension higher than $1$ display non-trivial behavior in the context of information theory. One aspect I want to point out explicitly, because it is vital for the rest of the paper: 

 \begin{observation}\label{PLANARINEFF}
Planar representations of linear datasets are inherently inefficient. 
\end{observation}
The surface of the plot in figure \ref{CANTORPLANEEFF} \emph{never} crosses the surface $z=1$. The embedding of $\mathbb{N}$ in $\mathbb{N}^2$ is \emph{inefficient} for \emph{every} element of those sets.  The lift to the surface $z=1$ is caused by the fact that we measure the input for the information efficiency function in terms of sets. If we use ordered pairs instead the surface will drop $1$ bit to $z=0$. This makes the cantor function  $\pi(x,y)=z$ prima facie a sort of computational \emph{perpetuum mobile} : in most cases $z$ contains more information than the elements of the ordered pair $(x,y)$. Ordered pairs have more structure than isolated numbers and this structure contains information. Specifically when $x \ll y$ or $y \ll x$ the function $\pi(x,y)=z$ does a bad job. In most cases these irregularities have no great impact on our computational processes, but as soon as the difference between $x$ and $y$ is exponential the effects are considerable, and relevant for our subject   For some data sets, search in $\mathbb{N}$ might be much more effective than search in $\mathbb{N}^2$ and vice versa.

The Cantor pairing functions play an important role in differential information theory since they allow us to develop information theories for multi-valued functions and richer sets. An example: Cantor himself already gave the the proof of the countability of the rational numbers.  A rational number has the form $\frac{a}{b}$ where $a,b \in \mathbb{N}$. It corresponds to an ordered pair $(a,b)$, which can be mapped on a two-dimensional cartesian grid. We can map the rational numbers onto the natural numbers by counting them along the counter diagonal. Georg Cantor himself used this construction to prove the countability of the set $\mathbb{Q}$, but it also provides us with an efficient theory of information measurement by the following definition: 

 \begin{equation}
 g_{\mathbb{Q}} (\frac{x}{y})= \pi(x,y)
\end{equation}

\begin{equation}\forall (x,y \in \mathbb{N}) (I(\frac{x}{y}) = \log(g_{\mathbb{Q}} (\frac{x}{y}))\end{equation} 

A visual impression of the information efficiency of this function is given in figure \ref{Cantor_Efficiency}. Following equation \ref{EffOrig} the information efficiency of this theory of measurement for a rational number $n= \frac{a}{b}$  is solely dependent on the value $\frac{a}{b}$, independent on the size of $n$. We analyse some limits that define the information efficiency of the function. On the line $y=x$ we get: 

\begin{equation}\label{EffDiag}  
\lim_{x \rightarrow \infty} \Delta \pi(x,x) = \lim_{x \rightarrow \infty}\log(\frac{1}{2} (2x + 1)(2x) ) - 2\log x =
\end{equation}

\[ \lim_{x \rightarrow \infty}\log \frac{ 2x^2 + x }{x^2} =  1 \] 

For the majority of the points in the space $\mathbb{N}^2$ the function $\pi$ has an information efficiency close to one bit. On every line through the origin $y=hx$ $(h>0)$ the information efficiency in the limit is constant: 

\begin{equation}\label{EffOrig}
\lim_{x \rightarrow \infty}  \Delta(\pi(x,hx)) = 
\end{equation}
\[  \lim_{x \rightarrow \infty}  \log (\frac{1}{2} (x+hx+1) (x+hx) + hx) - \log x - \log hx = \]  
\[\log(1/2(h+1)^2) - \log h \]

Yet on every line $y=c$ (and by symmetry $x=c$) the information efficiency is unbounded: 

\begin{equation}\label{EffConst}
 \lim_{x \rightarrow \infty}  \Delta(\pi(x,c)) = 
\end{equation}
\[  \lim_{x \rightarrow \infty}  \log (\frac{1}{2} (x+c+1) (x+c) + c) - \log x - \log c =  \infty \]  

Together the equations \ref{CANTORPAIRING}, \ref{CantEff}, \ref{EffDiag}, \ref{EffOrig} and  \ref{EffConst} characterize  the basic behavior of the information efficiency of the Cantor function.

\begin{figure}[ht!]
\centering
\includegraphics[width=120mm]{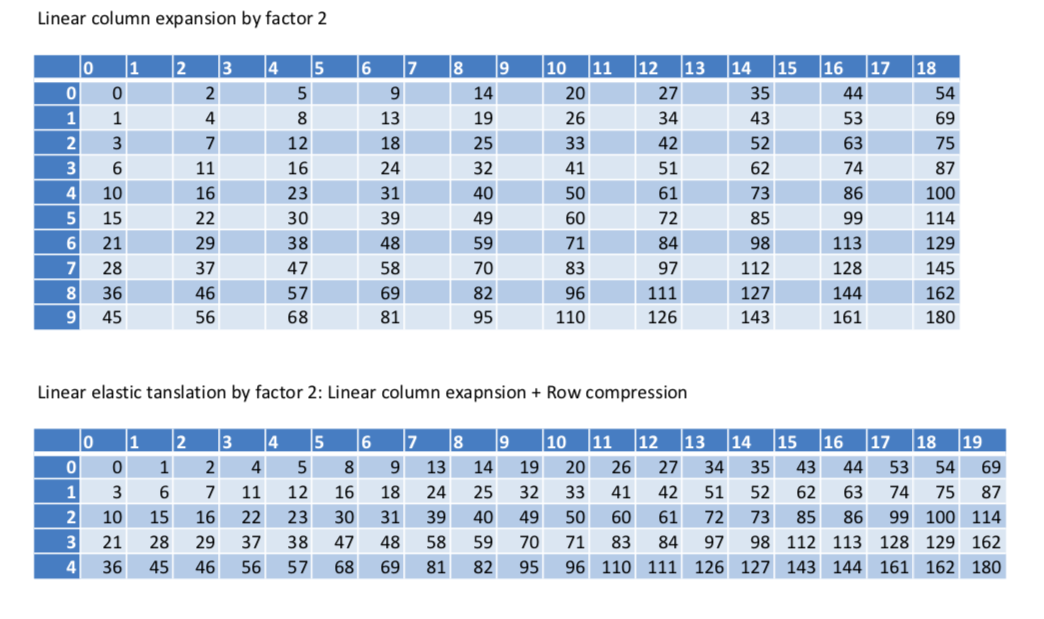}
\caption{Row expansion by  factor $2$ and dilation by a factor 2 for the segment in Figure  \ref{Cantor_Function}. For a computation of the exact information efficiency see figure \ref{ThreeDInformation_efficiency}.  \label{Cantor_Elasticity}}
\end{figure}

\begin{figure}[ht!]
\centering
\includegraphics[width=90mm]{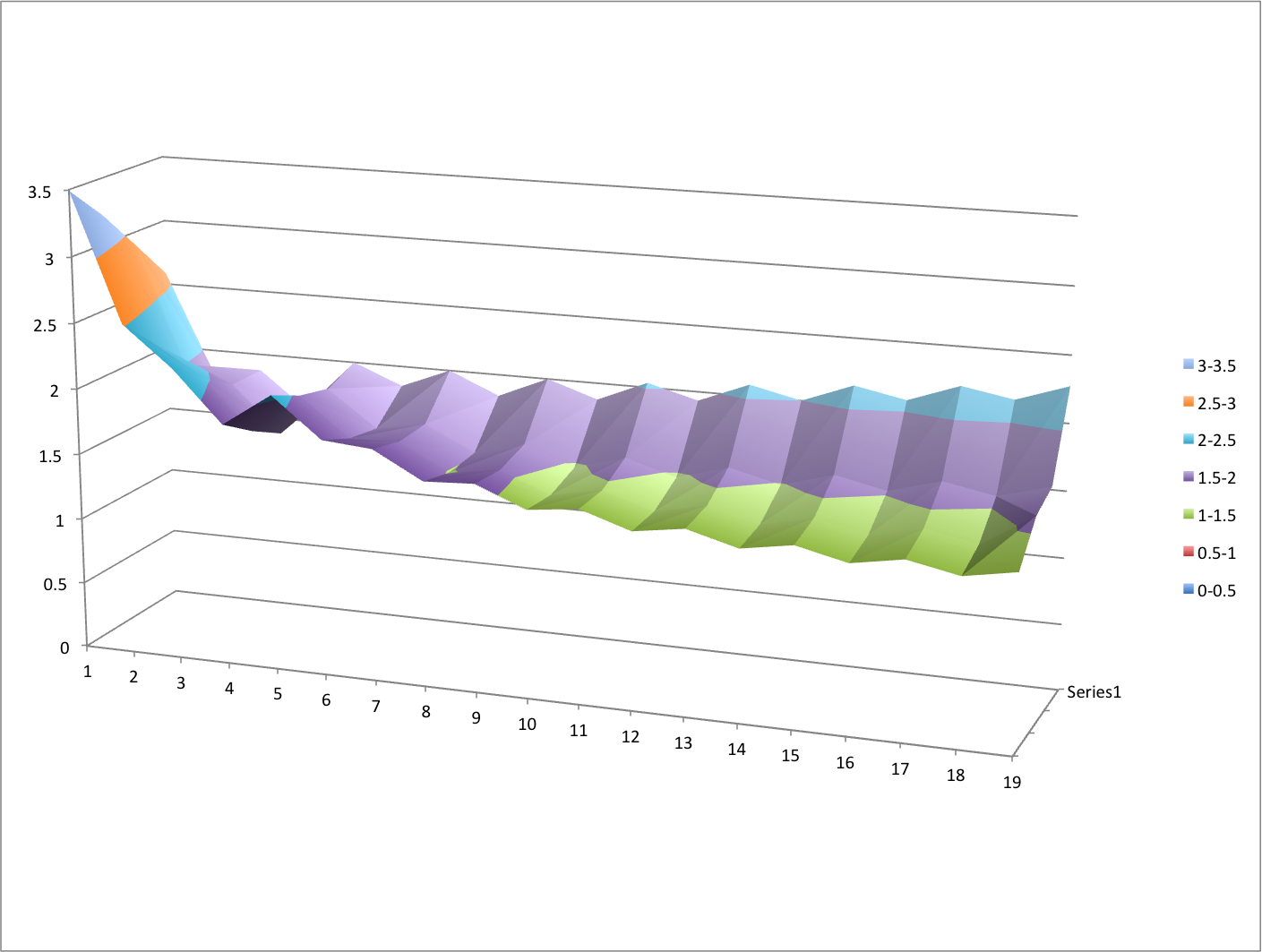}
\caption{An exact computation of the point by point information efficiency after an dilation by a factor $2$  for a segment (1-4 by 1-19) of the table in figure \ref{Cantor_Elasticity}. The existence of two seperate interleaving information efficiency functions is clearly visible.  \label{ThreeDInformation_efficiency}}
\end{figure}

\begin{figure}[ht!]
\centering
\includegraphics[width=90mm]{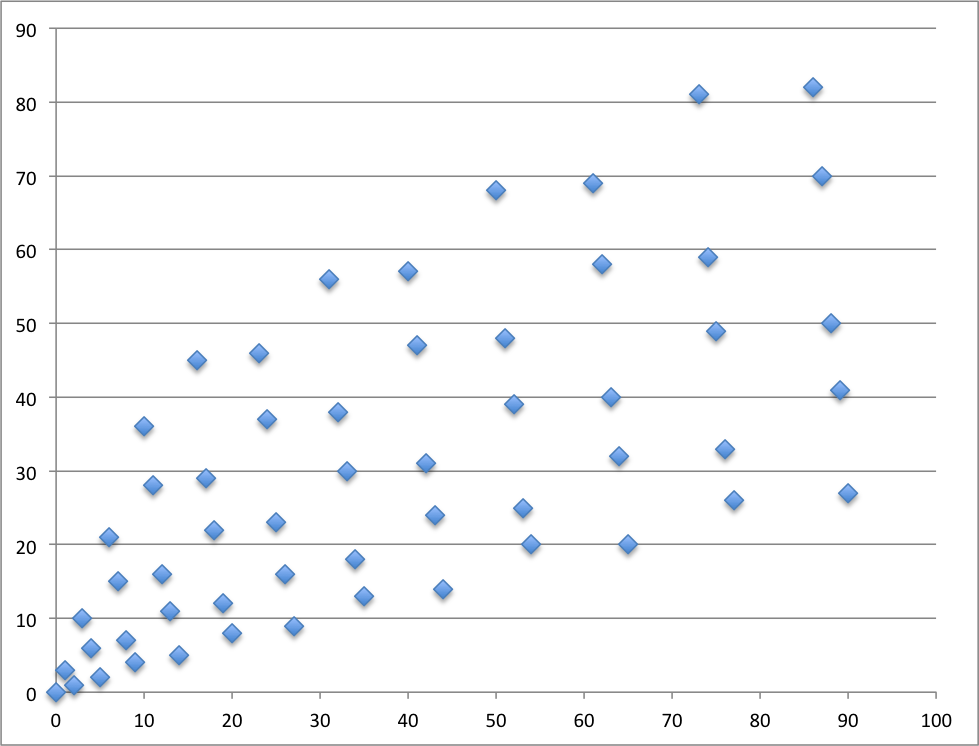}
\caption{A fragment of the bijection generated by an dilation by a factor $2$  for a segment (0-4 by 0-9) of the table in figure \ref{Cantor_Elasticity}. The bijection $\pi(\epsilon_2(\pi^{-1})): \mathbb{N} \rightarrow \mathbb{N}$ generates a cloud of compressible and expandable points. The compressible points are above the line $x=y$.  \label{bijection_shift2}}
\end{figure}

\subsection{Linear dilations of the Cantor space}

Now that we have a two-dimensional representation of the set of natural numbers we can start to develop, what one could call, \emph{dilation theory}: the study of the information effects of recursive operations on natural numbers  on this set in terms of topological deformations of the discrete plane.  We start with simple linear dilations over the $y$-axis that define bijections on $\mathbb{N}$.  

Equation  \ref{EffConst}  is responsible for remarkable behavior of the Cantor function under \emph{elastic dilations}. For such dilations we can compress the Cantor space along the $y$-axis by any constant without actually losing information. Visually one can inspect this counter intuitive phenomenon in figure \ref{Cantor_Efficiency} by observing the concave shape of the information efficiency function: at the edges ($x=0$, $y=0$) it has in the limit an unbounded amount of compressible information. The source of this compressibility in the set $\mathbb{N}$ is the set of numbers that is logarithmically close to sets $\frac{1}{2}x(x+1)$ and  $\frac{1}{2}y(y+1) + y)$. In terms of Kolmogorov complexity these sets of points define regular dips of depth $\frac{1}{2} \log x$ in the integer complexity function that in the limit provides an infinite source of highly compressible numbers.  In fact, when we would draw  figure \ref{Cantor_Efficiency} at any scale over all functions $f: \mathbb{N}^2 \rightarrow \mathbb{N}$ we would see a surface with all kinds of regular and irregular elevations related to the integer complexity function.

 Observe figure  \ref{Cantor_Elasticity}. The upper part shows a discrete translation over the $x$-axis by a factor $2$. This is an information expanding operation: we add the factor $2$, i.e. one bit of information, to each $x$ coordinate. Since we expand information, the density of the resulting set in $\mathbb{N}^2$ also changes by a factor 2. In the lower part we have distributed the values in the columns $0,2,4,\dots$ over the  columns $0-1,2-3,4-5,\dots$. We call this an dilation by a factor $2$. The space $10 \times 10 = 100 $ is transformed into a $5 \times 20 = 100$  space. The exact form of the translation is: $\epsilon_2(x,y) = (2x + (y\bmod{2}), \lfloor \frac{y}{2} \rfloor)$.

\begin{figure}[ht!]
\centering
\includegraphics[width=90mm]{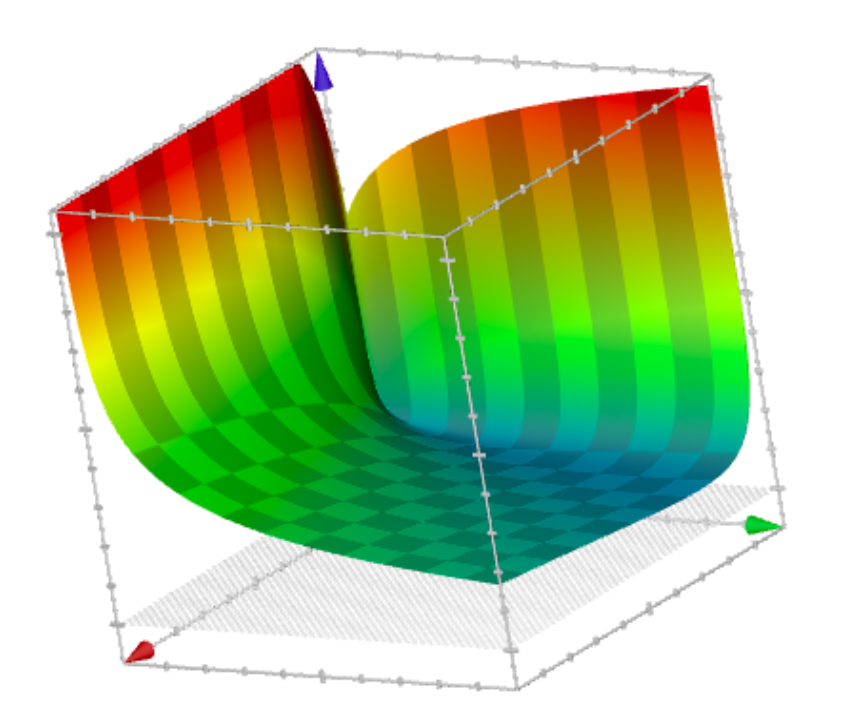}
\caption{The information efficiency of the reference function of the Cantor packing function on the same area as in figure \ref{Cantor_Efficiency}  after a dilation by a factor 100, $0< x < 10^9$, $0< y < 10^9$, $-1< z < 10.6$. The shaded area is the $z=0$  surface. \label{Elastic2}}
\end{figure}

 The effect of this dilation on the information efficiency on a local scale can be seen in figure \ref{ThreeDInformation_efficiency}.  After some erratic behavior close to the origin the effect of the translation evens out. There are traces of a phase transition: close to the origin the size of the $x$, $y$ coordinates is comparable to the size of the shift $c$, which influences the information efficiency considerably. From the wave pattern in the image it is clear that a linear dilation by a factor $c$ essentially behaves like a set of $c$ functions (in this case $2$), each with a markedly different information efficiency.

Even more interesting is the behavior, shown in figure \ref{bijection_shift2}, of the bijection:  
 
 \begin{equation}\label{B1}
\begin{tikzcd}
\mathbb{N}\arrow{r}{\pi^{-1}}   & \mathbb{N}^2\arrow{r}{\epsilon_2} & \mathbb{N}^2\arrow{r}{\pi} & \mathbb{N}
\end{tikzcd} 
\end{equation}

Although the functions $\pi$, $\pi^{-1}$ and $\epsilon_2$ are bijections and can be computed point wise in polynomial time, all correlations between the sets of numbers seems to have been lost. The reverse part of the bijection shown in formula \ref{B2} seems hard to compute, without computing large parts of \ref{B1} first. 
 
 \begin{equation}\label{B2}
 \begin{tikzcd}
\mathbb{N}\arrow{r}{\pi^{-1}}   & \mathbb{N}^2\arrow{r}{\epsilon_2^{-1}} & \mathbb{N}^2\arrow{r}{\pi} & \mathbb{N}
\end{tikzcd} 
\end{equation}

 \begin{observation}\label{Fold2}
Linear dilations introduce a second type of \emph{horizontal discontinuous folding operations over the  columns.}. These operations locally distort the smooth topology of the Cantor function into clouds of isolated points.
\end{observation}
 
 On a larger scale visible in figure \ref{Elastic2} we get a smooth surface. The distortion of the symmetry compared to figure \ref{Cantor_Efficiency} is clearly visible.  In accordance with equation  \ref{EffConst}, nowhere in the set $\mathbb{N}^2$ the information efficiency is negative. In fact the information efficiency is lifted over almost the whole surface. On the line $x=y$ the value in the limit is: 

\begin{equation}\label{EffDiag2}  
\lim_{x \rightarrow \infty} \Delta \pi(\epsilon_2((x,x))) =
\end{equation}
\[ \lim_{x \rightarrow \infty} \log (\frac{1}{2} (2x+x/2+1) (2x+x/2) + x/2) - \log 2x - \log x/2 = 2 \log 5 - 3\]

\begin{figure}[ht!]
\centering
\includegraphics[width=90mm]{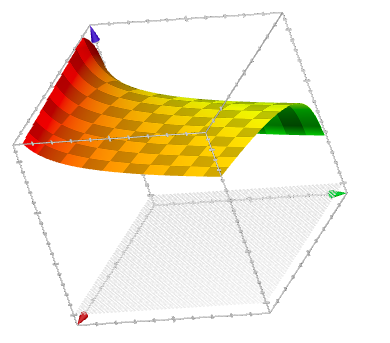}
\caption{The first information efficiency function of the Cantor packing function on the same area as in Figure \ref{Cantor_Efficiency}  after a dilation by a factor 100, $0< x < 10^9$, $0< y < 10^9$, $-1< z < 17$. The shaded area is the $z=0$  surface. For this transformation we have  $100$ different information efficiency functions. Only the first is shown.  \label{Elastic_Shift100}}
\end{figure}

A more extreme form of such a distortion can be seen in figure \ref{Elastic_Shift100} that shows the effect on the information efficiency after a dilation by factor $100$. Clearly the lift in information efficiency over the whole surface can be seen. We only show the first of $100$ different information efficiency functions here. Computed on a point by point basis we would see periodic saw-tooth fluctuations over the  $x$-axis with a period of $100$. This discussion shows that dilations of the Cantor space act as a kind of \emph{perpetuum mobile} of information creation. For every dilation by a constant $c$ the information efficiency in the limit is still positive:  

\begin{lemma}\label{Crux}
No compression by a constant factor $c$ along the $y$-axis (or $x$-axis, by symmetry) will generate a negative information efficiency in the limit. 
\end{lemma}
Proof: immediate consequence of equation  \ref{EffConst}. The information efficiency is unbounded in the limit on every line $x=c$ or $y=c$. $\Box$ 

\subsection{A general model of elastic dilations} 

In the following we will study a more general model of  \emph{dilations} of the space $\mathbb{N}^2$:

\begin{definition}\label{DEfElTrans}
The function $\epsilon_f: \mathbb{N}^2 \rightarrow \mathbb{N}^2$ defines an \emph{elastic dilation} by a function $f$ of the form: 

\begin{equation}\label{GeneralShift}
\epsilon_{r}(x,y) = (xf(x) + (y\bmod{r(x)}), \lfloor \frac{y}{r(x)} \rfloor)
\end{equation}

Such a dilation is \emph{super-elastic} when: 
\[\lim_{x \rightarrow \infty} r(x) = \infty\]
It is \emph{polynomial} when it preserves information about $x$ : 
\[\lim_{x \rightarrow \infty} r(x) = cx^k\]
It is \emph{linear} when:
 \[r(x)=c\]  
 The \emph{reference function} of the dilation is: 
\[\epsilon'_{r}(x,y) = (r(x)x, \frac{y}{r(x)})\]  

  We will assume that the function $f$ can be computed in time polynomial to the length of the input. 
\end{definition}

Observe that the reference function: $\epsilon'_{r}(x,y) = (r(x)x, \frac{y}{r(x)})$ is information neutral on the arguments: 

\[\log x + \log y - (\log r(x)x + \log \frac{y}{r(x)})  = 0\]

A dilation consists from an algorithmic point of view of two additional operations: 
\begin{enumerate}
\item An \emph{information discarding} operation on $y: \frac{y}{r(x)} \rightarrow  \lfloor \frac{y}{r(x)} \rfloor$. 
\item An \emph{information generating} operation on $x: r(x)x \rightarrow r(x)x + (y\bmod{r(x)})$.
\end{enumerate} 

\begin{figure}[ht!]
\centering
\includegraphics[width=90mm]{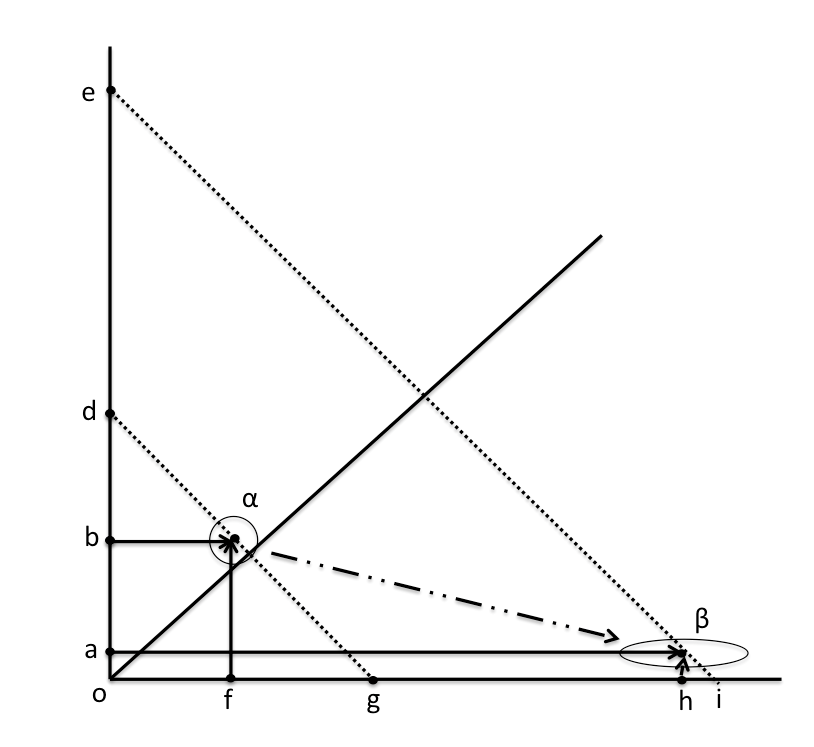}
\caption{A dilation from neighborhood $\alpha$ to neighborhood $\beta$.  \label{Cantor_Shift}}
\end{figure}

 A schematic overview of a linear dilation is given in figure \ref{Cantor_Shift}. Here the letters $a,\dots i$ are natural numbers. An arbitrary point  in neighborhood  $\alpha$ with coordinates $(b,f)$ close to the diagonal is translated to point $(a,h)$  in neighborhood $\beta$. The formula for the translation is given by definition \ref{DEfElTrans}: $ \epsilon_r(x,y) = (r(x)x + (y\bmod{r(x)}), \lfloor \frac{y}{r(x)} \rfloor)$.  We have $a = \lfloor \frac{b}{r(f)} \rfloor$ and $h = r(f)e + (b\bmod{r(f)})$.  
 
 The information efficiency of  an \emph{elastic dilation} is: 
 
\[\Delta(\pi(\epsilon_{r})(x,y)) =  \Delta(\pi( r(x)x + (y\bmod{r(x)}),  \lfloor \frac{y}{r(x)} \rfloor))\]

 The information efficiency of  a \emph{linear dilation} by a factor  $f(x)=c$  is: 
 
 \[\Delta(\pi( cx + (y\bmod c ),  \lfloor \frac{y}{c} \rfloor))\]

\begin{observation}
Dilations by a constant $c$ of the Cantor space replace the highly efficient Cantor packing function with $c$ different interleaving functions, each with a different information efficiency.  Equation \ref{GeneralShift} must be seen as a \emph{meta-function} or \emph{meta-program} that spawns off $c$ different new programs. 
\end{observation}

This is illustrated by the following lemma:  

 \begin{lemma}\label{L2}
Linear dilations generate information. 
\end{lemma}
Proof: This is an immediate effect of the use of the mod function. A \emph{dilation by a constant $c$} of the form: \[\epsilon_{c}(x,y) = (cx + (y\bmod c), \lfloor \frac{y}{c} \rfloor)\] has $c$ different information efficiency functions. Note that the function $x \bmod c$ produces all numbers $d < c$, including the incompressible ones that have no mutual information with $c$: $K(d|c)= \log d + O(1)$.For each value  $d = y\bmod c$ we get a function with different information efficieny: 
\[\Delta(\pi(\epsilon_{c,d}(x,y)))) = \log \pi (cx + d, \lfloor \frac{y}{c} \rfloor) -  \log x - \log y\]

On a point by point basis the number $d$ is part of the information computed by $\pi(\epsilon_{c,d}(x,y)))$. In other words the computation $\pi(\epsilon_{c,d}(x,y))$ adds information to the input for specific pairs $(x,y)$ that is not available in the formula for $\epsilon_{c}(x,y)$.  The effect is for linear dilations constant in the limit, so it is below the accuracy of Kolmogorov complexity.
$\Box$

For typical cells $(x,hx)$ on a line $y=hx$ the function  $\Delta(\pi(\epsilon'_{c}(x,y)))$ gives in the limit a constant shift which can be computed as: 

  \begin{equation}\label{E5} 
\lim_{x \rightarrow \infty}\frac{\pi(\epsilon'_{c}(x,y))}{\pi(x,y)}   =
\lim_{x \rightarrow \infty}\frac{ \frac{1}{2} (c x + \frac{h x}{c} + 1) (c x +  \frac{h x}{c}) +  \frac{h x}{c}}{ \frac{1}{2} (x + h x + 1) (x + h x) + h x} = \frac{(c^2 + h)^2}{c^2(1 + h)^2} \geq 1
\end{equation} 

Note that this value is only dependent on $h$ and $c$ for all $c \in \mathbb{N}^{+}$ and $r \in \mathbb{R}^{+}$. The general lift of the line $x=y$ for a dilation by a constant $c$ is: 

\begin{equation}\label{ConShift} 
 \lim_{x \rightarrow \infty} \log (\frac{1}{2} (cx+\frac{x}{c}+1) (cx+\frac{x}{c}) + \frac{x}{c})) - \log cx - \log \frac{x}{c} =
\end{equation}

\[- \log \frac{2}{c} - \log c +\log (2 + \frac{1}{c^2} + c^2)\]

 We get a better understanding of the extreme behavior of the reference function $\Delta(\pi(\epsilon'_{c}(x,y)))$ when we rewrite equation \ref{E5} as: 

  \[ \frac{(c^2 + h)^2}{c^2(1 + h)^2} = \frac{c^2} { 1 + 2h + h^2 }+  \frac{ 2h} {  1 + 2h + h^2 } +  \frac{ h^2}{ c^2 + 2c^2h + c^2h^2 } \geq 1\]
 
and take the following limit: 
  
     \[ \lim_{h \rightarrow \infty} \frac{(c^2 + h)^2}{c^2(1 + h)^2} =  c^{-2} \]  

If $c$ is constant then it has small effects on large $h$ in the limit. The reference function allows us to study the dynamics of well-behaved ``guide points'' independent of the local distortions generated by the information compression and expansion operations. Note that dilations start to generate unbounded amounts of information in each direction $y=hx$ in the limit on the basis of  equation \ref{E5}:

  \begin{equation}\label{E6} 
 \lim_{c \rightarrow \infty} \frac{(c^2 + h)^2}{c^2(1 + h)^2} = \infty
\end{equation} 
   
If $c$ grows unboundedly then the information efficiency of the corresponding reference functions goes to infinity for every value of $h$. Consequently,  when $c$ goes to infinity, the reference functions predict infinite information efficiency in $\mathbb{N}^2$ in all directions, i.e. we get infinite expansion of information in all regions without the existence of regions with information compression. This clearly contradicts central results of Komogorov complexity if we asume that dilations are defined in terms of a single program. The situation is clarified by the proof of lemma \ref{L2}: if $c$ goes to infinity we create an unbounded amount of new functions that generate an unbounded amount of information.

\subsection{Polynomial  transformations}    
The picture that emerges from the previous paragraphs is the following: we can define bijections on the set of natural numbers that generate information for almost all numbers. The mechanism involves the manipulation of clouds of points of the set $\mathbb{N}$: sets with density close to the origin are projected into sparse sets of points further removed from the origin. This process can continue indefinitely.

In this context we analyse polynomial translations. The simplest example is the dilation by the factor $x$:

\begin{figure}[ht!]
\centering
\includegraphics[width=120mm]{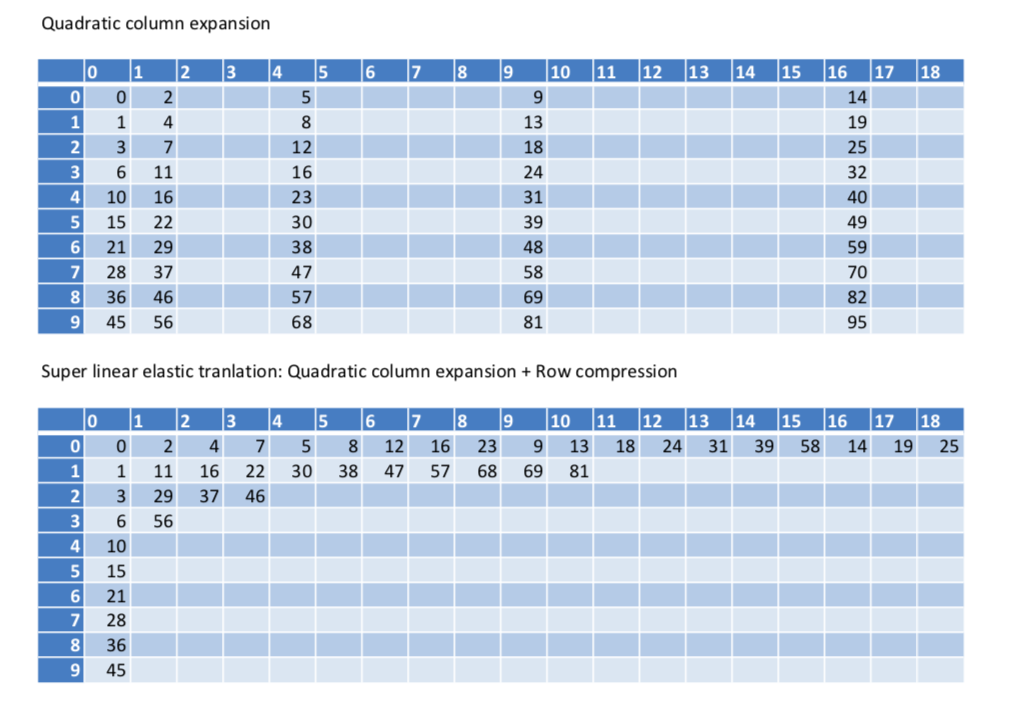}
\caption{Row expansion by a superlinear, $r(x)=x$ factor and a fragment of the corresponding  dilation for the segment in Figure  \ref{Cantor_Function} \label{Cantor_Super_Elasticity}}
\end{figure}

\begin{equation}\label{SupLin}
\epsilon_x(x,y) = (x^2 + (y\bmod{x}), \lfloor \frac{y}{x} \rfloor)
\end{equation}

\begin{figure}[ht!]
\centering
\includegraphics[width=120mm]{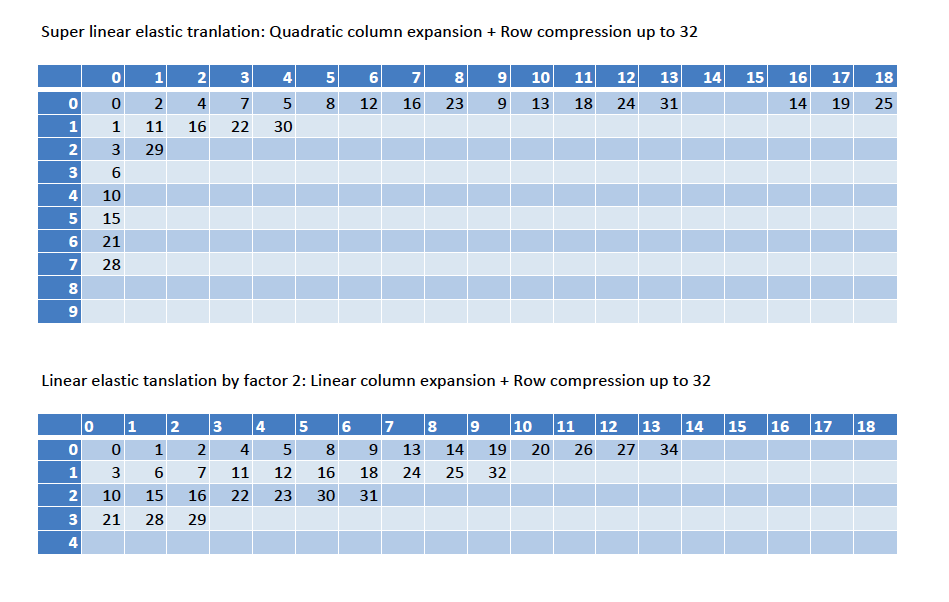}
\caption{Above: Super linear expansion by a function $r(x)=x$ up to 32.   Below: Linear dilation by a factor $2$ up to the number $32$.  \label{Discontinuous}}
\end{figure}

A tiny fragment of the effects is shown in figure \ref{Cantor_Super_Elasticity}. There does not seem to be a fundamental difference compared to the previous examples and there seems to be no difficulty in constructing such a translation along the lines suggested in the figure. However upon closer inspection things  are different as can be observed in figure \ref{Discontinuous}. The first table shows the computation of the function up to the number $32$. Observe that the columns $14$ and $15$ are empty, while column $16-18$ have a value. This effect does not appear in the second table. Linear dilations induce a change in the direction of the iso-information line $(d,g)$, but the image of the translation keeps a coherent topology at any stage of the computation.  

Polynomial translations on the other side are discontinuous. They tear the space apart into separate regions.  An appropriate metaphor would be the following: expansion away from the origin over the $x$-axis, sucks a \emph{vacuum} that must be filled by a contraction over the $y$-axis. Actually the creation of such a vacuum is an information discarding operation. The analysis above shows that the vacuum created by the shift described by equation \ref{SupLin} for cells on the line $y=0$ is bigger than the \emph{whole surface} of the triangle $(o,d,g)$. The effect is that the image of the translation becomes discontinuous.

\begin{theorem}\label{T1}
Polynomial dilations of $\mathbb{N}^2$: 
\begin{enumerate}
\item Discard  and expand  information on the line $y=0$ unboundedly. 
\item  Project a dense part of $\mathbb{N}^2$ and  $\mathbb{N}$ on $y=0$.
\item Generate an unbounded amount of information for typical points in $\mathbb{N}^2$ in the limit.   
\end{enumerate}
\end{theorem}
Proof: The formula for a polynomial translation is: 

\[\epsilon_{r}(x,y) = (cx^{k+1} + (y\bmod{cx}), \lfloor \frac{y}{cx} \rfloor)\]

Take $k= c=1$. Consider a typical point  $f$ in figure \ref{Cantor_Shift}. We may assume that the numbers $d$ and $g$ are typical (i.e. incompressible and thus $\pi(g,d)$ is inompressible too.   Remember that the Cantor function runs over the counter diagonal which makes the line $(d,g)$ an \emph{iso-information} line. 
\begin{enumerate}
\item Discard  and expand  information on the line $y=0$ unboundedly: 
\begin{itemize}
	\item Discard  information: Horizontally point $(f,0)$ will be shifted to location $(f^2,0)$ and $(f(+1),0)$ to $(f^2 + 2f + 1, 0)$. The cantor index for point $(f,0)$ is $\frac{1}{2}f(f-1) < f^2$. 
	\item Expand information: The cells from $(f,0)$ to $f(f,2f)$ will be ``padded'' in the strip between  $(f^2,0)$ and $(f^2 + 2f + 1, 0)$. But this operation ``steals'' a number of  $f$ cells from the domain above the line $(d,g)$. Now take a typical point $n$ on location  $(f,m)$. such that $f < m < 2f$. This point will land at  $(f^2 + m,0)$ somewhere between $(f^2,0)$ and $(f^2 + 2f + 1, 0)$, which gives: 
\[\pi(f,(f+m))= \frac{1}{2} (f + (f +m) +1)(f +(f+m)) + (f+m) = \]
\[ 2 f^2 + \frac{1}{2}fm + m^2 +  3f + 1\frac{1}{2} (m + fm)   \]
\[  \gg f^2 + 2f + 1 \]
\end{itemize}
 Note that the effects are dependent on $f$ and $m$, so they are unbounded in the limit.  Alternatively observe the fact that polynomial dilations are superelastic and apply equation \ref{E6}.
\item  Project a dense part of $\mathbb{N}^2$ and  $\mathbb{N}$ on $y=0$: For every point $f$ all the cells  $(f,0)$ to $f(f,2f)$  up to the line $y = 2x$ will end up on line $y=0$. since all points in this dense region are projected on the line $y=0$ most points $\pi(f,(f+m))$ are incompressible. 
\item Generate an unbounded amount of information for typical points in $\mathbb{N}^2$ in the limit. Take a typical point $(x,y)$ such that  $\log x \approx \log y$:

\begin{equation}\label{INFGROW}
\pi(\epsilon_{r}(x,y)) >  \pi(\epsilon_{r}(x,1)) = \frac{1}{2}( x^2+2)(x^2+1) +1 > \frac{1}{2}x^4
\end{equation}
\[\Delta (\pi(\epsilon_{r}(x,y))) >  \log  \frac{1}{2}x^4  - \log x - \log y \approx 4 \log x - 2 \log x = 2 \log x \]

which gives $K\pi(\epsilon_{r}(x,y)) > K(\pi(x,y)) + 2 \log x$. 
\end{enumerate}
This argument can easily be generalized to other values of $c$ and $k$. $\Box$ 

Polynomial shifts generate information above the asymptotic sensitivity level of Kolmogorov complexity.  Note that $\epsilon_{r}(x,y)$ is still  a computable bijection:  

 \[\forall (x,y)_{\in \mathbb{N}^2} \exists (u,v)_{\in \mathbb{N}^2} (\epsilon_{r}(u,v)= (x,y))\]   
 
 \[u^2 \leq x < (u+1)^2 \]  
 \[a = (u+1)^2 - u^2\]
 \[y = \lfloor \frac{v}{a} \rfloor \]
 
 This analysis holds for all values $\Delta (\pi(\epsilon_{r}(x,y))) \in \mathbb{N}$ including the values that are typical, i.e. incompressible. 

\begin{observation}An immediate consequence is the translation $\epsilon_{r}(x,y) = (cx^{k+1} + (y\bmod{cx}), \lfloor \frac{y}{cx} \rfloor)$ must be interpreted as a \emph{function scheme}, that produces a countable set of new functions. One for each column $x=c$. Actually $x$ can be seen as an \emph{index} of the function that is used to compute $\epsilon_{r}(u,v)= (x,y)$.
\end{observation}

\section{ Semi-Countable Sets} \label{SEMICOUNTSETS}

In this paragraph I investigate information theories for the set of finite sets of natural numbers. I wil show that this set is indeed semi-countable: there is no intrinsic theory of measurement for the set. This means that there is no objective answer to the question: how much information does a finite set of natural numbers contain? A useful concept in this context is the notion of \emph{combinatorial number systems}: 

\begin{definition}\label{COMBNUMBSYS}
The function $\sigma_k:\mathbb{N}^k \rightarrow \mathbb{N}$ defines for each element \[s = (s_k,\dots,s_2,s_1) \in \mathbb{N}^k\]  with the strict ordering $s_k > \dots s_2 > s_1 \geq 0$ its index in a $k$-dimensional \emph{combinatorial number system} as:

\begin{equation}\label{COMBSYST}
\sigma_k(s)= {s_k \choose k} + \dots + {s_2 \choose 2} + {s_1 \choose 1}
\end{equation}
\end{definition}

The function $\sigma_k$ defines for each set $s$ its index in the lexicographic ordering of all sets of numbers with the same cardinality $k$. The correspondence does not depend on the size $n$ of the set that the $k$-combinations are taken from, so it can be interpreted as a map from $\mathbb{N}$ to the $k$-combinations taken from $\mathbb{N}$. For singleton sets we have: $\sigma_1(x)= {x \choose 1}= x$, $x \geq 0$. For sets with cardinality $2$ we have:

\[\sigma_2((1,0))=  {1 \choose 2} + {0 \choose 1} = 0 \Leftrightarrow \{1,0\}\]
\[\sigma_2((2,0))=  {2 \choose 2} + {0 \choose 1} = 1\Leftrightarrow \{2,0\}\]
\[\sigma_2((2,1))=  {2 \choose 2} + {1 \choose 1} = 2\Leftrightarrow \{2,1\}\]
\[\sigma_2((3,0))=  {3 \choose 2} + {0 \choose 1} = 3 \Leftrightarrow \{3,0\}\]\[ \dots\]

We can use the notion of combinatorial number systems to prove the following result: 

\begin{theorem}\label{COUNTCARD}
There is a bijection $ \phi:  \mathfrak{P}(\mathbb{N}) \rightarrow  \mathbb{N}^2$ that can be computed  efficiently. 
\end{theorem}
Proof: Let $S_k$ be the subset of all elements $s \in \mathfrak{P}(\mathbb{N})$ with cardinality $|s| = k$. For each $k \in \mathbb{N}$ by definition \ref{COMBNUMBSYS} the set $S_k$ is described by a combinatorial number system of degree $k$. The function $\sigma_k:\mathbb{N}^{k} \rightarrow \mathbb{N}$ defines for each element $s = (s_k,\dots,s_2,s_1) \in \mathbb{N}^{k}$,  with the strict ordering $s_k > \dots s_2 > s_1 \geq 0$, its index in a $k$-dimensional combinatorial number system. By definition \ref{COMBNUMBSYS} the correspondence is a polynomial time computable bijection. Now define $\phi_{car}: \mathfrak{P}(\mathbb{N}) \rightarrow \mathbb{N}^2$ as: 

\begin{equation}\label{CARD}
\phi_{car}(s) =   ((|s| - 1),\sigma_{|s|}(s))
\end{equation}

We use the symbol $||$ to refer to the cardinality operation. Note that  both $\pi$ and $\sigma$ are computable bijections. When we have the set $s$ we can compute its cardinality $|s|$  in linear time and compute $\sigma_{|s|}(s)$ from $s$ in polynomial time. $\Box$ 

\begin{figure}[ht!]
\centering
\includegraphics[width=90mm]{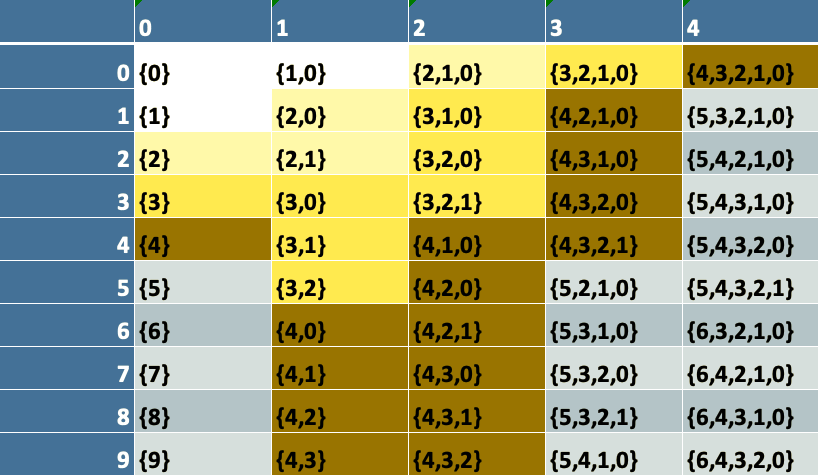}
\caption{An initial segment of the set  $\mathfrak{P}(\mathbb{N}) - \emptyset$ projected on the plane. It can be interpreted as an infinite planar representation of nested powersets indicated by shaded areas for the sets $\{1,0\}$, $\{2,1,0\}$, $\{3,2,1,0\}$  and $\{4,3,2,1,0\}$.
 \label{Cardinality-Grid}
 }
\end{figure}

The construction of the proof of theorem \ref{COUNTCARD} separates the set $\mathfrak{P}(\mathbb{N})$ in an infinite number of infinite countable partitions ordered in two dimensions: in the columns we find elements with the same cardinality, in the rows we have the elements with the same index. An elaborate example of the computation both ways is given in the appendix in paragraph \ref{A1}. An example of the mapping can be seen in figure \ref{Cardinality-Grid}.

Theorem \ref{COUNTCARD} not only gives us a proof of the countability of $\mathfrak{P}(\mathbb{N})$, but  we now also have an efficient theory of information measurement: 

\begin{equation}\label{CARDINFTH}\forall (s \in \mathfrak{P}(\mathbb{N})) I(s) = \log(\pi(\phi_{car}(s)))\end{equation} 

The bijection $\phi_{car}$ has interesting mathematical properties: 
\begin{itemize}
\item Note that $\emptyset$ is not represented in the plane. 
\item On the line $x=0$ we find sets of the form $\{0,1,\dots,k\}$ in column $k$. 
\item On the line $y=0$ we find the singletons containing the natural numbers. 
\item In column $k$ we find sets of cardinality $k+1$ ordered lexicografically. 
\item The first $n \choose k$ sets in column $k-1$ are exactly the subsets of cardinality $k$ of the set on location $(0,n)$. 
\item For the set on location $(0,k)$ we find the $2^k-1$ elements of the set $\mathcal{P}(\{0,1,2,\dots, k\}) - \emptyset$ in the columns $(0,x)$ with $x<k$. The length of the highlighted columns is given by the Pascal triangle at row $k$: 
${k \choose 1},  {k \choose 2}, \dots {k \choose k}$ again with omission of the empty set.
\item Note that ${k \choose 1} + {k \choose 2}+ \dots  + {k \choose k} = 2^k -1$, which coincides with a powerset of a set with $k$ elements minus the empty set $\emptyset$. 
\end{itemize}

\subsection{The binary number bijection $\phi_{bin}$.}

We first study the effect of the power sum operation. The power sum operation defines a bijection between $\mathbb{N}$ and $\mathfrak{P}(\mathbb{N})$. In fact this bijection is well-known since it is the basis for our binary number system. 

\begin{lemma}\label{UPS}
The function $\Upsilon: \mathfrak{P}(\mathbb{N}) \rightarrow \mathbb{N}$ with $\Upsilon(x) =  \Sigma_{i \in X}  2^i $ defines a two-way efficiently computable bijection. 
\end{lemma}
Proof: Suppose $\Upsilon(s)=n$ where $s= \{ s_1, s_2, \dots s_k\}$ is a set of natural numbers and $n$ an natural number. We have: 
\[\Upsilon(s) = \Sigma_{i \in s}  2^i    =2^{s_1}  +2^{s_2} + \dots + 2^{s_k} = n\]
This function is easy to compute. Observe that the natural number $n$ can be written as a unique binary number which can be computed efficiently:
\[ n =2^{a_1}  +2^{a_2} + \dots + 2^{a_m} \]
Since such a binary number is unique it is the case that 

\[ \{a_1, a_2, \dots ,a_m\} =  \{s_1, s_2, \dots ,s_k\} = s\]
$\Box$

\begin{figure}[ht!]
\centering
\includegraphics[width=90mm]{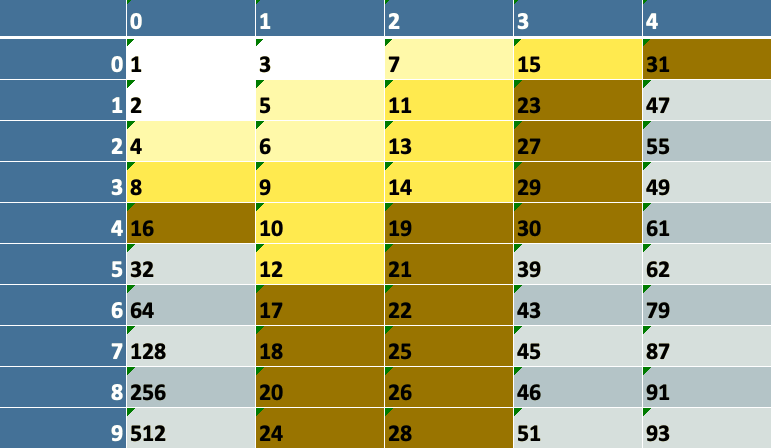}
\caption{The values of taking the sum of the exponentials of the elements of the sets in figure   \ref{Cardinality-Grid}.  If we take the empty set $\emptyset$ as representing $0$ we get an endomorphism on $\mathbb{N}$. On the line $x=0$ we find powers of two $2^i$. On the line $y=0$ we have $2^i-1$ and on the line $x=y$ we have $ 2 ( 2^i -1) -1$. Observe that the values in this bijection diverge considerably from the ones found in figure \ref{Cardinality-Grid}. 
  \label{Exponential-Sum}}
\end{figure}

An initial segment of the values of the function $\Upsilon$ is seen in figure \ref{Exponential-Sum}. Observe that we have three sets: 
\begin{itemize} 
\item The set of natural numbers: $\mathbb{N}$, 
\item The two-dimensional plane $\mathbb{N}^2$ and 
\item The class of all finite sets of natural numbers $\mathfrak{P}(\mathbb{N})$. 
\end{itemize}

with three corresponding efficiently computable bijections:

\begin{itemize} 
\item The Cantor function $\pi: \mathbb{N}^2 \rightarrow \mathbb{N}$,  
\item The cardinality  bijection  $\phi_{car}:  \mathfrak{P}(\mathbb{N}) \rightarrow  \mathbb{N}^2 $,   and 
\item The power sum operation  $\Upsilon: \mathfrak{P}(\mathbb{N}) \rightarrow \mathbb{N}$,  
\end{itemize}

Together these bijections characterize the interconnectedness of the  set of numbers and the set of finite sets on the basis of the following endomorphism:

 \begin{equation}\label{ENDMORF}
\begin{tikzcd}
\mathbb{N}\arrow{r}{\pi^{-1}}   & \mathbb{N}^2\arrow{r}{\phi_{car}^{-1}} & \mathfrak{P}(\mathbb{N}) \arrow{r}{\Upsilon} & \mathbb{N}
\end{tikzcd} 
\end{equation}

By definition the binary numbers define a second efficiciently computable  theory of information measurement: 

\begin{equation}\label{INFTHEORYPOWSUM}\forall (s \in \mathfrak{P}(\mathbb{N})) I(s) = \log(\Upsilon(s))\end{equation}

We define the corresponding bijection: 

\begin{equation}\label{EXPSUM}
\phi_{bin}(s) =  (\Upsilon(s),0)
\end{equation}

Metaphorically we could say that the power sum operation $\Upsilon$ pulls the sets in $ \mathfrak{P}(\mathbb{N})$ with exactly enough force to project each cell in $\mathbb{N}^2$ to a unique location on the line $y=0$. 

\subsection{Dilations of $\mathfrak{P}(\mathbb{N})$ on $\mathbb{N}^2$}\label{DILATHEOR}

In this paragraph we will study dilation on $ \phi_{car}$.  We will investigate the conditions under which these operation still lead to (efficiently) computable bijections. Since the grid is discrete the effect of such elastic translations is disruptive and involves discontinuous translations of cells over various distances. The force that ``pulls'' the cells in the $y$-axis direction will be an arithmetical operation on sets of numbers. There are four dilations that are of special interest to us: 

\begin{itemize}
\item The cardinality dilation: $\phi_{car}:  \mathfrak{P}(\mathbb{N}) \rightarrow  \mathbb{N}^2$
\item The sum dilation:  $\phi_{sum}:  \mathfrak{P}(\mathbb{N}) \rightarrow   \mathbb{N}^2$
\item The product dilation  $\phi_{prod}:  \mathfrak{P}(\mathbb{N})  \rightarrow  \mathbb{N}^2$
\item The binary number dilation  $\phi_{bin}:  \mathfrak{P}(\mathbb{N})  \rightarrow  \mathbb{N}^2$
\end{itemize}

Only the first is a real bijction. The other three are injections, although the last one gives us a direct bijection to $\mathbb{N}$. We give an example. 

\begin{example}\label{EX2}
Consider the set $s=\{1,4,6,8,10,11\}$. The cardinality of this set is $6$ so in figure \ref{Cardinality-Grid} it will be  placed in column $5$. The row of this set is $811$, the Cantor pairing function gives the index $\pi(\phi_{car}(s)) = \pi(5,811) = 334147$ (see appendix \ref{A1}). The operations of interest on this set are:

\begin{itemize}

\item Cardinality: $|\{1,4,6,8,10,11\}| = 6$,

\item Lexicographic rank in the class of sets with cardinality $6$: $\sigma_6(\{1,4,6,8,10,11\})= 811$,

\item Cantor index: $\pi(\phi_{car}(s)) = \pi((6-1),811) = 334147$,

\item Sum: $\Sigma_{i \in s} i = 1 + 4 +  6 +  8 +  10 +  11 = 40$

\item Product:  $\Pi_{i \in s} i = 1 \times 4 \times  6 \times  8 \times  10 \times  11 = 21120$

\item Binary number: $\Upsilon(s) =\Sigma_{i \in s}  2^i    =2^1  +2^4 + 2^6 + 2^8 + 2^{10} + 2^{11} = 3410$
\end{itemize}

We can study the effect of arithmetical dilations of $\phi_{car}$ on the plane. A dilation $\phi_{prod}$ will place the set $s$ in column $21120$, and a dilation  $\phi_{sum}$ will place it in column $40$.  A dilation $\phi_{bin}$ will place the set $s$ in column $3410$ on the plane. Note also that the set $s$ is typical: the binary representation of our example set is $110101010010$. We have $12$ bits with $6$ ones.  The three arithmetical operations define injections in $\mathbb{N}^2$.  Since the arithmetical functions define, in comparison to cardinality,  an information expansion function in the direction of the $y$-axis, there will be a corresponding ``compression'' force in the direction of the $x$-axis. Metaphorically speaking one could say that the points are attracted to the line $y=0$ and want to stay as close as possible to it. 
\end{example}

\subsection{A general theory of dilations for arithmetical functions}\label{GENTHEORDIL}
We define a general construction for the study of the information efficiency of arithmetical functions on sets of numbers:

\begin{definition}\label{SORTBIJ}
$\phi_{\zeta}:  \mathfrak{P}(\mathbb{N}) \rightarrow \mathbb{N}$, a \emph{injection sorted on $\zeta$}, is a mapping of the form:

\begin{equation}\label{GEN}
\phi_{\zeta}(s) = \pi(\zeta (s),\theta_{\zeta(s)}(s))
\end{equation}

where $\pi$ is the Cantor function and: 
\begin{itemize}
 \item $\zeta: \mathfrak{P}(\mathbb{N}) \rightarrow \mathbb{N}$ is a \emph{general arithmetical function operating on finite sets of numbers}. It can be interpreted as a type assignment function that assigns the elements of $\mathfrak{P}(\mathbb{N})$ to a type (column, sort) represented as a natural number.   
 \item   $\theta_k: \mathfrak{P}(\mathbb{N}) \rightarrow \mathbb{N}$ is an index function for each type $k$, that assigns an index to the set in column $k$.The equation $\theta_{\zeta(s)}(s)=n$ should be read as: $s$ is the $n$-th set for which $\zeta(s)=k$. 
\end{itemize}
\end{definition}

By theorem \ref{COUNTCARD} we have that $\phi_{car}$ is efficiently countable. We can use $\phi_{car}$  as a \emph{calibration device} to evaluate $\phi_{\zeta}$. If $\phi_{\zeta}$ is a sorted injection the following mappings exists $\phi_{car}^{-1}: \mathbb{N}  \rightarrow  \mathfrak{P}(\mathbb{N})$ and $\phi_{\zeta}^{-1}: \mathbb{N}  \rightarrow  \mathfrak{P}(\mathbb{N})$  such that $\phi_{car}^{-1}(\phi_{car}(x))$, $\phi_{\zeta}(\phi_{\zeta}^{-1}(x))$  are identities in $ \mathfrak{P}(\mathbb{N})$. Given this interconnectedness we can always use $\phi_{car}$ to construct $\phi_{\zeta}$ algorithimically:

\begin{algorithm}
\caption{Compute the sorted index function $\theta_{\zeta (s)}(s)$  using $\phi_{car}$.}
\begin{algorithmic}[1]
\STATE $ k \gets 0 $
\STATE $i \gets 0$
\WHILE {$k \leq \phi_{car}(s)$}
	\IF  {$\zeta (\phi_{car}^{-1}(k)) = \zeta(s)$}
	\STATE $k \gets k +1$
	\STATE $i \gets  i + 1 $ 
	\ELSE
	\STATE $k \gets k + 1 $
	\ENDIF
\ENDWHILE
\STATE $\theta_{\zeta (s)}(s) = i$  
\end{algorithmic}
\end{algorithm}

\begin{theorem}\label{PSEUDO}
If the function $\zeta$ exists and can be computed in polynomial time then sorted injections $\phi_{\zeta}:  \mathfrak{P}(\mathbb{N}) \rightarrow \mathbb{N}$ of the form $\phi_{\zeta}(s) = \pi(\zeta (s),\theta_{\zeta(s)}(s))$ exist and can be computed in time exponential to the representation of $s$.  
\end{theorem}
Proof: We have to compute $\phi_{\zeta}(s) = \pi(\zeta (s),\theta_{\zeta(s)}(s))$. The functions $\pi$ and $\zeta$ can be computed in polynomial time. The function $\theta$ can be computed using $\phi_{car}$ with algorithm 1.  This algorithm runs in time exponential in the representation of $k$ which is the index of the set $s$: $\phi_{car}^{-1}(k) = s$. 
$\Box$

 Observe that $\phi_{\zeta}$ can be interpreted as an operationalisation of the axiom of choice: 
 
 \begin{definition}[Axiom of Choice]
For every indexed family $(S_{i})_{i \in I}$ of nonempty sets there exists an indexed family $(x_{i})_{i\in I}$ of elements such that $x_{i}\in S_{i}$ for every $i\in I$. 
 \end{definition}
 
Here $\zeta: \mathfrak{P}(\mathbb{N}) \rightarrow \mathbb{N}$ indexes the elements of $ \mathfrak{P}(\mathbb{N})$ and $\theta_k: \mathfrak{P}(\mathbb{N}) \rightarrow \mathbb{N}$ selects the elements of the indexed classes. Dilation theory allows us to study the effects of the spectrum of choices on $\mathfrak{P}(\mathbb{N})$ associated with elementary recursive functions. Metaphorically one would envisage the spectrum in terms of a rubber sheet lying over the space  $\mathbb{N}^2$. We measure the information efficiency resulting from the translations. When one starts to pull the sheet over the $x$-axis it starts to shrink over the $y$-axis. As long as one pulls with linear force the notion of a smooth information efficiency associated with a computable bijection over the space is conserved. When we pull with monotonely increasing force of a polynomial function, a computable bijection is still available, but the space starts to fluctuate with increasing discontinuities and the tension in the limit is infinite. When one pulls superpolynomially, using functions on sets of numbers, the rubber sheet starts to  disintegrate at finite dimensions and stops to be a computable bijection. It becomes an injection that is only occasionally locally computable on a point to point basis, although every stage of the translation can be reconstructed from the origin in exponential time.  When we pull the sheet super exponentially the shift is so big that the place of the image on the x-axis starts to carry information about the origin of all sets for which the image is on the line $x=c$. This is the case for primes and square numbers that have only two generating sets. When all points in the space $\mathbb{N}^2$ are mapped densely to the line $y=0$ the translation becomes the trivial inverse of the Cantor packing function, which is by definition information efficient again. This is the case for $\phi_{bin}$.

\subsection{A formal analysis of  the object $\phi_{sum}$} \label{SUMANAL}

Define $\Sigma s = \Sigma_{s_i \in s} s_i$ as the addition function for sets. Suppose  $s$ is the set associated with the index $(x,y)$ in $\mathbb{N}^2$ and $\pi(x,y)$ in $\mathbb{N}$. We have the sum function for sets: $\Sigma: \mathfrak{P}(\mathbb{N}) \rightarrow \mathbb{N}$.  The function $\epsilon_{\Sigma}: \mathbb{N}^2 \rightarrow \mathbb{N}^2$ defines an \emph{dilation} by a function $\Sigma$ of the form: 

\begin{equation}\label{SUMShift}
\epsilon_{\Sigma}(x,y) = (\Sigma s,\theta_{\Sigma s}(s))
\end{equation}

Here $\theta_{\Sigma}: \mathfrak{P}(\mathbb{N}) \rightarrow \mathbb{N}$ is the function that enumerates sets with the same sum. We have the functions: $\pi: \mathbb{N}^2 \rightarrow \mathbb{N}$ and $\phi_{car}: \mathfrak{P}(\mathbb{N}) \rightarrow \mathbb{N}$.  We define an injection $\phi_{sum}:  \mathfrak{P}(\mathbb{N}) \rightarrow \mathbb{N}$ sorted on sum:

\begin{equation}\label{SUM}
\phi_{car}(\epsilon_{\Sigma}(s) )= \phi_{sum}(s) = \pi(\Sigma s,\theta_{\Sigma s}(s))
\end{equation}

Observe that the shift is an injection. For every natural number there is only a finite number of sets that add up to this number. The injection is dense when sampled over the counter diagonal.  

There are for each set $s_k = \{1,2,\dots  ,k\}$, with $\Sigma_{x \in s_k }  x = u$, according to equation \ref{Bell} exactly $B_k$ partitions that add up to $n$. This means that there is a super exponential number of sets that add up to $u$ which gives, when sampled over the counterdiagonal, using $\pi$, a density of $1$ in the limit. This fact is remarkable, at every line $x=c$ we compress an infinite set to a finite one, but the density of the resulting index stays close to $x - \log x$.
 
 \begin{figure}[!t]
\centering
\fbox{\includegraphics[ width=5in]{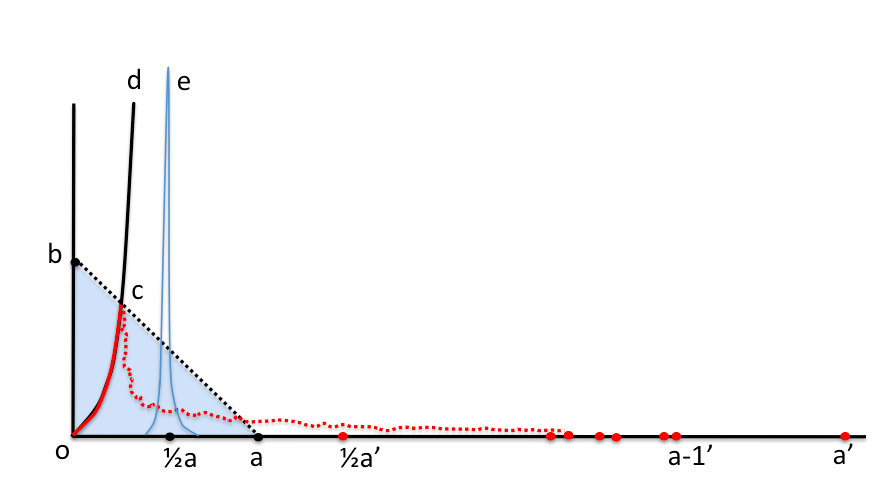}}
\caption{Schematic overview of the translation $\phi_{sum}$ generated by $\phi_{car}$ at a finite moment $t$. \label{SUMtranslation}}
\end{figure}

In figure \ref{SUMtranslation} we give a schematic overview of the translation $\phi_{sum}$ driven by the Cantor function for finite sets $\phi_{car}$ at a finite moment $t$. The area $(o,a,b)$ is the set of points that have been ``surveyed'' by $\phi_{car}$ up till now. The area below the red line illustrates the points that have been generated by  $\phi_{sum}$. The exponential function illustrated by the line $(0,c,d)$ marks the empty area where $\phi_{sum}$ will never generate images. Point $(a,0)$ is on line $y=0$. It codes the set $\{0,1,2,\dots a\}$ with the image $a'$ at $(\frac{1}{2}a(a-1),0)$. Point $a-1'$ is the image of $(a-1,0)$ at  $(\frac{1}{2}(a-1)(a-2),0)$. The gaussian distribution $e$ illustrates the density of the points around $\frac{1}{2}a$ that corresponds to sets of numbers, generated by a random selection of $\frac{1}{2}a$ elements from the set $\{0,1,2,\dots a\}$ that  will have an image in the neighborhood of $(\frac{1}{2} (\frac{1}{2}a (\frac{1}{2}a-1)),0)$. Note that only a tiny fragment of these sets has been surveyed by $\phi_{car}$. About al others we have no information at time $t$.

Observe that figure   \ref{SUMtranslation} describes the situation of the computation of  $\phi_{sum}(s)$ at any finite moment in time and that the line $(a,b)$ is a hard \emph{information boundary}. At any moment in time $t$ the characterization of the set  $\phi_{sum}$ by any finite segment of $\mathbb{N}$ is fundamentally incomplete. The images on the line $y=0$ form a discontinuous cloud of points. There is an exponential amount of relevant computations that we have not seen yet. 

\subsection{A formal analysis of  the object $\phi_{prod}$} \label{PRODANAL}

The information efficiency of multiplication is extensive. We do not lose information when we multiply although we lose the structure of the generating sets. This is an indication that $\phi_{prod}$ as a bijection is easier to compute than $\phi_{sum}$.  Define $\Pi s = \Pi_{s_i \in s} s_i$ as the multiplication function for sets. We define an injection $\phi_{prod}:  \mathfrak{P}(\mathbb{N}) \rightarrow \mathbb{N}$ sorted on product: 

\begin{equation}\label{PROD}
\phi_{prod}(s) = \pi(\Pi s,\theta_{\Pi s}(s))
\end{equation}

\begin{figure}[!t]
\centering
\fbox{\includegraphics[ width=5in]{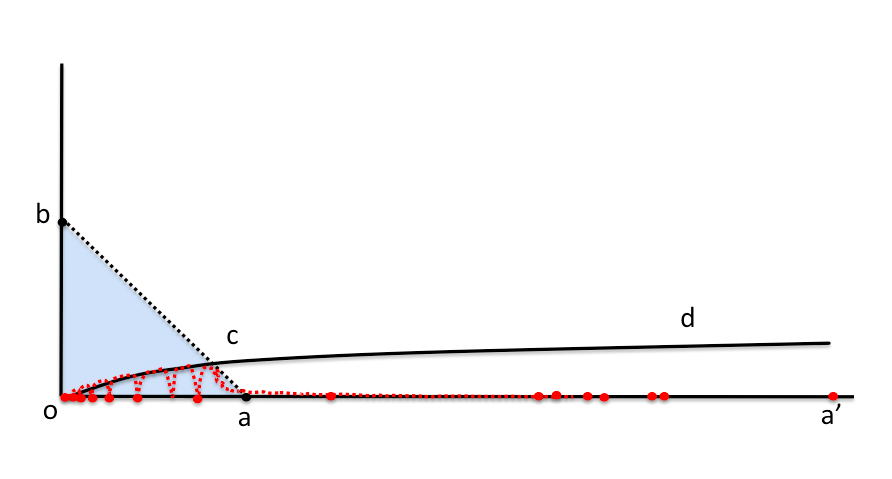}}
\caption{Schematic overview of the translation $\phi_{prod}$ generated by $\phi_{car}$ at a finite moment $t$. \label{PRODUCTtranslation}}
\end{figure}

Observe that this shift is also an injection. The complexity of $\phi_{prod}$ is fundamentally different from the complexity of $\phi_{sum}$.  For every natural number there is only a finite number of sets that multiply up to this number. There are for each set $s_k = \{1,2,\dots  ,k\}$, with $\Pi_{x \in s_k }  x = k!$, according to equation \ref{Bell} exactly $B_k$ partitions that multiply up to $k!$. This gives a density of the counter diagonal that is approximately logarithmic. 
 
 Since for every set $s$ the computation $\phi_{prod}(x) = n$ is different, the number $\phi_{car}(s) = \pi(x)$ is an index of the computation. Consequently we can only compute $\phi_{prod}^{-1}(n) = s$, if we already know $s$. This is exactly the case for prime numbers, they code the information about their own multiplicative history.  Since we know that primality is in $P$, the object  $\phi_{prod}$ is a bijection that at least partly, for certain columns $x=c$, can be computed in polynomial time both ways. 

\subsection{The axiom of choice at work: the divergence between $\pi(\phi_{car}(s))$  and $\Upsilon(s)$.\label{AXCHOICEWORK}}

Theorem \ref{NOTINSTRINSINF} tells us that there is no intrinsic information theory for  $\mathfrak{P}(\mathbb{N})$ and consequently there is no valid concept of a typical set in this domain. In this paragraph I show that different representational choices for $\mathfrak{P}(\mathbb{N})$ lead to diverging measurements of the information content of `typical' sets. The discovery was of vital importance in the definition of the notion of semi-countability, although now we have the concept the material  is only an illustration of the theory.

\begin{figure}[ht!]
\centering
\includegraphics[width=90mm]{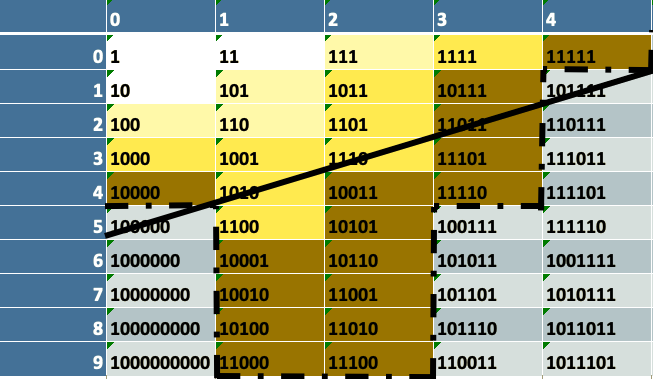}
\caption{The binary strings corresponding to the sets in figure   \ref{Cardinality-Grid}. The solid line indicates the iso-information border associated with the Cantor function that follows the counter diagonal. The dashed line indicates the iso-information border associated with the strings: this follows the binomial distribution and marks (in this case) the border between strings of length up to 5 and strings that are longer. 
  \label{Binary-Strings-Cardinality-Dilation}}
\end{figure}

Note that the values of the cardinality bijection and the power sum bijection vary considerably in the example above. We have  $\pi(\phi_{car}(s))  = 334147 \gg \Upsilon(s) = 3410$. This is interesting, since both operations define efficiently computable bijections to $\mathfrak{P}(\mathbb{N})$ and consequently  an endomorphism of $\mathbb{N}$. This endomorphism is illustrated in figure \ref{Binary-Strings-Cardinality-Dilation} where the binary strings related to the sets in figure   \ref{Cardinality-Grid} are depicted. We can draw iso-information lines that mark   borders between points with less and more information in the plots. The iso-information line of $\pi(\phi_{car}(x))$ follows the counter diagonal of the Cantor pairing function. The border that is associated with the length of the strings that code $\Upsilon(x)$ follows a binomial distribution given by the dashed line.

\begin{figure}[ht!]
\centering
\includegraphics[width=90mm]{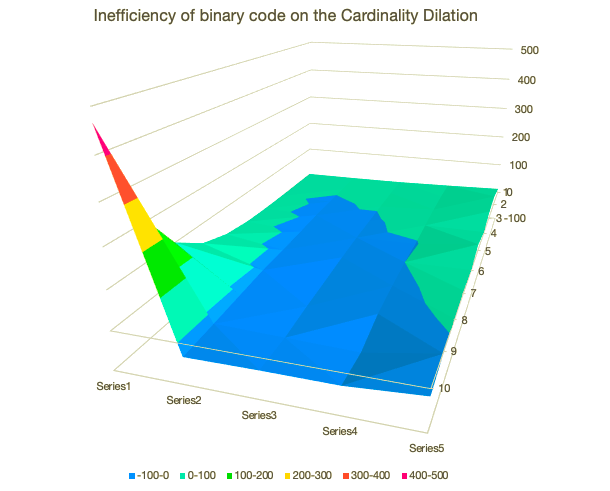}
\caption{A plot of the difference between the values generated by the Cardinality Dilation (see figure \ref{Cardinality-Grid}) and the Power Sum bijection (see figure \ref{Exponential-Sum}). The global difference between the two bijections is clearly visible. In the middle of the plot there is a valley in which the Power Sum bijection gives a more efficient coding of numbers. Around the valley are elevated parts of the landscape where the Cardinality bijection is more efficient. }
  \label{Difference-Plot-Exponential-Sum}
\end{figure}

Clearly visible in figure \ref{Binary-Strings-Cardinality-Dilation}  is the fact that the information coded in strings created  vaults of more efficient representations in the landscape of sets of numbers defined by $\phi_{car}$. A graphical overview of the difference between the two bijections is given in figure \ref{Difference-Plot-Exponential-Sum}. 

We now have two different representations of the set $ \mathfrak{P}(\mathbb{N})$, one as a bijection to $\mathbb{N}^2$ and one as a bijection to $\mathbb{N}$. Via the Cantor pairing function we also have an index for elements of $ \mathfrak{P}(\mathbb{N})$ in $\mathbb{N}$. The set from example \ref{EX2} seems to be typical, yet the index values vary greatly in the two representations: $334147$ versus $3410$.

We now investigate bijections for, what I call, \emph{string-typical sets}: sets of numbers based on random characteristic vectors. We show that the planar representation ordered on cardinality is inefficient for string-typical sets. 
  
 \begin{lemma} \label{POLYINV} $\phi_{car}$ is inefficient for typical sets. For a typical set $s$ with $\Upsilon(s) = 2^n$ we need at least $n - 2\log n - 5$ extra bits to represent $\phi_{car}(s)$. 
 \end{lemma}

Proof: We assume a typical set $s= \{s_1, \dots, s_{k}\}$ of numbers with cardinality $k$ for which the largest element has size $n=2k$. Furthermore we  assume that the elements of $s$ are ordered on numeric value $i<j \Leftrightarrow s_i > s_j$. Since we know that $s_1=2k$ and all other elements of $s$ are smaller than $2k$ we have the following upper bound for $\Upsilon(s)$:   

  \begin{equation}\label{UPSTYP}  
  \Upsilon(s) =\Sigma_{i \in s}  2^i    =2^{2k}  +2^{s_2} + \dots + 2^{2k}  < 2^{2k + 1}
  \end{equation}       
       
For the cardinality bijection we can compute a lower bound for the lexicographic index as:

 \begin{equation}\label{COMBSYSTTYP}
\sigma_k(s)= {2k \choose k} + \dots + {s_2 \choose 2} + {s_1 \choose 1} > {2k \choose k} 
\end{equation}
      
This gives the following lower bound for the Cantor index: 

 \begin{equation}\label{CNTRCOMBSYSTTYP}
 \phi_{car}(s) = \pi({2k \choose k} ,2k) := \frac{1}{2}({2k \choose k}  + 2k)({2k \choose k}  + 2k + 1)+2k >  \frac{1}{2}  {2k \choose k}^2 
 \end{equation}
 
 We have: 
   \begin{equation}\label{UPSTYPGAMMA}  
    {2k \choose k} =   \frac{ 2^{2k} \Gamma (k+ \frac{1}{2} )}
{\sqrt{\pi} \Gamma (k + 1)  }
>  \frac{ 2^{2k} \Gamma (k )}
{\sqrt{\pi} \Gamma (k + 1)  }
= \frac{2^{2 k}}
{\sqrt{\pi} k}
  \end{equation}

  Where $\Gamma(n) = (n-1)!$ is the gamma function.  Combining equations \ref{UPSTYP}, \ref{CNTRCOMBSYSTTYP} and \ref{UPSTYPGAMMA} we get:

   \begin{equation}\label{RATIO}
\frac{\phi_{car}(s)}{ \Upsilon(s)} >   \frac{  \frac{1}{2} (\frac{2^{2k}
    }
    {\sqrt{\pi} k
    })^2}{2^{2k + 1}} =  \frac{2^{2k - 1}
    }
    {4 \pi k^2} = \frac{2^{2k - 3}
    }
    {\pi k^2}
\end{equation}

Which gives for a set with an $\Upsilon$ index of length $n=2k$ bits (given equation \ref{UPSTYP}), an overhead for the $\phi_{car}$ index of  at least $2k -3 - \log_2 \pi - 2\log_2 k > n - 2\log_2 n - 5$ bits. $\Box$ 

We need almost twice as many bits to represent a typical set on the plane then as a binary number. This effect is exponential, so it is invariant under polynomially computable bijections of the plane. We make the following observation: 

\begin{observation}\label{TYPECALEDGEPLANAR}
This extreme behavior of string-typical sets is caused by a choice bias in their definition, which leads to an extreme unbalance in their planar coordinates in the cardinality bijection $\phi_{car}$. The phenomenon is illustrated in example \ref{EXCHOICE}. Since a set of size $k$ has $2^k$ subsets, for most sets the difference between their lexicographic index will be exponentially bigger than their cardinality, pushing their planar representation into edge regions of the plane in which planar representation is inherently inefficient (see par. \ref{PLANINF}). This illustrates the fact that a random selection of numbers from an initial segment of $\mathbb{N}$ under uniform distribution does not lead to a valid notion of a typical set. 
\end{observation}

The divergence between $\phi_{car}$ and $\phi_{bin}$ brings us a step closer to understanding conjecture \ref{ADHOCDES} and the effects of theorem \ref{NOTINSTRINSINF}. 

\begin{enumerate} 
\item The bijection $\phi_{bin}$ orders the set $\mathfrak{P}(\mathbb{N})$ on sums of the exponential functions in the one-dimensional space $\mathbb{N}$. 
\item The bijection $\phi_{car}$ orders the set $\mathfrak{P}(\mathbb{N})$ on cardinality in the two-dimensional space $\mathbb{N}^2$.  
 \item The bijection given in equation \ref{ENDMORF} between the two sets is efficiently computable but the information content of the data points that are connected by the bijection varies considerably. 
\item The embedding is topologically incoherent. There is in $\phi_{car}$ no neighborhood function that reflects the structure of $\phi_{bin}$. Sets with the same sum are scattered over the space $\phi_{car}$.
\item For string-typical sets the divergence can be estimated to be large, close to twice as many bits. See theorem \ref{POLYINV}. 
\item This gives two different information theories for sets  of numbers specified by equations \ref{CARDINFTH} and \ref{INFTHEORYPOWSUM}. In these two information theories the same set can have a different information content. 
\end{enumerate}
$\Box$

A binary number has two, what one could call, ``ontologies''; it is both a \emph{number} in the world of  $\phi_{bin}$ and a \emph{set of numbers} in the world  $\phi_{car}$. In the corresponding measurement theories the amount of information in \emph{the same object} varies. Lists code information inherently more efficient than isolated numbers, because they have a richer structure. The bijection $\phi_{car}$ gives one embedding of $\mathbb{N}$ in $\mathbb{N}^2$, the bijection $\phi_{bin}$ another.  For some numbers the set description is more effective, for some numbers it is the other way around. Specifically for finite typical subsets the number description is more effective. 

$\phi_{bin}$ and $\phi_{car}$ are two information theories that mutually expand and compress descriptive information in a topologically non-coherent way. $\phi_{bin}$ is linear, or one-dimensional;  $\phi_{car}$ is planar, or two-dimensional. $\phi_{bin}$ can be embedded in $\phi_{car}$ via its codification in terms of sets of numbers, but in this embedding the notion of topological coherence is lost. The elements of $\phi_{bin}$ are distributed over $\phi_{car}$ as a dense cloud of points, seemingly without any global structure. From the $\phi_{bin}$ perspective the  $\phi_{car}$ landscape looks like a chaotical structure of expandable and compressible isolated data points, and vice versa.

\subsection{Fractal dilations}\label{FRACDIL}

Given a number $r$ with a positional representation of length $n$ we have  $n = \lceil \log (r) \rceil$. 

\begin{definition}\label{SCALE}
For a natural number $n$ the scale of $n$ is defined as $\lceil \log_2 n \rceil$.
\end{definition}

The scale is equal to the number of bits we need to represent the number. Equation \ref{ENDMORF} describes an endomorphism on the set of natural numbers. For reasons of readability I repeat it here: 

\[\begin{tikzcd}
\mathbb{N}\arrow{r}{\pi^{-1}}   & \mathbb{N}^2\arrow{r}{\phi_{car}^{-1}} & \mathfrak{P}(\mathbb{N}) \arrow{r}{\Upsilon} & \mathbb{N}
\end{tikzcd} \]

Here $\Upsilon(x) =  \Sigma_{i \in X}  2^i $, with $i \in \mathbb{N}$ (see lemma \ref{UPS}). The construction of the endomorphism `travels' over the representation of the plane, via a bijection $\phi_{car}^{-1}$ to $\mathfrak{P}(\mathbb{N})$ with density $1$, to a bijection $\Upsilon$ to the set  $\mathbb{N}$ with density zero in the plane. 

The transformation defines a \emph{phase transition} from a \emph{planar} to a \emph{numerical} representation of information that in principle can be continued in unbounded number of cycles, leading to ever increasing topological distortions of the discreate plane. In this paragraph I investigate dilations that are `close' to $\Upsilon$ in terms of information efficiency and show that they lead to data sets with a fractal nature. Observe that $\Upsilon$ is based on powers of two. Data sets `close' to the one generated by $\Upsilon$ are based on \emph{injured} or \emph{randomized} natural numbers selected by the \emph{intervals} defined by powers of two. I define:

\begin{definition}\label{RND}
The function $rnd(2^i)$ selects a random number from the interval $[0,2^i]$ under uniform distribution. 
\end{definition}

\begin{definition}
A codebook \emph{template}\footnote{The term ``template'' was suggested to me by Daan van den Berg.} is a finite or infinite totally ordered set of numbers. The \emph{canonical template} is the infinite set of powers of two $P = [2^0,2^1,\dots]$.
\end{definition}

\begin{definition}\label{SCFENC}
A \emph{randomized Scale-free Codebook} based on the canonical template is a totally ordered set of unique numbers of the form $a_i = rnd(2^{i})$ with $i \in \mathbb{N}$. 
\end{definition}

 \begin{figure}[ht!]
\centering
\includegraphics[width=110mm]{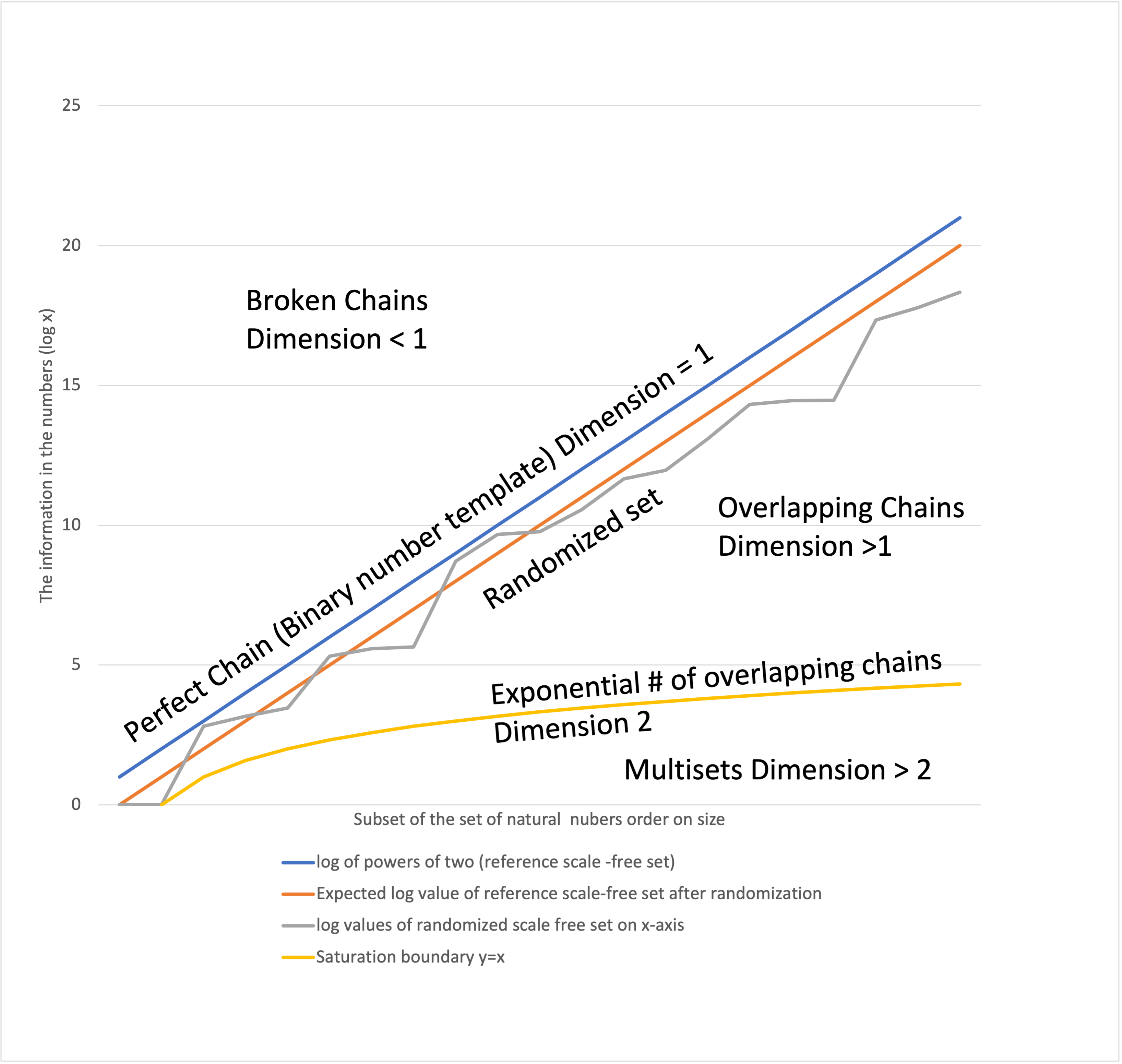}
\caption{The densities of $\phi_{sum}$ dilations of various types of subsets of $\mathbb{N}$ ordered by the information in the elements of the sets. The gray line in the figure is  based on a small scale-free randomized set that is analyzed extensively in the appendix in paragraph \ref{SCALEFSSPROB}.
 \label{INFSCALEFREEDENS}}
\end{figure}

Figure  \ref{INFSCALEFREEDENS} gives an overview of the fractal landscape that emerges from this analysis:
\begin{itemize}
\item  The blue line, $y=\log_2 x$, gives the information in the canonical code book  $P = [2^0,2^1,\dots ]$. The sum dilation of this set again generated the totally ordered chain of natural numbers and thus has dimension $1$ in the discrete plane. 
\item  Scale-free code books are self-similar to their own set of expected values: \[P^{\frac{1}{2}} = [\{0,1\},2^0, 2^1\dots ]\] where we replace $2^0$ by zero or one and $2^i$ by $2^{i -1}$. The resulting codebook has two representations for each number. 
\item The scale-free structure of sum dilations of such codebooks is clear from the fact that the codebook $P$ is \emph{self-similar} under multiplication with powers of two.
\item For the sum diliation of any subset of $\mathbb{N}$ that stays asymptotically above the $y=\log_2 x$ line, the chain will be broken. The fractal dimension of such a dilation in the discrete plane will be $<1$. 
\item For a bit string associated with a binary number $n$ the expected value of corresponding sum, using a  semi-countable code book generated under uniform distribution, will be $1/2 n$, which is represented by the red line $y=(\log_2 x)-1$ in figure \ref{INFSCALEFREEDENS}. 
\item The expected density of the sum dilation of such a  semi-countable code book generated under uniform distribution is $2$: i.e. each number is expected to be the sum of two sets.    
\item The most dense dilation of all semi-countable sets is $\phi_{sum}$, that simply maps the sums of all elements of $\mathfrak{P}(\mathbb{N})$. The information in this set is given by the orange line $y=\log_2 x$ in figure \ref{INFSCALEFREEDENS}. Any dilation on sums of this set almost fills the whole space $\mathbb{N}^2$, since each, sufficiently large, natural numbers have an exponential number of sets that add up to it. 
\item Below the orange line $y=\log_2 x$ the elements of the set have insuffcient information to be distinguished from other numbers. It is the region where infinite multisets of numbers reside. 
\end{itemize}
  
The fractal dimension of the scale-free sets described in this chapter can be derived from definition \ref{SCFENC}: at each scale the expectation of the amount of numbers we remove from the solution space $\mathbb{N}$ is $\frac{1}{2}$. We cannot use the standard definition of fractal dimension because the set  is countable. It has Hausdorff dimension zero, which gives $\frac{\log_2 1}{\log_2 2} = 0$. If we allow for estimates of traces of fractals in countable domains we can use, in some cases, the \emph{density} as the basis for the  fractal dimension. Let $A(n)$ be the solution set of a scale-free subset sum problem. If the density of the set is defined then according to equation \ref{DENSITY} we have: 

\begin{equation}\label{FRACDENSITY}
d(A) = \lim_{n \rightarrow \infty} \frac{c_{A}(n)}{n} = \frac{1}{2}
\end{equation}

The related dimension is $\log_2 d(A) = \log_2 \frac{1}{2} = -1$. For a finite subset sum problem the density is always defined so the fractal dimension helps us to estimate the expected density of the search space of scale-free problems. An analysis and an elaborate example of scale-free subset sum problems are given in appendix \ref{SCALEFSSPROB}. Similar results have been published in \cite{BERGADR21}.

\subsection{Discussion}
\begin{figure}[!t]
\centering
\fbox{\includegraphics[ width=5in]{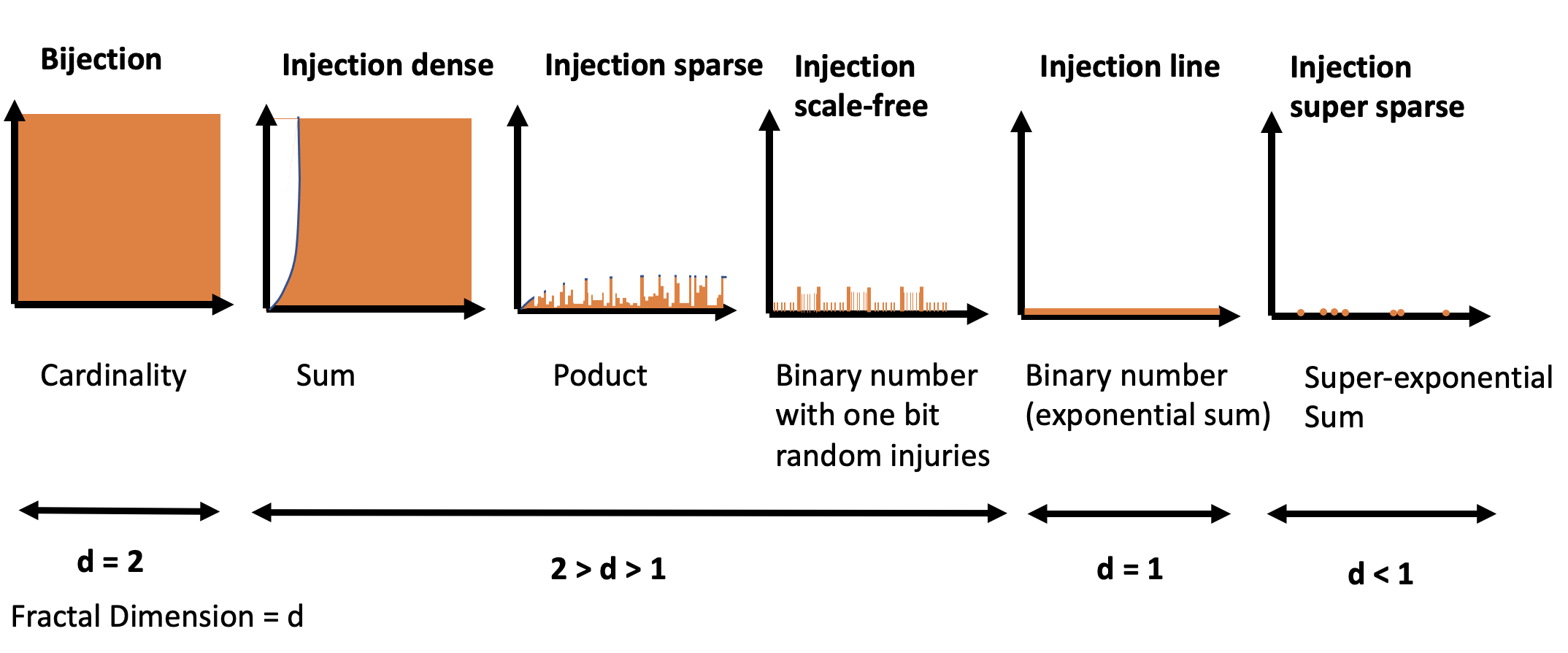}}
\caption{Schematic overview of the phase transitions of the semi-countable set $\mathfrak{P}(\mathbb{N})$ and their fractal dimension under recursive dilations of increasing information efficiency: \emph{cardinality}, \emph{sum}, \emph{product}, \emph{injured binary number}, \emph{binary number} (exponential sum)  and super exponential sum. The binary number dilation has a computable bijection to the plane again (see Equation \ref{ENDMORF}), which generates a class of efficiently computable unbounded topological deformations of the space. Consequently $\mathfrak{P}(\mathbb{N})$ has no intrinsic information theory.  \label{SEMICOUNTPHASETRANS}}
\end{figure}

 Semi-countable sets inhabit the (literally) chaotic region between the discrete plane $\mathbb{N}^2$ and the discrete number line $\mathbb{N}$ (see figure \ref{SEMICOUNTPHASETRANS}).  The objects $\phi_{sum}$ and $\phi_{prod}$ belong to the most complex constructions we will ever encounter in mathematics. $\phi_{sum}$ encodes everything there is to know about addition and is associated with the $P$ vs. $NP$ problem. $\phi_{prod}$ encodes all we can ever know about multiplication including the locations of the primes. It is assiciated with the \emph{factorization} problem. The class $\phi_{prod}$ is easier to compute than $\phi_{sum}$. The last class is exactly in the sweet spot of dilations where the tension between compression (information loss on the $y$-axis)  and expansion (information generation over the $x$-axis) is maximal. Only when we apply exponential transformations, the deformations become so extreme that the translations start to carry information about the original cells in their new location on the $x$-axis and we come close to the easier class $\phi_{prod}$. The  dilation becomes trivial for $\phi_{bin}$, but sets generated by operations close to this set have a \emph{fractal} structure. Some observations:  
\begin{itemize}
\item The Cantor function maps $\mathbb{N}$ onto $\mathbb{N}^2$. The objects  $\phi_{sum}$ and $\phi_{prod}$ correspond to an infinite compression of $\mathbb{N}^2$ over the $y$-axis that still has a dense representation in $\mathbb{N}$ when ``sampled'' via the Cantor function. 
\item Consider figures \ref{Cantor_Shift}, \ref{SUMtranslation} and \ref{PRODUCTtranslation}. Note that all natural numbers  and measurements in these schematic geometrical images have logarithmic representations. Consequently any measurable neighborhood $\epsilon > 0$ has exponential size in the limit. We cannot search any area or distance with measure $>0$ efficiently in the limit. 
\item By the same reasoning any measurable neighborhood that is more than $\epsilon >0$ removed from $x=0$ or $y=0$ will contain an exponential number of  cells with incompressible indexes in the limit. 
 \item Observe figures \ref{SUMtranslation} and \ref{PRODUCTtranslation}. At any finite stage of the construction of these images there is an exponential  amount of points on the line $y=0$ for which the corresponding image in the domain has not been found. These cells correspond to the descriptions ``the first set $x$ that adds/multiplies up to  $k$. 
\end{itemize}

The semi-countable sets form an natural link between the countable and uncountable sets: $\mathbb{N}$ is countable, $\mathcal{P}(\mathbb{N})$ is uncountable. The class  $\mathfrak{P}(\mathbb{N})$ could be seen as the class that stretches  countability to the maximum. The elaborate scheme to map the class  $\mathfrak{P}(\mathbb{N})$ to $\mathbb{N}$ via the Cantor function and combinatorial number systems is in fact an application of the axiom of choice: we define an infinite set of different subclasses that allows us to make an infinite set of choices for the first element.  If we then define super polynomial information generating operations on such a mapping, the fabric of countability is torn apart and we enter the chaotic universe of semi-countable sets. 

\subsection{Data structures and information}\label{DATASTRUANDINF}

\begin{figure}
    \centering
    \begin{minipage}{0.40\textwidth}
 \centering
\includegraphics[width=50mm]{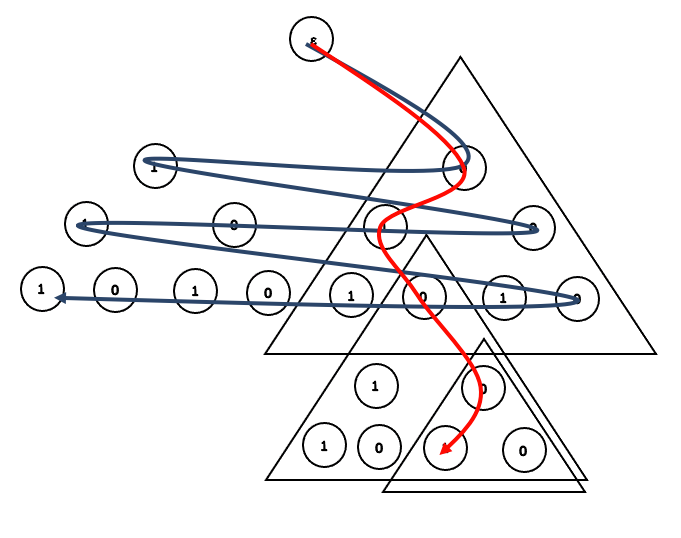}
\caption{Self-similarity in the infinite binary tree: $\{0,1\}^{\infty}$\label{SELFSIMBINTREE}}     
    \end{minipage}\hfill
    \begin{minipage}{0.45\textwidth}
  \centering
        \includegraphics[width=13mm]{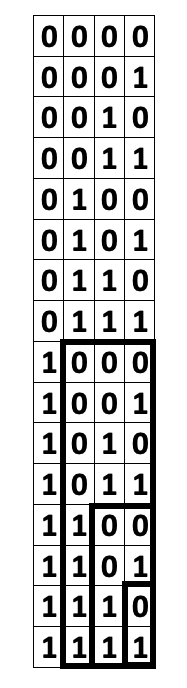} 
        \caption{Partial self-similarity in the set of binary strings of length $4$: $\{0,1\}^{4}$ \label{SELFSIMBINSTRNG}}        
    \end{minipage}
\end{figure}

A  function is recursive if the function itself is invoked in its definition, so by definition recursive functions operate on data structures that are self-similar. Self-similar data sets contain identical copies of themselves, so by definition they are infinite. An example of such a data  structure is the binary tree (see figure  \ref{SELFSIMBINTREE}). We can count the nodes of a binary tree breadth first (blue arrow). This is a simple process that can easily be implemented on a deterministic computer. At any stage of the counting process we investigate a set of nodes of the tree with the same path length $k$. Such a set can be characterized by a set of binary strings of length $k$ in which traces of the self-similar structure of the tree still exist (see figure \ref{SELFSIMBINSTRNG}). If we want to search the string depth first (red arrow) the situation is different. We have to make a choice between two options at each step and so with each step we generate one bit of information. This can not be done by a determinstic computer. We can say that there is a standard method to search the tree breadth first, but there is no such method for an exploration of the tree depth first. 

Note that both methods generate the same amount of information when identifying a node. In order to arrive at the node $100110$ the depth first process has to take $6$ binary decisions which generate $6$ bits of information in $6$ steps. The determinstic process has to check $2^5 + 2^2 +2^1 = 38$ cases. We have $log_2 38 \approx 5,25$ bits, which gives in discrete bits $\lceil 5,25 \rceil = 6$. The deterministic process creates in exponential time the same amount of information as the non-deterministic process in linear time.

The self-similar structures of these data sets suggest that we can analyze them as fractals, but this analogy is not straightforward. A well-known fractal is the Cantor set. It is constructed from the real interval $[0,1]$ by removing all the segments of numbers that contain the digit $1$ when expressed as a ternary number fraction with base $\{0,1,2\}$. Although this may appear to be a harmless operation it involves the manipulation of infinite objects and as such it lies beyond the jurisdiction of information theory. In fact almost all empirical reasearch into fractal data sets involves symbolic computation on finite representations. The concept of a fractal arises from the fact that in finite domains we observe repetative patterns created by recursive computational processes that can generate chaotic behaviour at any stage of their evolution.The estimates of the fractal dimensions of these data sets are by necessity made in finite discrete domains for which the Hausmann dimension is zero. 

What we study in this way is the notion of unbounded recursion in countable domains. In this context we can say that countable sets have a fractal structure and estimate their dimension. The Cantor set then consists of all ternary numbers that do not contain the number $1$. This set can be expressed as $\{0,2\}^{\infty}$. Consequently the cardinality of the set is equal to that of the set of binary numbers associated with $\{0,1\}^{\infty}$, i.e. it is uncountable. The density of the countable subset of the Cantor set consisting of rational numbers can be estimated for each $k$ as $\frac{2^k}{3^k}$, which gives the difference in expressivity between a ternary number system and a binary number system in terms of the length of their representations in a string of lenght $k$: 

\begin{equation}
\frac{\log 2^k}{\log 3^k}= \frac{\log 2}{\log 3} \approx 0.63093
\end{equation}

Since the set $\mathbb{Q}$ is countable, the formal Hausdorff dimension is $0$, but the analysis shows that we may for countable subsets of fratal sets use counting methods to estimate the fractal dimension.

We can use the elements of  $\{0,1\}^{\ast}$ to encode subsets of finite sets and to measure the information in these subsets

\begin{figure}
    \centering
    \begin{minipage}{0.55\textwidth}
 \centering
\includegraphics[width=80mm]{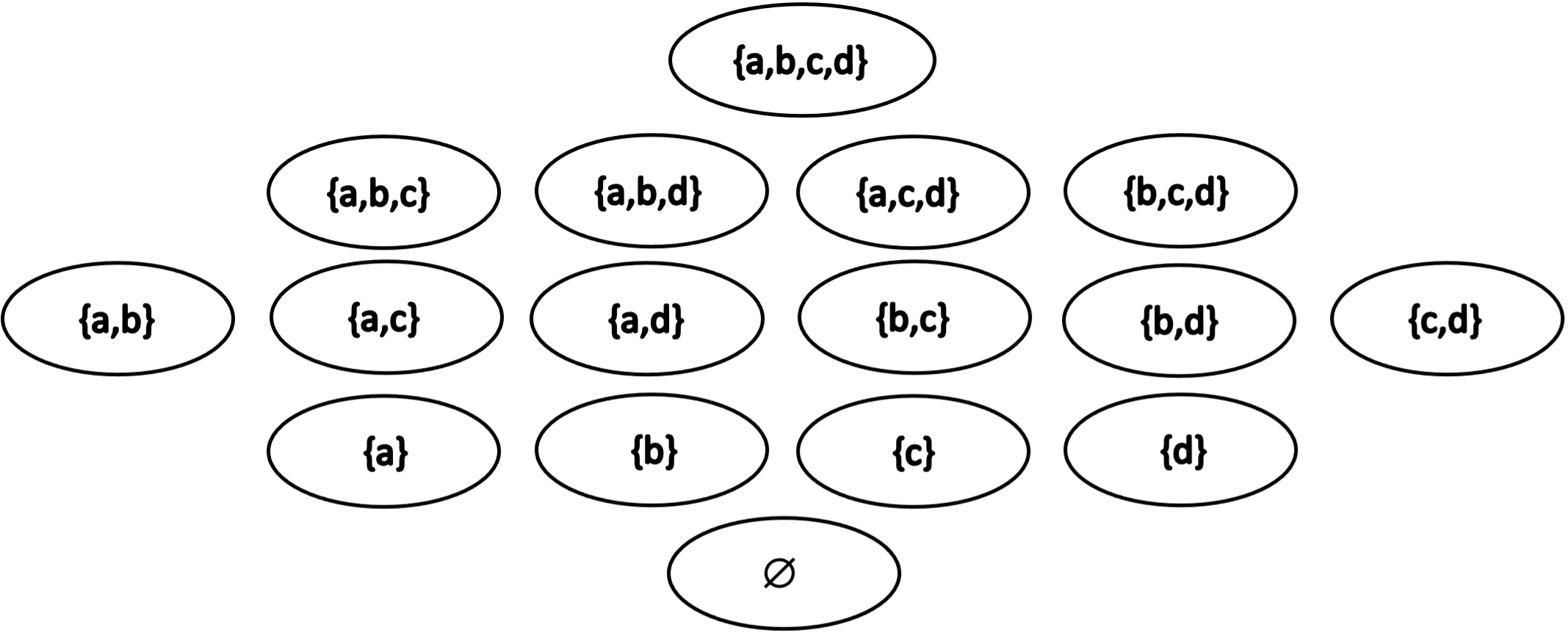}
\caption{Power set $\mathcal{P}(\{a,b,c,d\})$ ordered on cardinality (vertical) and lexicographically horizontal.\label{POWERSETLATTICE}}     
    \end{minipage}\hfill
    \begin{minipage}{0.35\textwidth}
  \centering
        \includegraphics[width=13mm]{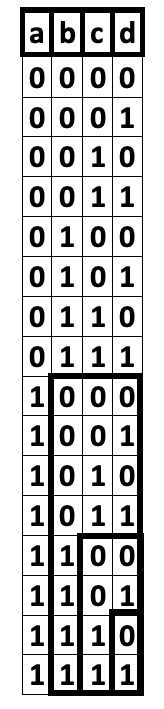} 
        \caption{The information structure of the set $\mathcal{P}(\{a,b,c,d\})$ \label{POWERSETCHARSTRING}}        
    \end{minipage}
\end{figure}

\begin{definition}[Conditional information in subsets]\label{CONDINFSUBS}
Let $S$ be a totally ordered finite set with cardinality $n$, where $s_i$ is the $i$-th element of the set. A subset of $S' \subset S$ is coded by a \emph{characteristic string}: a binary string $a$ of length $n$ such that $a_i=1$ iff $s_i \in S'$ and $s_i = 0$ otherwise. The conditional information in $S'$ with cardinality $k \leq n$ given $S$ is: 
\begin{equation}\label{INFINSUBSETS}
I(S'|S) = \log_2 {n \choose k}
\end{equation}
This is the value of the Hartley function (see equation \ref{HARTLEY}) applied to the number of ways we can choose $k$ elements from a set of $n$ elements. The set of binary strings of length $k$, written as $\{0,1\}^k$, gives the information structure of the power set of any finite set with $k$ elements. See figure \ref{POWERSETLATTICE} and \ref{POWERSETCHARSTRING}.  
\end{definition}

\begin{example}\label{EXCHOICE}The set $s = \{2,3,5\}$ is coded as a subset of a set of cardinality $6$ by the string $0010101$  and as a subset of a set of cardinality $10$ by the string $00101010000$. In the first representation the set is coded as an element of the powerset of $7$ elements and in the second as an element of the power set of $11$ elements. The conditonal information in the set $s$ in the first representation is $\log_2 {6 \choose 2} \approx 2.0$ and in the second representation $\log_2 {15 \choose 2} \approx 6.7$. There are infinitely many other strings that describe $s$. The effects of the bias generated by these different options is analyzed in section \ref{AXCHOICEWORK}.
\end{example}

For infinite sets this approach is not unproblematic. $\mathcal{P}(\mathbb{N})$ is  the powerset or set of subsets of $\mathbb{N}$. This set is associated with the object $\{0,1\}^{\infty}$ or coded by the infinite binary tree and its elements are \emph{transcendent}. We cannot manipulate them in our world. But even for sets that we do encounter in daily life measuring information is not always trivial. Consider the \emph{set of finite subsets of $\mathbb{N}$}, which I denote as $\mathfrak{P}(\mathbb{N})$. 

\begin{theorem}\label{NOTINSTRINSINF}
There is no intrisic definition of the amount of conditional information for elements of $\mathfrak{P}(\mathbb{N})$.
\end{theorem}
Proof: Definition \ref{CONDINFSUBS} does not apply. Since the domain of $\mathfrak{P}(\mathbb{N})$ is infinite, equation \ref{INFINSUBSETS} gives no finite value. For any $x \in \mathfrak{P}(\mathbb{N})$ the conditional information $I(x|\mathbb{N})$ is undefined.  $\Box$

I define a separate class for this object: 

\begin{definition}
$\mathfrak{P}(\mathbb{N})$ is \emph{semi-countable}. 
\end{definition}
The set has many interesting qualities: 
\begin{itemize}
\item The set is formally countable, but the proofs rely on the notorious \emph{axiom of choice}. Choices generate information, so if we need to make choices to count an infinite set of objects, then the individual elements of the counted set contain more information then the numbers we use to count them. 
\item $\mathfrak{P}(\mathbb{N})$ has a clear partial order on the subset relationship, but information theory is not monotone over set theoretical operations (see pargraph \ref{TYPSET} and \cite{Adri20}, par. 6.2) so this partial order cannot be used to prove countability. 
\item There is no natural mapping between the countable set $\{0,1\}^{\ast}$ and  $\mathfrak{P}(\mathbb{N})$ in terms of characteristic vectors. The set of $\{0,1\}^k$ characterizes the power set of finite sets of cardinality $k$ (see example \ref{EXCHOICE}).  
\end{itemize}

\begin{figure}[ht!]
\centering
\includegraphics[width=65mm]{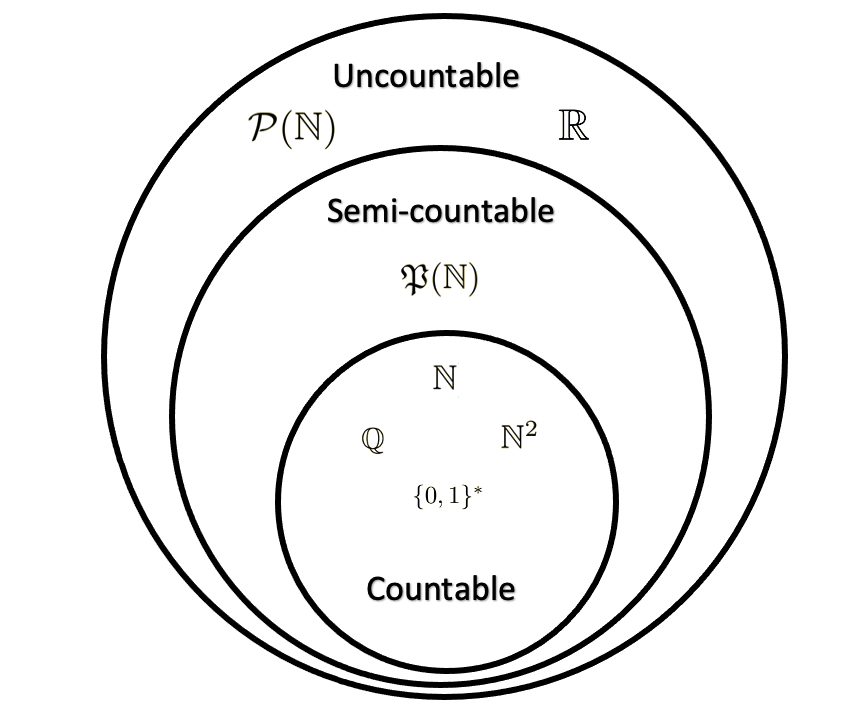}
\caption{A hierarchy of countability classes.\label{COUNTCLASS}}
\end{figure}

This analysis proves the existence of an infinite class of objects that is placed \emph{in between} the domain of the countable and the uncountable: semi-countable sets (see figure \ref{COUNTCLASS}). The fact that we need the axiom of choice to count these classes implies that semi-countable classes are more expressive: their elements contain \emph{more} information than normal countable objects. Through the choices we make we \emph{add} information to the elements which make them accessible. Observe that semi-countable sets are very common in mathematics. Any problem domain in which we consider finite subsets of totally ordered domains is semi-countable: e.g. sets of litterals, nodes in a graph, etc.

\section{The class $NP$ \label{THEORYI}}

The class $NP$ can  be characterized in terms of problems for which there exists an efficient checking function: the validity of a solution can be checked by a deterministic computer in time polynomial to the length of the input. Research over the past decades has shown that the problems in this class are a diverse lot and in general the application of information theory to these problems is difficult.  Two aspects are important for an analysis: 1) the domain of the problem class (e.g. graphs, k-dimensional spaces, logical formulas, sets of numbers, sets of strings etc.) and 2) the nature of the operation invoked by the checking function. 

We get a better insight in the nature of the problem in $NP$ when we apply the results of section \ref{DILATHEOR} to the analysis  finite semi-countable sets. We investigate dilations of finite power sets of  four operations in increasing order of expressiveness:  

\begin{enumerate}
\item Addition modulo $2$ reduces to one constant bit of information. This operation only preserves the parity of the sum of the set, so it gives maximum information reduction with maximal discrimination (since half of the natural numbers are even): 
   
  \[\Delta (x \oplus y) = 1  - ( \log_2 x + \log_2 y) \ll 0\]  
  
An example of the operation on the elements of the power set of $\{1,2,3,4\}$ is given in figure  \ref{DILSUMMODPROD}. All sets are compressed to the multiset: 

\[\{0,0,0,0,0,0,1,1,1,1,1,1,1,1,1\}\]

The associated set is: $\{0,1\}$. 

\begin{figure}[ht!]
\centering
\includegraphics[width=100mm]{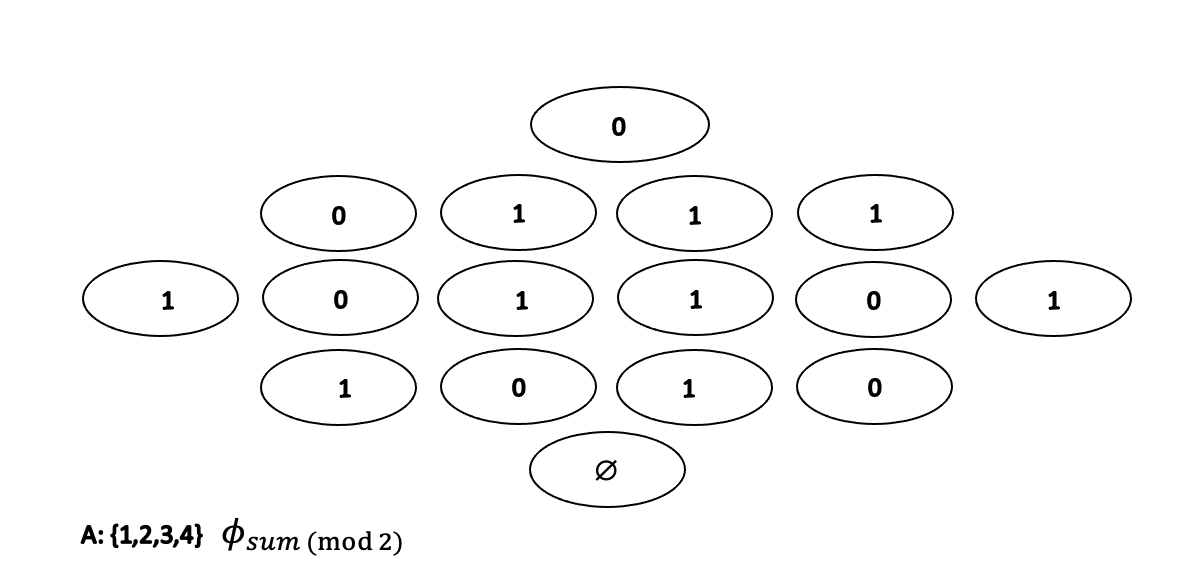}
\caption{The sum modulo $2$ dilation of the power set of the set $\{1,2,3,4\}$.}
  \label{DILSUMPROD}
\end{figure}

\item  Addition is information discarding:  
  
 \[\Delta (x + y) = \log_2 (x + y)  - ( \log_2 x + \log_2 y) < 0\]
 
 In figure \ref{DILSUMPROD} we see dilations of the set $\{1,2,3,4\}$. All sets are compressed to the multiset: 

\[\{1,2,3,3,4,4,5,5,6,6,7,7,8,9,10\}\]

The associated set is: $\{1,2,3,4,5,6,7,8,9,10\}$. 

\begin{figure}[ht!]
\centering
\includegraphics[width=100mm]{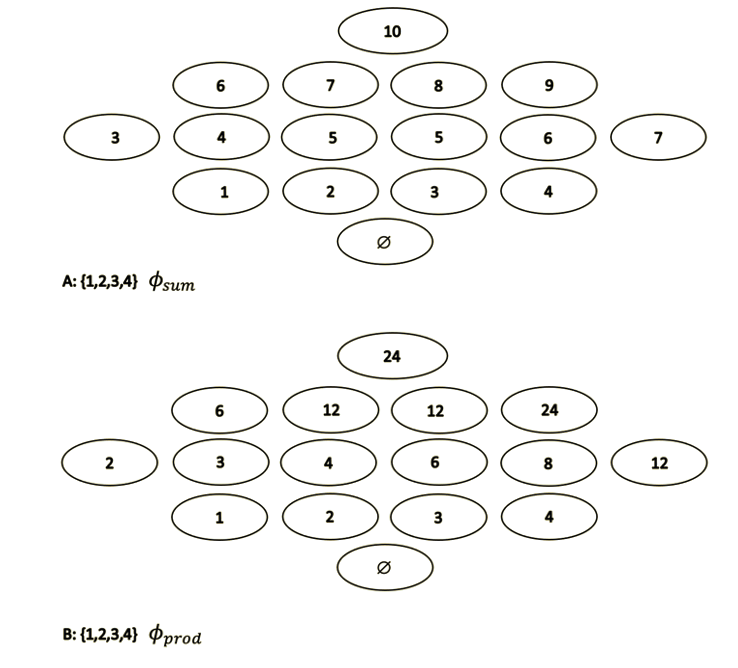}
\caption{Two dilations of the power set of the set $\{1,2,3,4\}$. On top the sum dilation, at the bottom the product dilation.}
  \label{DILSUMMODPROD}
\end{figure}
\item  Multiplication is information efficient in terms of the log measurement, but it loses the structure of the set. I prefer to call it \emph{extensive}: 

  \[\Delta (x \times y) = \log_2 x + \log_2 y - ( \log_2 x + \log_2 y) = 0\]

All sets are compressed to the multiset: 

\[\{1,2,2,3,3,4,4,6,6,8,12,12,12,24,24\}\]

The associated set is sparse: $\{1,2,3,4,6,8,12,24\}$. 
 
\item  The binary numbers operation preserves the whole structure of the input set:   

\[\Delta (bin(x,y)) = \log_2 2^x + \log_2 2^y - ( \log_2 x + \log_2 y) = x -\log_2 x + y - \log_2 y\]

The binary number operation, that computes the sum of the powers of two specified by the set is completely information preserving (see figure \ref{DILSUMPROD}): 

\[\{1,2,3,4,5,6,7,8,9,10,11,12,13,14,15\}\]

This is the result of applying the sum operation to the scale-free set $\{2^0,2^1,2^2, 2^3\}$.  The second set,  $\{0,1,3,6\}$, is random scale-free. It is generated by randomizing numbers on intervals given by: $2^0=1$ , $2^1=2$, $2^2=4$, $2^3=8$ given by the set $\{0,1,2,3\}$. The resulting multiset of numbers is chaotic: 

\[\{0,1,1,3,3,4,4,6,6,7,7,9,9,10,10\}\]

The associated set is irregular: $\{0,1,3,4,6,7,9,10\}$.

\begin{figure}[ht!]
\centering
\includegraphics[width=100mm]{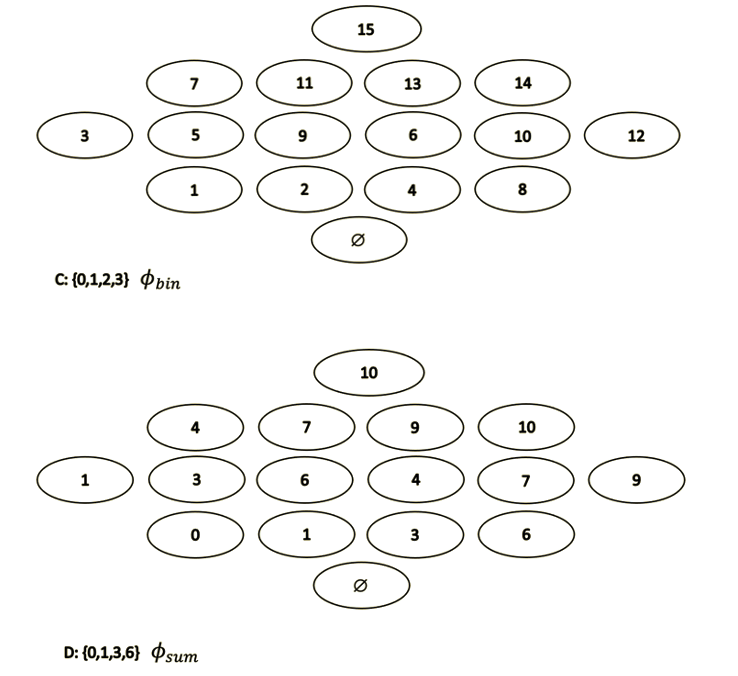}
\caption{On top the binary number dilation of the power set of the set $\{0,1,2,3\}$. This operation conserves the information in the original sets. At the bottom the sum dilation of thepowerset of a related scale free set $\{0,1,3,6\}$.The resulting information measurement is chaotic.  } \label{DILBINSUM}
\end{figure}

\end{enumerate}

 The abstract information structure of the powerset is the same for all examples and is  given by figure \ref{POWERSETCHARSTRING}. In line with the results of section \ref{DILATHEOR} there is no clear relation between cardinality of the sets and with typicality of the sets. For the sum operation there are many double values and there are values that are produced by both typical and non typical sets. There is no direct correlation between the values produced and the conditional information in the sets. The product operation conserves the extensiveness of information but discards structure. The values are further apart but have more tendency to cluster. The relation between the computed value of the subsets and their condittional information is lost. The corresponding decision problems are: 

\begin{problem}
Subset Sum modulo 2
\begin{itemize}
\item Instance: A finite set $A \subset \mathbb{N}$.
\item Question: Is there a subset $A' \subseteq A$ such that the sum modulo 2 of the elements of $A'$ is $1$? 
\end{itemize}
\end{problem}

Subset Sum modulo 2 is trivial: if there is at least one uneven number in the set the answer is yes.

\begin{problem}
Subset Sum 
\begin{itemize}
\item Instance: A finite set $A \subset \mathbb{N}$ and a number $n \in \mathbb{N}$.  
\item Question: Is there a subset $A' \subseteq A$ such that the sum of the elements of $A'$ is exactly $n$? 
\end{itemize}
\end{problem}

\begin{problem}
Subset Product 
\begin{itemize}
\item Instance: A finite set $A \subset \mathbb{N}$ and a number $n \in \mathbb{N}$.  
\item Question: Is there a subset $A' \subseteq A$ such that the product of the elements of $A'$ is exactly $n$? 
\end{itemize}
\end{problem}

Note that Subset product can also be formulated as Subset Sum on information in numbers on the basis of $x \times y = z \Leftrightarrow I(x) + I(y) = I(z) \Leftrightarrow \log x + \log y = \log z$.  Factorisation and Primality are special cases of Subset Product on maximally dense domains:  

\begin{problem}
Subset Product on dense sets (Factorization)
\begin{itemize}
\item Instance: A  number $n \in \mathbb{N}$ and a finite set $A = \{x \in \mathbb{N}^+| x < n\}$. 
\item Question: Is there a subset $A' \subseteq A$ such that the product of the elements of $A'$ is exactly $n$? 
\end{itemize}
\end{problem}

\begin{table}
\begin{center}
\begin{tabular}{ |c|c|c|c|c|} 
\hline
Problem class    & Output Information  & Efficiency & Range on $[0,n]$ & Solution Density  \\ 
\hline
Subset Product & $\log_2 x + \log_2 y$ & $\Delta(x \times y) = 0 $  &   $[0,n!]$ & Sparse:  $\frac{n^2}{n!}$  \\ 
Subset Sum &  $\log_2 (x +  y)$ & $\Delta(x + y) < 0 $ &    $[0,\frac{1}{2} n (n+1)]$ & Dense:  $\frac{n^2}{\frac{1}{2}n(n+1)}$ \\ 
Subset Sum mod 2&  $1$ & $\Delta(x \oplus y) \ll 0 $ &    $[0,1]$ & Dense:  $\frac{n^2}{2}$ \\ 
 \hline
\end{tabular}
\end{center}
\caption{Elementary mathematical operations and their information efficiency and solution density}
 \label{TAB0}
 \end{table}

An overview of the corresponding information efficiencies is given in table \ref{TAB0}. There is an interplay between the density of the domain and the information efficiency of the checking function. If we multiply all numbers from $1$ to $n$ we get value $n!$, but there are only $2^n$ possible subsets of values we can construct from the set of numbers. The set of numbers smaller than $n!$ we can generate on the basis of numbers smaller than $n$ is sparse. The reverse holds for addition. If we add all numbers from $1$ to $n$ we get $\frac{1}{2}n(n+1)$, which is way smaller than $2^n$. The set of solutions is dense. 
Factorization is both in $NP$ and in co-$NP$. Since the Subset Product problem can be seen as Subset Sum on the information of numbers, the difference in expressivity of multiplication and addition plays a minor role in the structure of the class of $NP$-complete problems. If the checking function reduces the information in the input to a constant, then the related problem is trivial.  In this way we get an elegant structure of  classes of decision problems based on additive checking functions (see table \ref{TABDENSE}).

\begin{table}
\begin{center}
\begin{tabular}{ |c|c|c|} 
\hline
Checking function  & Maximally dense domain   & Sparse domain \\ 
\hline
Addition of Information:  & Factorization  & Subset Product\\ 
 $ \Sigma_{x_i \in X} \log x_i = \log n$ & $NP$ \& Co-$NP$ & $NP$-complete \\ 
Information extensive &   & \\ 
 \hline
Addition of values:  & Subset Sum &Subset Sum  \\ 
 $ \Sigma_{x_i \in X} x_i =  n$ & $P$ & $NP$-complete \\ 
Information discarding &   & \\ 
 \hline
 Addition of values mod 2:  & Subset Sum mod 2 &Subset Sum  mod 2 \\ 
 $ \Sigma_{x_i \in X} x_i =  n (mod\ 2)$ & $P$ & $P$ \\ 
 Information reduced to constant &   & \\ 
 \hline
\end{tabular}
\end{center}
\caption{Elementary mathematical checking functions and their complexity classes over dense and sparse domains}
 \label{TABDENSE}
 \end{table}

We see two effects in table \ref{TABDENSE}. 
\begin{enumerate}
\item Problems on maximally dense domains have lower complexity than the ones defined on sparse domains. 
\item Problems with information discarding checking functions are easier than  the ones based on more efficient functions. 
\end{enumerate}
 
 This suggests that the hardness of problems in $NP$ is associated with a  balance between two forces: density of the domain and information efficiency of the checking function.\footnote{ Comparable issues are at stake when we analyze decision problems based on sets of literals. A literal is an atomic formula containing one variable, either negated or non-negated. The relevant operations are OR, AND, XOR. We can easily extend this to sets of literals but here the conditions for satisfaction are less transparent. A sequence of AND operations on a set of literals will have no satisfying truth assignment if it contains  a variable and its negation. A sequence of OR operations on a set of literals always has a satisfying truth assignment in this case. For a sequence of XOR operations the situation is more complicated. The interesting set of problems are the mixed sets of literals with multiple nested logical operations. The computational complexity varies: 
\begin{itemize}
\item CNF (conditonal normal form): A disjuntion of AND clauses. The computational complexity is in linear time since only one of the clauses needs to be satisfied.
\item DNF (disjunctive normal form): A conjuntion of OR clauses. The problem in NP. 
\item ANF (algebraic normal form): A conjuction of XOR clauses. The computational complexity is polynomial using gaussian elimination. 
\end{itemize}} Important is the following result:
 
\begin{theorem}\label{NOCOMPSEMI}
$NP$-complete problems are defined on semi-countable domains. 
\end{theorem}
Proof: immediate consequence of the fact that subset sum is $NP$-complete and the fact that $\mathfrak{P}(\mathbb{N})$ is semi-countable. $\Box$

All $NP$-complete problems can be reduced to each other and consequently the domains on which these problems are defined have at least the same expressivity as  $\mathfrak{P}(\mathbb{N})$. This implies that there is no intrinsic information theory for the domains of $NP$-complete problems and consequently no effective search algorithms can be developed. Information theories for semi-countable sets under effeciently computable transformations gives rise to wildly varying information measurements on the data structures and  no coherent neighbourhood structures are preserved. It is no wonder that known proof techniques do not work for semi-countable domains (See \cite{BAGI75}, \cite{RARU97}.)

\subsection{Subset Bitwise XOR: crossing the border to countable domains}

In this paragraph I will analyse a problem called Subset Bitwise XOR. The problem is interesting from a theoretical point of view since it is close to the subset sum problem. It could actually be described as subset sum modulo 2. It is defined on finite domains,  $\{0,1\}^{k}$ for each $k \in \mathbb{N}$, and be solved in polynomial time using Gaussian elimination.\footnote{I thank Peter Van Emde Boas and Patrick Weigert for pointing this out to me. } 

Shannon information theory gives a good theory to analyse the effect of recursive operations on random strings (See appendix in paragraph \ref{INFEFFRNDINP}). The application of the bitwise AND operation on two random strings will generate a string with roughly a quarter of ones, and thus destroy the randomness condition for the output. For the OR operation this will be roughly a quarter of zeros. Repeated application of bitwise AND operations on sets of bit strings will lead to highly compressible output. The same holds for the OR operation.  The sensitivity of XOR remains when we consider bitwise operations on sets of bit strings.\footnote{Historically the special character and irregular behavior of the XOR operation has been known ever since Jackson Wright discovered the munching squares on the PDP-1 in 1962. (see also, \cite{Wolfram2002} pg. 871). } We define:

\begin{definition}
Let $S$ be a set of $n$ binary strings of length $k$ then:
\[\oplus S =  (s_1 \oplus( \dots \oplus s_n) \dots)\]
 is the result of applying bitwise XOR to all elements $s_i \in S$. 
\end{definition}
Note that, since $\oplus$ is associative and commutative, the order of the operation is not relevant so we can write $\oplus S =  s_1 \oplus \dots \oplus s_n \dots$. By equation \ref{ENTSETSBITWBITVRND3} we have under random input:\footnote{See appendix in paragraph \ref{INFEFFRNDINP}} 

\begin{equation} 
H(\oplus S) = k
\end{equation}

and by equation \ref{INEFFSETSBITWBITVRND3}we have under random input:\footnote{See appendix in paragraph \ref{INFEFFRNDINP}} 

\begin{equation} 
\Delta(\oplus S) = k - nk
\end{equation}

This implies that the bitwise XOR operation lies in terms of information efficiency between the cardinality and the sum operation close to the place where the transition between problem in $P$ and $NP$ must be (See figure \ref{SEMICOUNTPHASETRANS}).   We can now define the related decision problem on the finite domain $\{0,1\}^k$

\begin{problem}
Subset Bitwise XOR (SB-XOR)
\begin{itemize}
\item Instance:  A finite set $A$ of $n$ binary strings of length $k$ and a target binary string $t$ of length $k$
\item Question: Is there a subset $A' \subseteq A$ such that the $\oplus A' = t$? 
\end{itemize}
\end{problem}

The relation between $n$ and $k$ defines the density of the problem space (see table \ref{XORDENSE}).  The value $k=n$ marks a phase shift in the search space. Below that value there are less subsets of vectors than there are binary strings of length $n$. So not all strings can be coded. For the value $k= c$ the problem becomes trivially tractable. 
\begin{table}
\begin{center}
\begin{tabular}{ |c|c|c|c|} 
\hline
$\#$-vectors & $n=2^k$ & $n=k$ & $n=c$ \\
\hline
Length input  & $k(2^k + 1)$ & $k(k+1)$ & $k(c + 1)$ \\
\hline
Density  & Dense &Polynomial & Sparse \\
 \hline
\end{tabular}
\end{center}
\caption{Density effects of Subset Bitwise XOR}
 \label{XORDENSE}
 \end{table}

The value $n=k$ marks the sweet spot of the harder cases.  Because of the noise in the vectors the solution space will have a sort of fractal distribution over the $2^n$ possible strings: some sets can be constructed by only one subset, some by more than one. According to the pigeon hole  principle, since there are only $2^n$ possible subsets and exactly $2^n$ possible strings to be encoded we have at least $d-1$ sets that have no solution for every vector that can be coded by $d$ subsets. This motivates the following definition: 

\begin{problem}
Canonical Subset Bitwise XOR (CSB-XOR)
\begin{itemize}
\item Instance:  A finite set $A$ of $n$ independent random binary strings of length $n$ and an independent target binary string $t$ of length $n$
\item Question: Is there a subset $A' \subseteq A$ such that the $\oplus A' = t$? 
\end{itemize}
\end{problem}

Bitwise XOR on binary strings is a weaker operation than Subset Sum on binary numbers. It is in fact  Subset Sum modulo 2 on binary numbers. There are many ways to illustrate the difference between Subset Bitwise XOR and $NP$-complete problems: 
\begin{itemize}
\item The information efficiency of Bitwise XOR is constant and consequently lower than that of addition, but higher than cardinality. 
\item Bitwise XOR is in $P$ via Gaussian elimination. The columns and rows of an instance of subset bitwise XOR can be permuted at liberty. This is also the case for the cardinality operation, but not for the sum operation. It seems this internal `coherence' preserved by the carry-overs of the sum operation is important for the hardness of the subset sum problem class. 
\item Bitwise XOR is defined on finite domains. Addition is closed for the whole set $\mathbb{N}$. 
\item The problem class is \emph{constraint-free}.  Any (random) string of adequate length defines a problem instance. Any binary string of length $p \times q$ defines a normalized SB-XOR problem of $p$ vectors of length $q$. There are $2^{p \times q}$ of such problems and if one wants to be sure there are no double vectors, the size of the set of problems is somewhat smaller but still astronomical ${2^q \choose p}$. 
\item It destroys information at \emph{maximum speed} of one bit per operation. 
\item It is sensitive under \emph{maximum entropy}: the yes- no-instances are densely distributed and one bit removed from each other. 
\end{itemize} 

 It is clear that there might be an abundance of circumstances in which problems like Subset Bitwise XOR (SB-XOR) occur in nature. Any situation in which we have $n$-independent Bernoulli processes generates SB-XOR problems. If we let the $n$ processes run for $nk$ steps we have a set of $nk$ binary vectors of length $n$ and we can ask the question wether a subset of these vectors absorb another vector.

\section{A taxonomy for decision problems}\label{TAXDECPRO}

The results of the previous paragraphs enable us to formulate a taxonomy of problems in $NP$. Three dimensions are important: expressiveness of the domain, density of the domain, information efficiency of the checking function (see table \ref{THREEDIM}): 

\begin{table}
\begin{center}
\begin{tabular}{ |c|c|} 
\hline
Expressiveness  & Countable $<$ Semi-countable $<$ Uncountable  \\
\hline
Density   & Sparse $<$ Maximal \\
\hline
Information efficiency & Constant $<$ Discarding $<$  Extensive $<$ Conserving \\
\hline
\end{tabular}
\end{center}
\caption{Categorisation of decision problems in three ordered dimensions: expressiveness of the domain, density of the domain, information efficiency of the checking function.   }
 \label{THREEDIM}
 \end{table}

Table \ref{NPTAX} gives an overview of some important problems in this taxonomic space. I make some observations: 
\begin{itemize}
\item The overview supports the insight that decision problems in $NP$ are hard when the checking functions do not conserve information and search domain is sparse.
\item For each of these problems there is a phase transition between sparse and complete representations 
\item  For all the hard problems the \emph{information structure} of the problem is \emph{independent} of the \emph{set theoretical structure}. 
\item For Subset bitwise XOR this is because the information efficiency of the checking function is constant. It conveys no information about the number of sets involved in the computation. For the other problems this is caused by the semi-countable nature of the domain and the lack of an intrinsic information theory. 
\item The semi-countable domain is over-expressive. It is easy to compute efficient transformations beween semi-countable representations but it is not possible to measure the information effects of these transformations precisely. 
\item When the checking function conserves information, as in the binary number dilation, the set theoretical structure of the problem is irrelevant. We can simply reconstruct each set from its computed value. There is no phase transition. 
\item Close to the binary sum dilation the search space has a fractal structure, which illustrates the fact that the related decision problems are hard. 
\item For simple multiplication the discriminate power is still insufficient to convey the structure of the generating set. It is in fact a variant of subset sum of the log values of the numbers.The nature of the phase transition to factorisation is unclear. 
\end{itemize}
\begin{table}
\begin{center}
\begin{tabular}{ |c||c|c|c|c|c|} 
\hline
Problem class &Domain & Expressiveness  & Density &  Efficiency & Class  \\
\hline \hline

Subset Bitwise XOR   & $\{0,1\}^{\ast}$ &  Countable & Sparse & Constant & $P$\\
\hline

Subset Bitwise XOR   & $\{0,1\}^{\ast}$ &  Countable & Maximal & Constant & $P$\\
\hline
Subset sum   & $\mathfrak{P}(\mathbb{N})$ &  Semi-countable & Sparse & Discarding & $NP$-complete \\
\hline
Subset sum   & $\mathfrak{P}(\mathbb{N})$ &  Semi-countable & Maximal & Discarding & $P$ \\
\hline
Subset product  & $\mathfrak{P}(\mathbb{N})$ &  Semi-countable & Sparse & Extensive & $NP$-complete\\
\hline
Subset product   & $\mathfrak{P}(\mathbb{N})$ &  Semi-countable & Maximal & Extensive & ? \\
\hline
Subset binary   & $\mathfrak{P}(\mathbb{N})$ &  Semi-countable & Sparse & Conserving & $P$ \\
\hline
Subset binary   & $\mathfrak{P}(\mathbb{N})$ &  Semi-countable & Maximal & Conserving & $P$ \\
\hline
\end{tabular}
\end{center}
\caption{Categorisation of various decision problems in increasing order of expressiveness of the domain and information efficiency of their checking functions. Note that subset product on complete domains is factorisation. Its status is unclear.  }
 \label{NPTAX}
 \end{table}

 \section{Conclusion}
DIT is non-stochastic but via the concept of non-deterministic computing its relation with Shannon information is relatively unproblematic. It is conceptually closer to Kolmogorov complexity in the sense that it studies deterministic computational functions. The difference is that Kolmogorov complexity quantifies over all computable functions with unbounded time and therefore is uncomputable. Differential information theory investigates the other end of the spectrum: it studies efficiently computable bijections between infinite data sets and the natural numbers. It is known that Kolmogorov complexity can be approximated from below and differential information theory can be said to explore initial segments of these approximations.

The main motivation for the development of differential information theory is the formulation of a theory of information measurement that is 1) \emph{non-stochastic},  2) \emph{non-asymptotic}, and 3)\emph{efficiently computable}. The choice for recursive functions on natural numbers helps us to reach these objectives.  We focus on the \emph{generation} of information through computation.  Since recursion theory is axiomatic, we do not have to bother about the selection of a reference universal Turing machine: the set of recursive functions is our natural choice.   An additional bonus is the fact that recursive functions have been studied to a much larger extend than programs for Turing machines and there is a rich history of many deep results that we can use in our analysis. Referring to the issues mentioned in the Introduction of this paper we can draw the following conclusions: 

 \begin{enumerate}
 \item We have (at least partly) unified the theory of quantitative information measurement by developing a theory of pure information measurement defined on recursive functions. A soon as this theory is applied to existing quantitative theories in space and time we get a clear picture of the interaction between information and computation. A non-stochastic version of Shannon information is associated with non-deterministic
 walks in discrete spaces. Such an approach can easily be enriched with probability theory, but this is not essential for the development of pure \emph{generative} information theory. At the other end of the spectrum Turing machines are not essential for the development of a \emph{descriptive} theory of information. A recursive function enumerating all possible recursive functions will do the trick equally well. The information production of deterministic Turing machines decays logarithmically with the time. Non-deterministic Turing machines cover the middle ground between Shannon information and deterministic machines. 
\item The interaction between information and computation can be studied on the basis  the theorems \ref{INFEFFRECFUNC} and \ref{ASSYM}: the time efficiency of a Turing machine computing a primitive recursive function can be estimated from the information efficiency of the related recursive function. For general recursive functions this is not possible because the information deficiency is not defined. For search and decision problems this implies that we can either implement a search algorithm as a primitive recursive function performing exhaustive search (for $i=1$ to $n$ do $\dots$) or as a more efficient general recursive function (do until $\dots$). The first option is by definition inefficient. The performance of the second solution depends on the amount of information generated by the solution of the search problem. 
 \item A proof of separation between the classes $p$ and $NP$ in an information theoretical context is in principle simple: if we show that the solution of a problem in $NP$ implies the generation of $k$ bits of information then a deterministic Turing needs at least $2^k$ computation steps to solve it. For such an approach to work, we need to make an extensive analysis of the information efficiency of primitive recursive checking functions like addition and multiplication, the exact measurement of information in sets of numbers and related decision problems like subset sum, subset product and factorisation. We will analyse these issues in the context if DIT in more detail in a separate publication.  Extensions to continuous time and space might be developed along the lines of the material presented in \cite{chirikjian2009stochastic}. 
  \item Using equation \ref{VOLBIN} we can map the set of natural numbers to multidimensional spaces, like $\mathbb{N}^k$, but this gives only a superficial information theoretical characterisation of these sets. The Cantor pairing functions (equation \ref{CANTPAIRFUNC}) are a special case for $k=2$. The Fueter - P\'{o}lya theorem \cite{FP23} states that the Cantor pairing function and its symmetric counterpart $\pi'^2(x,y)=\pi^2(y,x)$ are the only possible quadratic pairing functions. The original proof by Fueter and P\'{o}lya is complex, but a simpler version was published in \cite{vsemirnov2002two} (cf. \cite{nathanson2016cantor}). 
  The  Fueter - P\'{o}lya \emph{conjecture} states that there are no other polynomial functions that define such a bijection. The conjecture is open for more than 100 years. In the context if DIT the issue is relevant because it reveals a lack of understanding of spatial mappings using primitive recursive functions.   We will analyse this question and related issues in a separate publication. 
  \item  The operation of taking the logarithm takes us from a mathematical world in which our intuitions are straightforward to a world in which the behaviour of objects is in many ways counter-intuititive. For instance, in the limit the identity function $f(x) = x$ and the logarithm $f(x) = \log x$, according to common mathematical practice, are considered to grow to the same ``size'':
\begin{equation}\lim_{x \rightarrow \infty} \log x = \lim_{x \rightarrow \infty}  x = \infty, \end{equation}
but their derivatives are fundamentally different: 
\begin{equation}\lim_{x \rightarrow \infty} \frac{d}{dx} \log x = 0 \neq\lim_{x \rightarrow \infty} \frac{d}{dx}  x = 1 \end{equation}
It is quite embarrassing that the behaviour in the limit of the  function we use to measure information is ill understood. We believe that the nature of the exponentiation and the logarithmic function as \emph{type conversions} is key to clearing this issue and it is in principle possible to develop a (fragment of) theory of \emph{transfinite information measurement}. 
 \end{enumerate}
 
Information is a natural epiphenomenon of our mathematical descriptions of the world.  Whenever in physics we are interested in numbers as a characteristic of extensive properties, like volume, mass, entropy, information, size or distance,  we study measurements in their logarithmic representation.   Every mathematical model of the world is also a theory of information about the world. 

\section{Acknowledgements}
I thank Peter Van Emde Boas for many inspiring discussions during a period of more than thirty years. The exchange of ideas with David Oliver, Duncan Foley and Amos Golan has been a great inspiration, as were the visits to the Info-Metrics Institute in the past years. I thank the University of Amsterdam and especially Cees de Laat for allowing me the leeway to pursue my research interests. I thank Daan van den Berg and his student Ruben Horn for the empirical analysis of scale-free sets, Patrick Weigert for his analysis of subset bitwise XOR and Rini Adriaans for a life time of continuous support.

\bibliographystyle{plain}
\bibliography{Differential_Information_Theory}

\section{Appendix: The information structure of data domains}\label{INFSTRUDADO}

\begin{definition}Suppose $m, n, o, p, \dots$  are symbols and $\oplus$
 is a tensor or concatenation operator. We define the class of sequences:
\begin{enumerate}
\item Any symbol is a sequence.
\item If $alpha$ and $\beta$ are sequences then $\alpha \oplus \beta$ 
is a sequence.
\end{enumerate}
\end{definition}
Sequences can be ordered on their structural qualities (See \cite{Adri20}, par. 5.1.2.).  We have the folllowing information discarding operations on sequences: 
\begin{enumerate}
\item Contraction:
\[(m \oplus m) = m\]
Contraction destroys information about frequency in the sequence.
\item Commutativity:
\[m \oplus n = n \oplus m\]
Commutativity destroys information about order in the sequence. 
\item Associativity:
\[(p \oplus (q \oplus r)) = ((p \oplus q) \oplus r)\]
Associativity destroys information about nesting in the sequence. 
\end{enumerate}

There are eight different tensor operations. Table \ref{TENSORRANK} gives some that are of interest to us. Systems of sequences with \emph{contraction, commutativity and associativity} behave like \emph{sets}. If  we leave out \emph{contraction} we get \emph{multisets}. When we remove associativity we get strings.

\begin{table}
\begin{center}
\begin{tabular}{ |c|c|c|c|c|} 
\hline
Operator  & Contraction   & Commutativity & Associativity & Domain \\ 
\hline
\hline
 $\oplus_{\not c, \not c, \not a}$ & N & N & N & Sequence \\
 \hline
  $\oplus_{\not c, \not c, a}$ & N & N & Y & String \\
  \hline
   $\oplus_{\not c, c, a}$ & N & Y &Y & multiset \\
 \hline
 $\oplus_{c, c, a}$ & Y & Y &Y & Set \\
 \hline
\end{tabular}
\end{center}
\caption{Tensor operator ranked in terms of decreasing information expressivity under the influence of structural rules}
 \label{TENSORRANK}
 \end{table}

 Measuring information in these various types of data sets is non trivial. The number of ways to write down $n$ pairs of balanced brackets is given by the \emph{Catalan} number: 
\begin{equation}\label{Catalan}
C_n = \frac{1}{n-1}{2n \choose n }
\end{equation}

The \emph{Stirling number of the second kind} is the number of ways a set of $n$ numbers can be partioned into $k$ non-empty subsets: 
\begin{equation}\label{Stirling}
\stirling{a}{b}= \frac{1}{k!} \Sigma_{i=0}^{k} (-1)^i{k \choose i }(k-i)^n
\end{equation}

The \emph{Bell} number counts the number of possible partitions of a set: 
\begin{equation}\label{Bell}
B_n= \Sigma_{k=0}^{n}\stirling{n}{k} 
\end{equation}

Let $\oplus_{\not c, \not c, \not a}: \mathbb{N}^2 \rightarrow \mathbb{N}$ be a tensor operator that is \emph{non-contractive}, \emph{non-commutative} and \emph{non-associative}.  Suppose $\# \oplus_{\not c, \not c, \not a}$ is the number of different ways we can write $n-1$ pairs of brackets in $n!$ different sequences of strings: 

\begin{equation}\label{NcNa}
\#\oplus_{\not c, \not c, \not a}(n) = n!C_{n-1} = n! \frac{1}{n}{2(n-1) \choose (n-1) }
\end{equation}

Suppose $\oplus_{\not c, \not c, a}$ is \emph{commutative} and \emph{non-associative}, then, to our knowlegde, there is no formula that describes the number of different computations for $n$ objects, but it is, for large enough $n$, certainly bigger than the total number of unbracketed partitions, which is given by the Bell number, and thus  smaller than $\#\oplus_{\not c, \not c, \not a}(n)$ but still super exponential: 
\begin{equation}\label{cNa}
\#\oplus_{\not c, \not c, \not a}(n) > \#\oplus_{\not c, \not c, a}(n) > B_n > 2^n
\end{equation}

If $\oplus_{\not c,  c, a}$ is \emph{commutative} and \emph{associative}, then any sequence of $n$ objects collapses into a multiset.  Which gives the following correlation between structural rules and the number of computations defined on sets: 
\begin{equation}\label{ca}
\#\oplus_{\not c, \not c, \not a}(n) = n!C_{n-1}  > \#\oplus_{\not c, \not c, a}(n) > B_n > 2^n > \#\\oplus_{\not c,  c, a}(n) = 1
\end{equation}

\subsection{Appendix: The $\mu$-operator for unbounded search}\label{MUOPRECURSIVE}

In the world of Turing machines this device coincides with infinite loops associated with undefined variables. It is defined as follows in \cite{Odi16}:

\begin{definition}\label{DEFMU} For every 2-place function $f(x,y)$ one can define a new function, $g(x) = \mu y[f(x,y)=0 ] $, where $g(x)$ returns the smallest number y such that $f(x,y) = 0.$  
\end{definition}
Defined in this way $\mu$ is a partial function. One way to think about $\mu$ is in terms of an operator that tries to compute in succession all the values $f(x,0)$, $f(x,1)$, $f(x,2)$, ... until for some $m$ $f(x,m)$ returns $0$, in which case such an $m$ is returned. In this interpretation, if $m$ is the first value for which $f(x,m) = 0 $ and thus $g(x) = m$, the expression  $\mu y[f(x,y)=0 ]$ is associated with a routine that performs exactly  $m$ successive test computations of the form $f(x,y)$ before finding $m$. Since the $\mu$-operator is unbounded $m$ can have any value. Consequently the descriptive complexity of the operator in the context of primitive recursive functions is unbounded. 

In the context of recursive functions an appropriate stop criterion can generate information that is not present in the deterministic program computing it. Suppose we build a simple Turing machine that  counts from one to $2^k$, where $k$ is longer than the length of the index of our Turing machine. We let it run for a time in the order of $2^{k-1}$ and then we stop it at an ad hoc point in time and look at the value of the counter. With high probability the  value is a random number. So in a certain way the machine created new information. But it is clear that the decision to stop the machine comes from outside. We \emph{created} the information by choosing the time to stop it. It was never in the program.

I have baptized the related conundrum the \emph{Paradox of Systematic search} (see \cite{AB2011}, par. 6.1):  ``\emph{Any information that can be found by means of systematic search has no value, since we are certain to find it, given enough time}. In this context I have formulated the following conjecture of the existence of data compressing \emph{effective ad hoc descriptions}:

\begin{conjecture}\label{ADHOCDES}
``There exist numbers that are compressed by non-constructive unique effective descriptions, i.e. the validity of the description can be checked effectively given the number, but the number cannot be constructed effectively from the description, except by means of systematic search.'' (see \cite{AB2011}, par. 6.1)
\end{conjecture}

The $\mu$ operation becomes a total function when two conditions are added: 
\begin{enumerate}
\item there actually exists at least one $z$ such that $f(x,z) = 0$ (this is in fact Russell's solution in his theory of descriptions) and 
\item for every $y\prime \leq y$, the value $f(x,y\prime)$ exists and is positive (i.e. $f$ is a total function). 
\end{enumerate}

We are now in a position to specify the difference between construction and search in more detail: 

\begin{itemize}
\item \emph{Descriptive power versus computation time.} The more powerful your theory of computing is, the more efficient your descriptions of objects become. Primitive recursive functions are provably weaker than $\mu$-recursive functions so a description of an object in the former formalism will possibly be longer, say $k$ bits. The shorter descriptions in the context of $\mu$-recursive functions do not imply reduction in time necessary to compute the object: we still have to do a sequence $2^k$ of computations which takes exponential time. This explains the Meno paradox in computation. Search is fundamentally different from construction. The introduction of unbounded search operations in a computational system causes the descriptive complexity of the objects to decrease while the  time to compute those objects increases with the representational efficiency gained. 

\item \emph{General function schemes versus particular function names.} The name $g$ does not refer to a function but to a function-scheme. The $x$ in the expression $g(x)$ is not an argument of a function but the index of a function name $f_x(y) \Leftrightarrow f(x,y)$. We can interpret the $\mu$-operator as a meta-operator that has access to an infinite number of primitive recursive functions. In this interpretation there is no such thing as a general search routine.  Each search function is specific: searching for your glasses is different from searching for your wallet, even if you look for them in the same places. 
\end{itemize}

It is easy to define a version of Kolmogorov complexity in the context of differential information theory. All we need is an \emph{information theory for the set of general recursive functions}. Suppose $A$ is the set of recursive functions and we have a computable function $f: \mathbb{N} \rightarrow A$  such that $f(i) = r_i)$, then the amount of descriptive information in a number $x \in \mathbb{N}$ condtional to a number $y$ is:   
\begin{equation}\label{KOLDIFF}
I_f(x|y)= min_i \{ \log_2 i | f(i)=r_i \wedge r_i(y)=x  \}
\end{equation}

The unconditional version  is $I_f(x)= I_f(x|0)$. If we restrict ourselves to the primitive recursive functions, then we get a weaker constructive notion of descriptive complexity.

\section{Appendix: Computing the bijection between $\mathfrak{P}(\mathbb{N})$ and $\mathbb{N}$ via $\mathbb{N}^2$ using the Cantor pairing function: an example \label{A1}} 

\subsection{From $\mathfrak{P}(\mathbb{N})$ to $\mathbb{N}$}
Take a random finite set of natural numbers:  
\[\{1,4,6,8,10,11\}\]

\subsubsection{Compute the index in a combinatorial number system of order $6$}

\[\sigma_6(s)=  {11\choose 6} + {10 \choose 5}+ {8 \choose 4} + {6 \choose 3}+ {4 \choose 2} + {1 \choose 1} =\]

\[462 + 252 + 70 + 20 + 6 + 1 = 811\]

\subsubsection{Compute the Cantor index for  location $(5,811)$}

\[\pi(6,811) = 1/2 ((5 + 811  + 1) (5 + 811 )) + 811 = 334147\]

 \subsection{From $\mathbb{N}$ to $\mathfrak{P}(\mathbb{N})$}

Take the  index: $\pi(x,y) = 334147= z$. 

 \subsubsection{Compute the location for $334147$}

\[\pi(x,y)=1/2(x+y+1)(x+y)+y\]

Define\footnote{We follow the computation given in https://en.wikipedia.org/wiki/Pairing\_function, retrieved February 27, 2018. }.

\[w = x + y\]

\[t = \frac{w(w+1)}{2} = \frac{w^2 + w}{2}\]

\[z = t+y\]

Here $t$  is the triangular number of  $w$. We solve: 

\[w^2 + w -2t = 0\]

\[w = \frac{\sqrt{8 t + 1} - 1 }{2}\]

which is a strictly increasing and continuous function when $t$ is a non-negative real. Since

\[t \leq z =t+y < t+(w+1) = \frac{(w+1)^2 + (w+1)}{2}\]

we get that 

\[w \leq  \frac{\sqrt{ 8 z + 1} - 1}{2} < w + 1\]

and thus

\[w = \lfloor \frac{\sqrt{ (8 \times 334147) + 1} - 1}{2}\rfloor = 816\]

\[t = (817^2 + 817)/2 = 333336\]

\[y = z – t\]

\[y = 334147 - 333336 = 811\]

\[x = 816 - 811 = 5  \]

The set has cardinality  $5+1=6$ and index $811$. 

\subsubsection{Compute the set associated with location $(5,811)$.}

We reconstruct the number sequence from $\mathbb{N}^+$. Compute: ${x \choose (5+1)} < 811 = 11.802 \dots$. The biggest number is $11$ 

\[811 -  {11 \choose 6} = 349 \]

Compute: ${x \choose 5} < 349 = 10.521 \dots$

Next number  is $10$. 

\[349 -  {10 \choose 5} = 97\]

\[97 - {8 \choose 4} = 27\]

\[27 - {6 \choose 3} = 7\]

\[7 - {4 \choose 2} = 1\]

\[{1 \choose 1} = 1\]

The sequence is $1,4,6,8,10,11$.

\section{Appendix: Scale-free Subset Sum problems}\label{SCALEFSSPROB}

Suppose $A \subseteq \mathbb{N}$ is infinite and we want to know whether we can find a subset of $A$ that adds up to $n$. We can easily discard all elements of $A$ that are bigger than $n$, so for a specific instance we will only have to analyze an initial segment of $A$. Consequently an infinite subset sum problem defines an infinite number of finite versions. The dilation $\phi_{bin}$ forms the foundation of such an infinite version of subset sum based on the set of powers of two. These finite versions of subset sum are all easily computable, because they are based on the interpretation of binary numbers.\footnote{An empirical analysis of the distribution of information in subset sum problems is given in \cite{BERGADR21}.} This motivates the definition of the concept of a \emph{Canonical Subset Sum Problem}:

\begin{definition}\label{CANONICALSS}
A \emph{Canonical Subset Sum} problem $(P,n)$ has the following form: 
\begin{itemize}
\item $P = [2^0,2^1,\dots ]$ is the reference set of all powers of two,
\item $M$ is a finite typical subset of $P$ with index set $S$  and
\item  $n = \Sigma_{i \in M_2}i$ is a natural number. 
\end{itemize}
\end{definition}

A solution to the problem $(P,n)$  is a set $M \subset P$ with index function $S$ such that:   

\begin{equation}\Upsilon(S) = \Sigma_{i \in M} i =  n\label{CHECKCANSS}\end{equation}

Here $\Sigma_{i \in M} i $ is the checking function for $(P,n)$.  For canonical subset sum problems this equation specifies a bijection that is efficiently computable both ways. Yet the index of both sides varies considerably.  I give an example:

\begin{example}\label{SSEX1}
We generate a canonical subset sum problem $(P,n)$, with the following characteristics: 
\begin{itemize}
\item The set $P = [2^0,2^1,\dots ]$. 
\item The characteristic string $s = 0000010100010111110101$ which is generated by flipping a perfect coin $22$ times,
\item The  element $S = \{5,7,11,13,14,15,16,17,19,21\}$  of $\mathfrak{P}(\mathbb{N})$ which is the \emph{index set} defined by the characteristic string $s$. 
\item The associated subset of $P$ which is:  \[M = \{32, 128, 2048, 8192, 16384, 32768, 65536, 131072, 524288, 2097152\}\] 
Note that $M$ is with high probability a typical subset, i.e. a random subset of a set with maximal entropy (see par. \ref{TYPSET}). 
\item The sum of $M$ which is the number $n$ associated with $S$: $2877600$
\end{itemize}
For a typical set $M$ with $\Upsilon(S) = n$ and the scale of $n$ is $k$ we need according to theorem \ref{POLYINV} at least $k - 2\log k - 5$ extra bits to represent $\phi_{car}(S)$.    We check this computation for our example. We can compute the lexicographic index of the set $S = \{5,7,11,13,14,15,16,17,19,21\}$ using equation \ref{COMBSYST}: 

\begin{equation}\label{COMBSYSTEX}
\sigma_k(S)= {21 \choose 10} + \dots + {7 \choose 2} + {5 \choose 1}= 488757
\end{equation}
This results in the Cantor index $\pi(488757,10)= 119446834538$  with scale $36$. Which gives  $36 - 21 = 15$ extra bits. This is in line with the results of theorem \ref{POLYINV} that predicts that the Cantor index in this case will be at least 7 bits longer than the binary representation: $21 - 2\log 21 - 5 \approx 7$.
\end{example}

This shows that the checking function is not trivial: equation \ref{CHECKCANSS} defines a complex endomorphism for both $\mathbb{N}$ and $\mathbb{N}^2$. This explains part of the difficulties in exploring the search space of a subset sum problem. As soon as we start to manipulate sets in our search algorithms, we operate in a domain with an inherently inefficient information representation. This does not prove in itself that problems in $NP$ are hard, but the fact that this phenomenon already can be observed in the context of efficiently computable bijections, shows how fundamental this insight is. There is no direct connection between the representation of sets of numbers and the sums of these numbers, even if these values are derived from the \emph{same} representation.

\begin{definition}\label{SCALEFREESS}
A \emph{scale-free} subset sum problem $(C_{cb},m)$ based on the template $(P,n)$ is generated in the following way:
\begin{enumerate}
\item We create a randomized scale free-codebook $C_{cb}$ on the basis of the first $k$ powers of two, using the function $rnd(2^i)$ to generate element $i$ of $C_{cb}$.
\item We generate a typical subset $M \subset C_{cb}$ of size approximately $\frac{1}{2}k$ based on the index set $S$ of problem $(P,n)$.
\item We compute the sum $m$ of elements of $M$.
\item The resulting problem is $(C_{cb}, m)$.
\end{enumerate} 
\end{definition}

\begin{example}\label{SSEX2}
We generate a new problem on the basis of the efficiently solvable problem  $(P, n)$ from example \ref{SSEX1}. We transform the set $P = [2^0,2^1,\dots, 2^{21}]$  into a code book using the randomization function $rnd(2^i)$ from definition \ref{RND}. We get a codebook as specified in definition \ref{SCFENC}:

 \[C_{cb}= [0,1,2,7,12,9,40,50,48,420,874,1511,813,3987,8740,20506,26753,3244,\]\[22545,226247,331136,166612]\]
 
 Note that the elements in the list are not ordered on size. Using the code book $C_{cb}$ and the characteristic string $s = 0000010100010111110101$ from example \ref{SSEX1} we create a scrambled version of the original message set: 
\[M_2: \{9, 50, 1511, 3987, 8740, 20506, 26753, 3244, 226247, 166612\}\]
With a new sum m $= 457659$.

\begin{figure}[ht!]
\centering
\includegraphics[width=110mm]{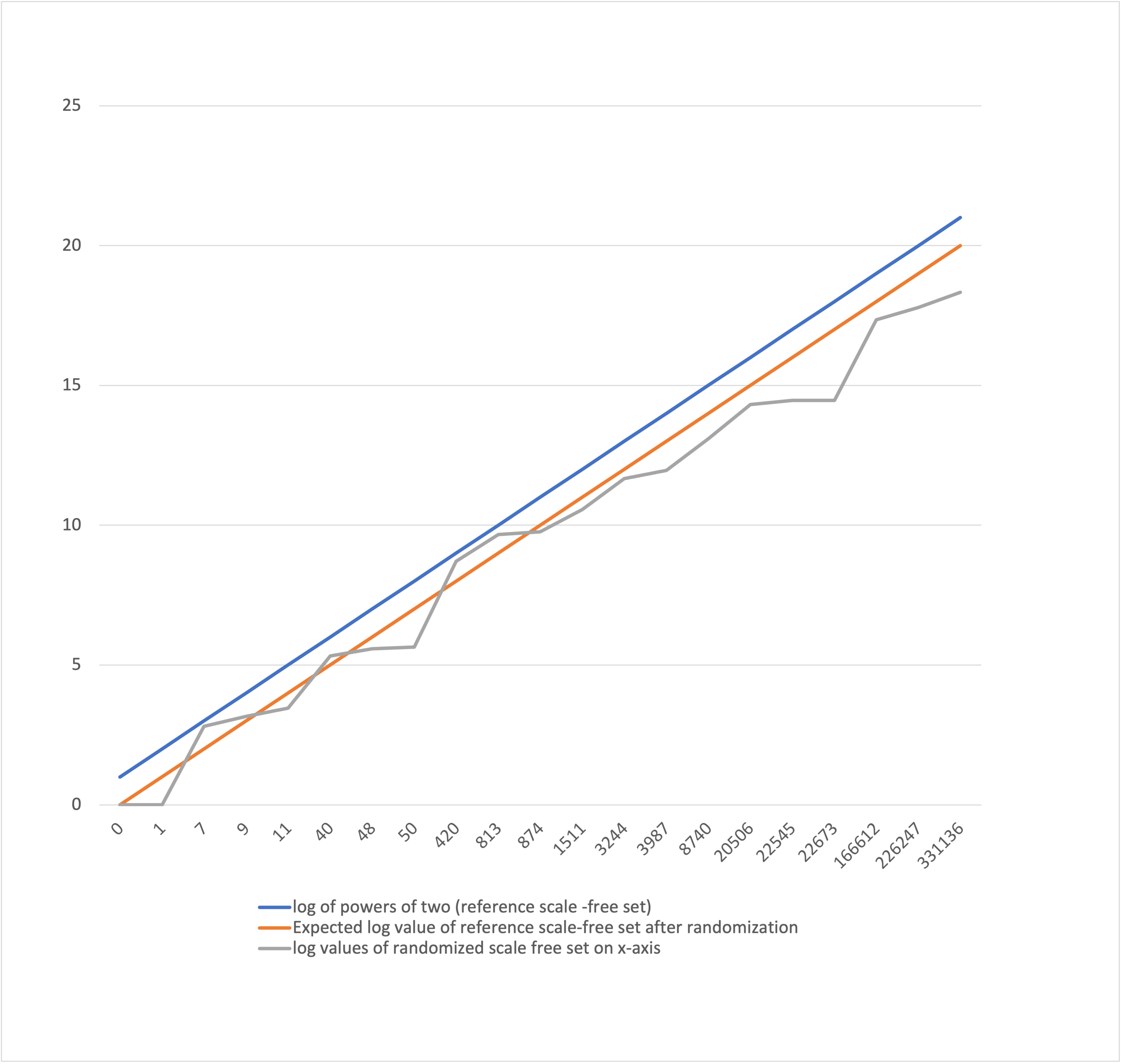}
\caption{The information in the elements of the scale-free set of example \ref{SSEX2} in ascending order (grey line). The upper diagonal line gives the information in the corresponding set of powers of two (blue line). The lower diagonal line gives the expected value of the scale free set selected under uniform distribution of the interval $[0,2^i]$.
 \label{INFSCALEFREE}}
\end{figure}

At first sight this appears to be just another example of a subset sum problem, but the special character of scale-free subset sum problems becomes clear when we trace the relationship with the original sums defined in terms of binary numbers. This is apparent  when we evaluate the scales of the underlying codebook: 
      \[C_{scale}= [undefined,0,1,2,3,3,5,5,5,8,9,10,9,11,13,14,14,11,14,17,18,17]\]
and the difference in scale with the original codebook based on powers of two: 
    \[C_{dif}= [-1,-1,-1,-1,-1,-2,-1,-2,-3 ,-1,-1,-1,-1,-2,-1,-1,-2,-5,\]\[-4,-2,-2,-4]\] 
With high probability the randomization process has destroyed at least one bit of information at every scale (see figure \ref{INFSCALEFREE}). Which leads tot the following claim: 

\begin{claim}
With high probability $m$ is an \emph{effective ad hoc description} in the sense of conjecture \ref{ADHOCDES} of the set $M_2$ given $C_{cb}$. 
\end{claim}

The intuitive motivations for the claim are clear: The expected value of $m$ is $\frac{1}{2}n$. The scale of the sum $m$ is $18$, just two bits below that of $n$, but the conditional description of the set $M_2$ is still given by the index set $S$ with an index of scale  $36$ and the characteristic string $s$ of length $22$. The objects $n$, the string $s$ and the associated set $S$ on which the computable bijection $\phi_{bin}$ depends are not known and any trace of their information has been destroyed at each information scale. 
\end{example}

\begin{figure}[ht!]
\centering
\includegraphics[width=110mm]{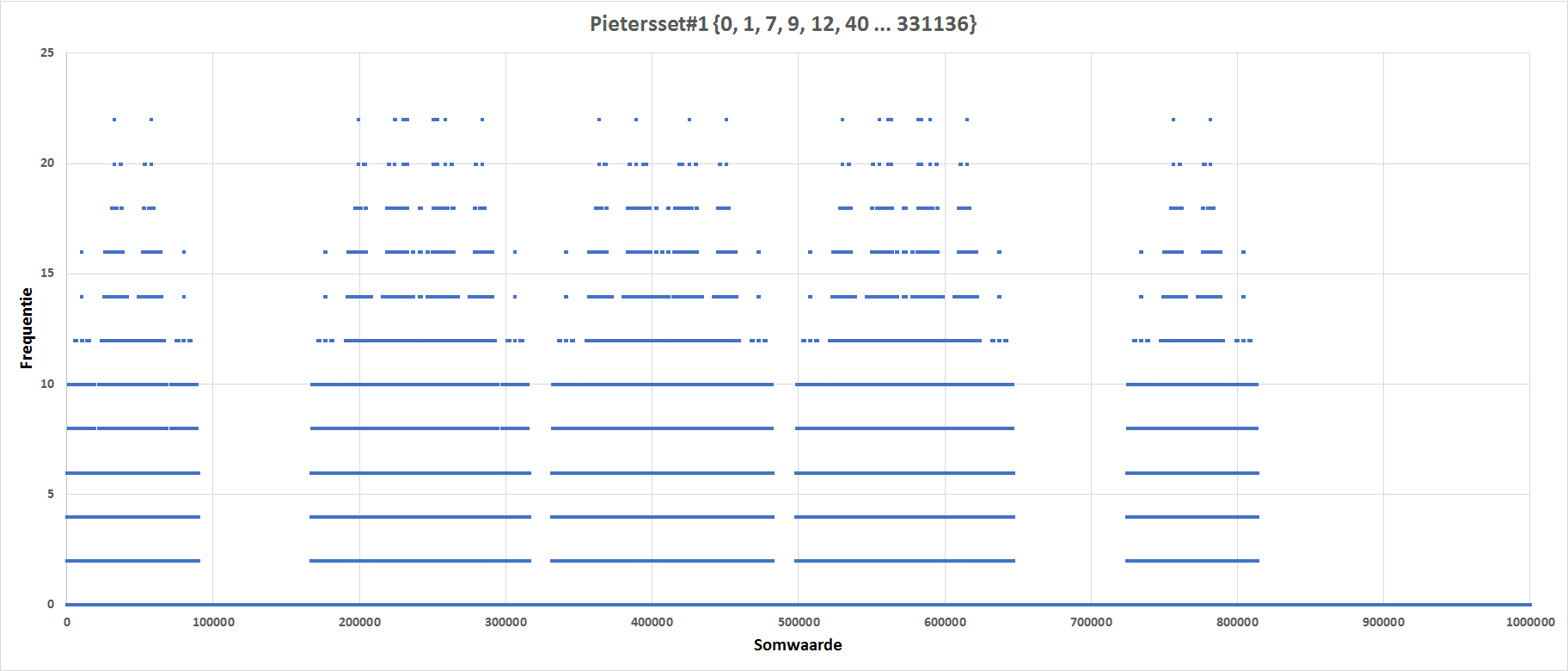}
\caption{An illustration of the fractal nature of the solution space of scale-free subset sum problems. The total solution space of the subset sum problem of example  \ref{SSEX2}: the number of solutions per sum from $0$ to $100000$.  Details at lower scales are given in figures \ref{SEG1SOLSP}, \ref{INTZOOMSOLSP} and \ref{INTZOOM1SOLSP}. \label{TOTALSOLSP}}
\end{figure}

\begin{figure}[ht!]
\centering
\includegraphics[width=110mm]{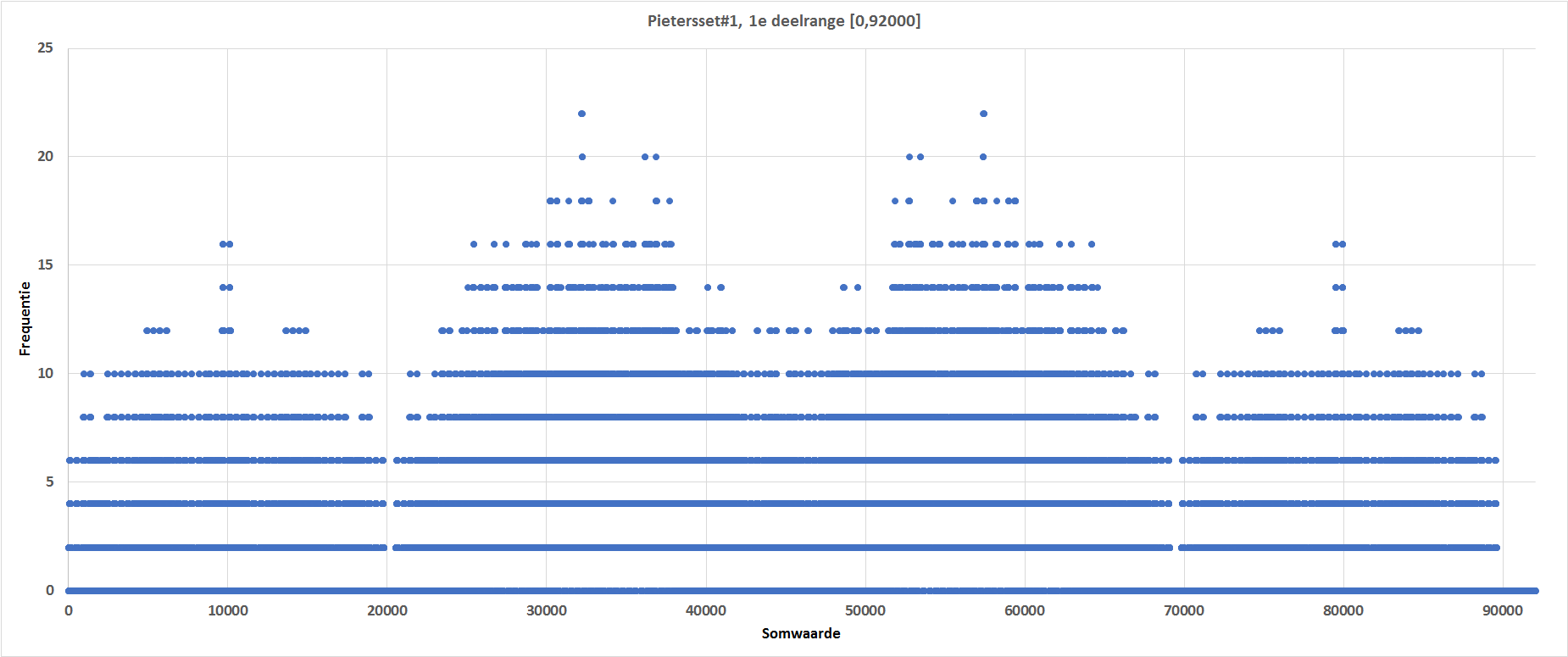}
\caption{Detail of figure \ref{TOTALSOLSP}. An overview of a segment solution space of the subset sum problem of example  \ref{SSEX2}: the number of solutions per sum from $0$ to $90000$.   \label{SEG1SOLSP}}
\end{figure}

\begin{figure}[ht!]
\centering
\includegraphics[width=110mm]{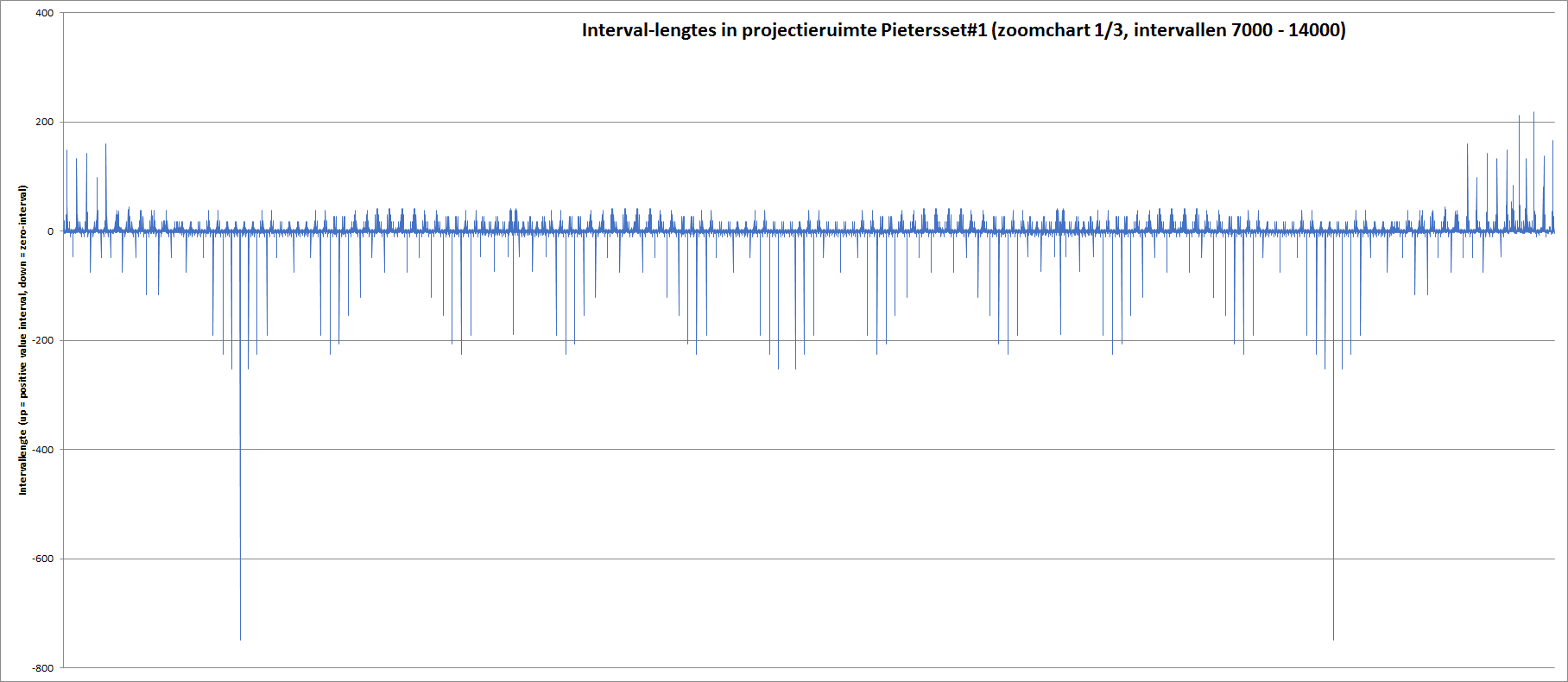}
\caption{An overview of a segment of the solution space of the subset sum problem of example  \ref{SSEX2}. Detail of figure \ref{SEG1SOLSP}: the length of the intervals between solutions from $7000$ to $14000$.   \label{INTZOOMSOLSP}}
\end{figure}

\begin{figure}[ht!]

\centering
\includegraphics[width=110mm]{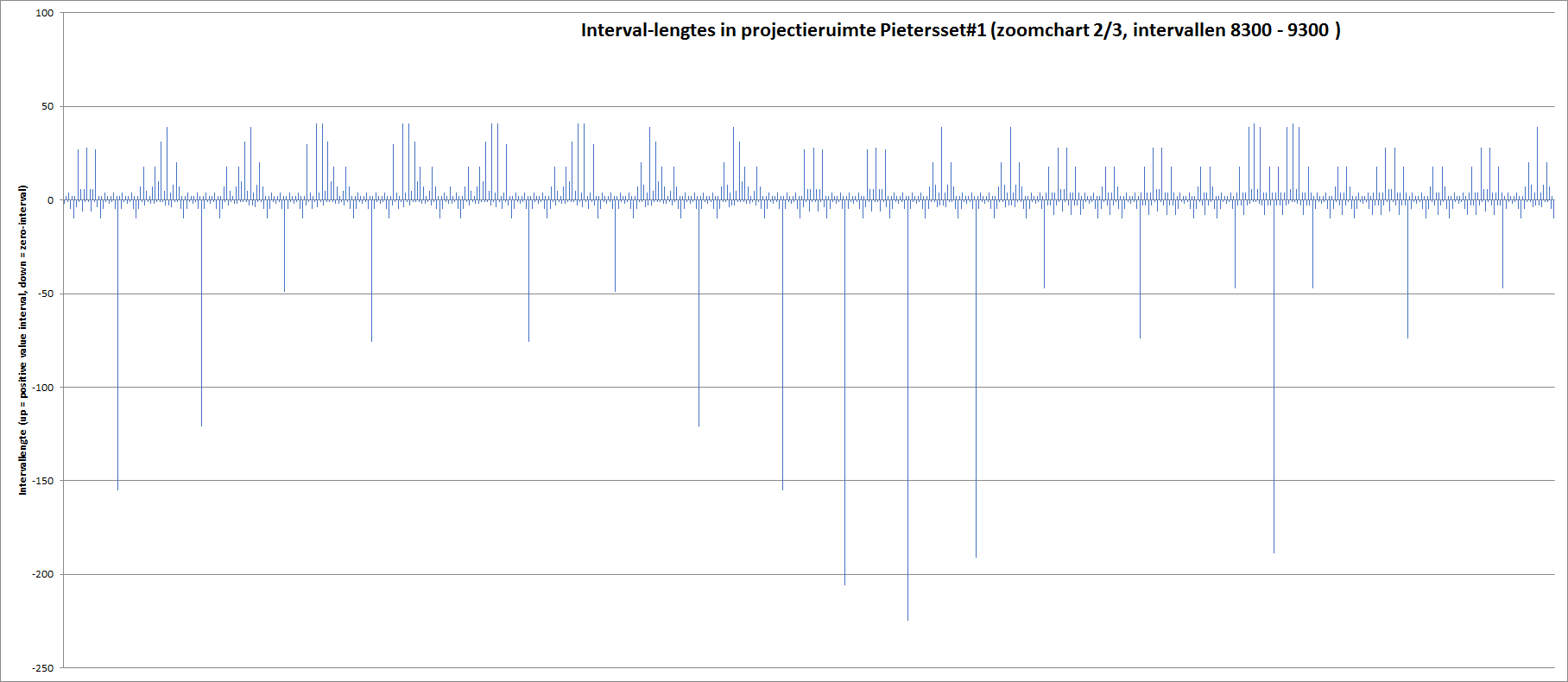}
\caption{An overview of the total solution space of the subset sum problem of example  \ref{SSEX2}. Detail of figure \ref{INTZOOMSOLSP}:the length of the intervals between solutions from $8300$ to $9300$ \label{INTZOOM1SOLSP}}
\end{figure}

The scale free nature of the solution space of the problem in example \ref{SSEX2} is illustrated in figures   \ref{TOTALSOLSP}, \ref{SEG1SOLSP}, \ref{INTZOOMSOLSP} and \ref{INTZOOM1SOLSP}.\footnote{I thank Daan van den Berg for implementing the example and generating these images. }



\section{Appendix: The entropy and information efficiency of logical operations on bits under maximal entropy of the input} \label{INFEFFRNDINP}
 
A useful proof technique is the analysis of the information efficiency of deterministic computational processes under maximum entropy input. In some cases we can compute the resulting information efficiencies exactly.  I give an example for logical operations.

\begin{example}\label{EXBITXOR} (Information efficiency of logical operation under max entropy input)\footnote{For a systematic overview of information efficiency of logical functions under random input see the appendix in paragraph \ref{INFEFFRNDINP}.} We will use the $\oplus$ sign to represent the logical XOR operation. The OR, AND and XOR functions are defined in table \ref{TABLOG}. The operations are \emph{commutative} and \emph{associative}. Under random input we can interpret these logical operations as systems of messages in Shannon's sense and compute the entropy and the information efficiency. Both the  AND and the OR operation have an inbalance between zeros and ones in the output and therefore discard more information than the XOR function.  Using Shannon's function, $H = - \Sigma_{i} p_i \log_2 (p_i)$, the entropy of logical operations on bits under random input is:

\begin{table}
\begin{center}
\begin{tabular}{ |c|c|c|} 
 \hline
 A & B & $\vee$ \\
\hline
1 & 1 & 1  \\ 
1 & 0 & 1  \\ 
0 & 1 & 1\\ 
 0 & 0 & 0 \\ 
 \hline
\end{tabular}
\quad 
\begin{tabular}{ |c|c|c|} 
 \hline
 A & B & $\wedge$ \\
\hline
1 & 1 & 1  \\ 
1 & 0 & 0  \\ 
0 & 1 & 0\\ 
 0 & 0 & 0 \\ 
 \hline
\end{tabular}
\quad
\begin{tabular}{ |c|c|c|} 
 \hline
 A & B & $\oplus$ \\
\hline
1 & 1 & 0  \\ 
1 & 0 & 1  \\ 
0 & 1 & 1\\ 
 0 & 0 & 0 \\ 
 \hline
\end{tabular}
\end{center}
\caption{The , OR, AND and XOR operation.}
 \label{TABLOG}
 \end{table}

\begin{equation}\label{ENTLOGRND1}
H(x \vee y))=H(x \wedge y)= (\frac{3}{4} \log_2 \frac{3}{4}) + (\frac{1}{4} \log_2 \frac{1}{4}) \approx 0.8
\end{equation}

\begin{equation}\label{ENTLOGRND3}
H(x \oplus y)= 2 (\frac{1}{2} \log_2 \frac{1}{2}) ) = 1
\end{equation}

The information efficiency of logical operations on bits under random input is: 

\begin{equation}\label{INEFFLOGRND1}
\Delta(x \vee y)= \Delta(x \wedge y)  \approx 0.8 -  2 \log_2 \frac{1}{2}=  -1.2
\end{equation}

\begin{equation}\label{INEFFLOGRND3}
\Delta(x \oplus y)=  1 - 2 \log_2 \frac{1}{2}=  -1
\end{equation}

Under random input we lose roughly $1.2$ bits when we perform the AND and the OR operation and $1$ bit when we perform the XOR operation. Since the operations are commutative and associative, they can be interpreted as operations on multisets of bits, irrespective of the order of the bits in a specific string. This implies that the first two operations performed on random strings are trivial. We know that random strings have roughly the same amount of zero and one bits. A sequence of OR operations on a random string will always be true, even if we flip a considerable number of bits. The same holds for a sequence of AND operations on a random string: it will always be false. Only the XOR operation remains maximally sensitive under bit flipping, even if the input is random.
\end{example}

\begin{equation}\label{ENTLOGRND1a}
H(x \vee y))= (\frac{3}{4} \log_2 \frac{3}{4}) + (\frac{1}{4} \log_2 \frac{1}{4}) \approx 0.8
\end{equation}

\begin{equation}\label{ENTLOGRND2}
H(x \wedge y)= H(x \vee y)
\end{equation}

\begin{equation}\label{ENTLOGRND3a}
H(x \oplus y)= 2 (\frac{1}{2} \log_2 \frac{1}{2}) ) = 1
\end{equation}

The information efficiency of logical operations on bits under random input is:  

\begin{equation}\label{INEFFLOGRND1a}
\Delta(x \vee y) \approx 0.8 -  2 \log_2 \frac{1}{2}=  -1.2
\end{equation}

\begin{equation}\label{INEFFLOGRND2}
\Delta(x \wedge y) = \Delta(x \vee y)
\end{equation}

\begin{equation}\label{INEFFLOGRND3a}
\Delta(x \oplus y)=  1 - 2 \log_2 \frac{1}{2}=  -1
\end{equation}

Entropy of logical operations on bit vectors under random input: 

\begin{equation}\label{ENTBITVGRND1}
H(x_1  \vee x_2 \vee \dots \vee x_k)= \frac{2^{k}-1}{2^k} \log_2 \frac{2^{k}-1}{2^k} +  \frac{1}{2^k} \log_2 \frac{1}{2^k} \approx \frac{k}{2^k} \approx 0
\end{equation}

\begin{equation}\label{ENTBITVRND2}
H(x_1  \wedge x_2 \wedge \dots \wedge x_k)=  H(x_1  \vee x_2 \vee \dots \vee x_k
\end{equation}

\begin{equation}\label{ENTBITVRND3}
H(x_1 \oplus x_2 \oplus \dots \oplus x_k)= 2 (\frac{1}{2} \log_2 \frac{1}{2}) ) = 1
\end{equation}

Information efficiency of logical operations on bit vectors under random input: 

\begin{equation}\label{INEFFBITVRND1}
\Delta(x_1  \vee x_2 \vee \dots \vee x_k) = H(x_1  \vee x_2 \vee \dots \vee x_k) - k \approx \frac{k}{2^k} - k \approx - k
\end{equation}

\begin{equation}\label{INEFFBITVRND2}
\Delta((x_1  \wedge x_2 \wedge \dots \wedge x_k))= \Delta(x_1  \vee x_2 \vee \dots \vee x_k)
\end{equation}

\begin{equation}\label{INEFFBITVRND3}
\Delta(x_1 \oplus x_2 \oplus \dots \oplus x_k) = H(x_1 \oplus x_2 \oplus \dots \oplus x_k)  - k = 1 - k
\end{equation}

Entropy of logical bitwise operations on bit vectors of length k under random input: 

\begin{equation}\label{ENTBITWBITVGRND1}
H(\overline{x} \vee \overline{y}) \approx  0.8k
\end{equation}

\begin{equation}\label{ENTBITWBITVRND2}
H(\overline{x} \vee \overline{y})= H(\overline{x} \wedge \overline{y}))
\end{equation}

\begin{equation}\label{ENTBITWBITVRND3}
H(\overline{x} \oplus \overline{y})= k
\end{equation}

Information efficiency of logical bitwise operations on bit vectors of length k under random input: 

\begin{equation}\label{INEFFBITWBITVRND1}
\Delta(\overline{x} \vee \overline{y}) = H(\overline{x} \vee \overline{y}) - 2k  \approx  - 1.2k
\end{equation}

\begin{equation}\label{INEFFBITWBITVRND2}
\Delta(\overline{x} \vee \overline{y})= \Delta(\overline{x} \wedge \overline{y}))
\end{equation}

\begin{equation}\label{INEFFBITWBITVRND3}
\Delta(\overline{x} \oplus\overline{y})= H(\overline{x} \oplus\overline{y}) -2k = -k
\end{equation}

The operations can be extended to bit strings of equal length: let $a$ and $b$ be such strings, the expression $(a \ \texttt{XOR} \ b) = c$ indicates that $c$ is the result of applying the $XOR$ operation pairwise to the bits of $a$ and $b$.

 Entropy of logical bitwise operations on sets of bit vectors of length k of cardinality n under random input:

\begin{equation}\label{ENTSETBITWBITVGRND1}
H(\overline{x_1} \vee \overline{x_2} \vee \dots \vee \overline{x_n})  \approx k \frac{n}{2^n}  \approx 0
\end{equation}

\begin{equation}\label{ENTSETBITWBITVRND2}
H(\overline{x_1} \vee \overline{x_2} \vee \dots \vee \overline{x_n}) = H(\overline{x_1} \wedge \overline{x_2} \wedge \dots \wedge \overline{x_n}) )
\end{equation}

\begin{equation}\label{ENTSETSBITWBITVRND3}
H(\overline{x_1} \oplus \overline{x_2} \oplus  \dots \oplus \overline{x_n})= k
\end{equation}

Information efficiency of logical bitwise operations on sets of bit vectors of length k of cardinality n under random input:

\begin{equation}\label{INEFFSETSBITWBITVRND1}
\Delta(\overline{x_1} \vee \overline{x_2} \vee \dots \vee \overline{x_n})  =  H(\overline{x_1} \vee \overline{x_2} \vee \dots \vee \overline{x_n})  - kn \approx -kn 
\end{equation}

\begin{equation}\label{INEFFSETSBITWBITVRND2}
\Delta(\overline{x} \vee \overline{y})= \Delta(\overline{x} \wedge \overline{y}))
\end{equation}

\begin{equation}\label{INEFFSETSBITWBITVRND3}
\Delta(\overline{x_1} \oplus \overline{x_2} \oplus  \dots \oplus \overline{x_n}) =  H(\overline{x_1} \oplus \overline{x_2} \oplus  \dots \oplus \overline{x_n}) -nk = k - kn 
\end{equation}

\end{document}